\documentclass[11pt,a4paper]{article}

\usepackage[utf8]{inputenc}
\usepackage{amsmath,amssymb,amsthm}
\usepackage{geometry}
\usepackage{graphicx}
\usepackage{subcaption}
\usepackage{booktabs}

\newcommand{\panelwidth}{0.86\textwidth}
\newcommand{\panelheight}{0.2\textheight}
\newcommand{\panel}[1]{%
  \includegraphics[width=\panelwidth,height=\panelheight,keepaspectratio]{#1}}

\newcommand{\emppanelwidth}{0.95\textwidth}
\newcommand{\emppanel}[1]{%
  \includegraphics[width=\emppanelwidth,height=0.265\textheight,keepaspectratio]{#1}}

\graphicspath{{convergenceTest/}{approximationTest/}{shiftedGammaTest/}{empiricalTest/}}

\newtheorem{theorem}{Theorem}[section]
\newtheorem{proposition}[theorem]{Proposition}
\newtheorem{lemma}[theorem]{Lemma}
\newtheorem{corollary}[theorem]{Corollary}
\newtheorem{remark}{Remark}[section]
\newtheorem{definition}{Definition}[section]

\DeclareMathOperator{\Var}{Var}

\newcommand{\E}{\mathbb{E}}
\newcommand{\Q}{\mathbb{Q}}
\newcommand{\R}{\mathbb{R}}
\newcommand{\one}{\mathbf{1}}

\title{Beyond Strike Remapping:\\Conditional-Gamma Pricing of Leveraged ETF Options under Local Volatility}
\author{%
Guolin Chen\thanks{Opinions expressed in this paper are those of the authors, and do not necessarily reflect the view of JP Morgan.} \quad Jiang Zhou\footnotemark[1] \quad Richard Ma\footnotemark[1]\\
Xia Sheng\footnotemark[1] \quad Anson Tse\footnotemark[1] \quad Terry Wei\footnotemark[1]\\[5pt]
\textit{Quantitative Trading \& Research, JPMorganChase}\\[3pt]
{\small\texttt{collin.chen@jpmorgan.com} \quad \texttt{jiang.zhou@jpmorgan.com}}\\
{\small\texttt{richard.j.ma@jpmorgan.com} \quad \texttt{vincent.sheng@jpmorgan.com}}\\
{\small\texttt{anson.ht.tse@jpmorgan.com} \quad \texttt{terry.x.wei@jpmorgan.com}}}
\date{\today}

\begin{document}
\maketitle

\begin{abstract}
Options on leveraged ETFs (LETFs) are priced and hedged off the LETF volatility surface, but path-dependent variance drag makes that surface difficult to build consistently with the underlying index smile. We study three methods for constructing it: a closed-form first-order strike remapping, a new conditional-Gamma method, and a two-dimensional joint-law PDE that makes no distributional assumption and serves as the benchmark. The conditional-Gamma method is our main contribution. It conditions on the terminal underlying return, which preserves the state dependence of variance drag that a single remapped strike discards. Three one-dimensional forward PDEs under a calibrated local-volatility model give the conditional mean and variance of integrated variance, and a Gamma law matched to these moments keeps realised variance non-negative and prices the option in closed form. A shifted-Gamma variant that also matches the conditional third moment gives its clearest gains for inverse products at longer maturities, and a Gamma-mixture extension covers the full range of conditional skewness compatible with non-negative realised variance. On S\&P~500 and Nikkei~225 surfaces the conditional-Gamma smile tracks the benchmark closely at every leverage tested, including the wings where the strike remapping drifts. A full LETF smile takes $0.85$--$1.42$ seconds against $197$--$224$ seconds for the benchmark, a 158- to 236-fold speed-up. Against listed TQQQ quotes the conditional-Gamma smile lies inside the market bid--ask at most strikes.
\end{abstract}

%==============================================================================
\section{Introduction}
%==============================================================================

Leveraged exchange-traded funds (LETFs) target a constant daily multiple $\beta$ of the return of an underlying index. In the continuous-rebalancing idealisation, maintaining that leverage generates a path-dependent \emph{variance drag}: the terminal LETF value is not a simple power of the terminal underlying but carries an exponential of realised variance. Concretely, after deterministic financing and carry have been removed, the forward-normalised LETF log-return obeys the exact identity
\[
R_L=\beta R_S-\gamma I_T,
\qquad \gamma=\frac{\beta(\beta-1)}{2},
\]
where $R_S$ is the underlying forward log-return and $I_T=\int_0^T\sigma_{\mathrm{inst}}^2\,dt$ is realised integrated variance. This factorisation, established by Avellaneda and Zhang \cite{avellaneda_zhang2010}, is the starting point for pricing options on LETFs. The practical question is how the LETF implied-volatility smile relates to the smile of the underlying, given a calibrated local-volatility (LV) model. Discrete rebalancing, tracking error, jumps, and random fees are outside the present idealisation.

\subsection{Related work and contribution}

The first-order relationship is well understood. Leung and Sircar \cite{leung_sircar2015} and Leung and Santoli \cite{leung_santoli2016} derive moneyness-scaling relations that map the LETF smile to the underlying smile. Ahn, Haugh and Jain \cite{ahn_haugh_jain2015} study model-consistent LETF option pricing using transforms, moment matching, and saddlepoint methods, while Leung, Lorig and Pascucci \cite{leung_lorig_pascucci2017} derive price and implied-volatility expansions under general local-stochastic volatility dynamics. Section~\ref{sec:first-LV} recovers the leading-order strike remapping under local volatility by a Brownian-bridge argument built on Bergomi's exact gamma-weighted implied-variance representation and its first-order expansion \cite[Ch.~2]{bergomi_smiledynamics}, together with Gatheral's most-probable-path approximation \cite[Ch.~3]{gatheral2006}; the derivation requires no separability or parametric assumption on the local-volatility surface and is collected in Appendix~\ref{app:first-order-proof}.

Our main construction conditions on the terminal underlying forward log-return, denoted $X_T$ (so $X_T=R_S$), and models the remaining variance drag through the conditional law of $I_T$ given $X_T$. Since $R_L=\beta R_S-\gamma I_T$, fixing $X_T$ leaves only the drag random, so conditioning keeps the link between each LETF strike and the underlying outcomes that determine its payoff. Conditional integrated variance is used in affine-model simulation and approximation; see Broadie and Kaya \cite{broadie_kaya2006} and the Gamma expansion of Glasserman and Kim \cite{glasserman_kim2011}. Here, under a calibrated LV model, the conditional moments of $I_T$ given $X_T$ follow from one-dimensional forward PDEs for the partial moments $\E[I_T^n\one_{\{X_T\in dx\}}]$, a moment hierarchy rooted in Kac's treatment of additive functionals \cite{kac1949}. A Gamma distribution matched to the first two moments keeps $I_T$ non-negative and, entering the LETF payoff directly, gives the conditional price in closed form. The LETF identity \cite{avellaneda_zhang2010}, the gamma-weighted implied-variance representation with its most-probable-path approximation \cite{gatheral2006,bergomi_smiledynamics}, density recovery from option prices \cite{breeden_litzenberger1978}, conditional moments of integrated variance and Gamma closures are each established tools. Our contribution is to combine them under a calibrated LV surface, which yields an LETF smile that retains the state dependence of variance drag at the cost of a few one-dimensional PDE solves per maturity.

\subsection{Structure of the paper}

Sections~\ref{sec:setup}--\ref{sec:first-LV} establish the factorisation and the first-order strike map, and Appendix~\ref{app:first-order-proof} contains its Brownian-bridge derivation. Sections~\ref{sec:conditional-setup}--\ref{sec:joint-law} derive the conditional pricing formulas, the forward moment equations and the joint-law benchmark, and Section~\ref{sec:algorithm} summarises the implementation. Section~\ref{sec:numerics} tests the numerical solver, compares the approximations, tests them against listed TQQQ quotes and reports their cost, and Section~\ref{sec:conclusion} concludes. Appendix~\ref{app:gamma-mixture} discusses an alternative fit for conditional skewness outside the shifted-Gamma admissible region.

%==============================================================================
\section{Setup and Notation}\label{sec:setup}
%==============================================================================

\subsection{Underlying Asset}

Throughout, asset prices are strictly positive, all expectations are under the pricing measure $\Q$, and rates, carry, and contractual financing terms are deterministic. We assume sufficient regularity and integrability for the stochastic exponentials below to be true martingales and for all displayed moments to be finite.

\begin{definition}[Local-volatility dynamics]
Let $S_t$ denote the underlying asset price satisfying
\begin{equation}\label{eq:LV}
\frac{dS_t}{S_t} = \mu_t^S\,dt + \sigma(t,S_t)\,dW_t,
\end{equation}
where $\mu_t^S$ is a deterministic drift (determined by rates, dividends, and borrow costs), $\sigma(t,s)$ is the local volatility function, and $W_t$ is a standard Brownian motion.
\end{definition}

\subsection{Leveraged ETF}

A leveraged ETF with constant leverage ratio $\beta$ satisfies \cite{avellaneda_zhang2010,leung_sircar2015,leung_santoli2016}
\begin{equation}\label{eq:LETF}
\frac{dL_t}{L_t}
=\beta\,\frac{dS_t}{S_t}+\phi_t\,dt,
\end{equation}
where $\phi_t$ collects deterministic financing costs and fees. Thus the arithmetic-return drift is $\mu_t^L=\beta\mu_t^S+\phi_t$. The It\^o variance correction is \emph{not} included in $\phi_t$: it appears only after taking logarithms and is precisely the random variance-drag term in the factorisation below.

\subsection{Forward Levels and Normalisation}

Set $F_0^S=S_0$ and $F_0^L=L_0$, and define deterministic forward levels by
\begin{equation}
\frac{dF_t^S}{F_t^S} = \mu_t^S\,dt, \qquad \frac{dF_t^L}{F_t^L} = \mu_t^L\,dt,
\end{equation}
with forward-normalised processes
\begin{equation}
\widetilde S_t=\frac{S_t}{F_t^S},
\qquad
\widetilde L_t=\frac{L_t}{F_t^L}.
\end{equation}
By the standing true-martingale assumption,
\begin{equation}
\E[\widetilde S_t]=\E[\widetilde L_t]=1,
\end{equation}
so $F_t^S$ and $F_t^L$ are the corresponding time-zero forward levels. Let $D_t$ denote the deterministic discount factor from time $t$ to time zero, with $D_0=1$.

\subsection{Volatility Objects}

We distinguish:
\begin{itemize}
\item $\sigma(t,s)$: local volatility of the underlying (function of time and spot),
\item $\sigma_S(T,K)$: Black implied volatility of the underlying at maturity $T$ and strike $K$,
\item $\hat{\sigma}_L(T,K)$: Black implied volatility of the LETF at maturity $T$ and strike $K$.
\end{itemize}

\subsection{Log-Moneyness}

\begin{equation}
x_K = \log\!\left(\frac{K}{F_T^L}\right), \qquad x_K^S = \log\!\left(\frac{K}{F_T^S}\right).
\end{equation}

%==============================================================================
\section{Exact LETF Factorisation}\label{sec:factorisation}
%==============================================================================

\begin{proposition}[Pathwise factorisation]\label{prop:factor}
Under the LV dynamics described above,
\begin{equation}\label{eq:factor}
\widetilde L_t=\widetilde S_t^{\;\beta}V_t,
\end{equation}
where
\begin{equation}\label{eq:Vt}
V_t = \exp\!\left\{-\frac{\beta(\beta-1)}{2}\int_0^t \sigma_{\mathrm{inst}}^2(u)\,du\right\},
\end{equation}
with $\sigma_{\mathrm{inst}}^2(u)=\sigma^2(u,S_u)$.
\end{proposition}

\begin{proof}
The forward-normalised processes satisfy
\[
\frac{d\widetilde S_t}{\widetilde S_t}=\sigma_{\mathrm{inst}}(t)\,dW_t,
\qquad
\frac{d\widetilde L_t}{\widetilde L_t}=\beta\sigma_{\mathrm{inst}}(t)\,dW_t.
\]
By It\^o's lemma applied to $f(x) = x^\beta$:
\[
\frac{d(\widetilde S_t^{\;\beta})}{\widetilde S_t^{\;\beta}}
=\beta\sigma_{\mathrm{inst}}\,dW_t
+\frac{\beta(\beta-1)}{2}\sigma_{\mathrm{inst}}^2\,dt.
\]
Define $V_t$ by \eqref{eq:Vt}, so that $dV_t/V_t = -\frac{\beta(\beta-1)}{2}\sigma_{\mathrm{inst}}^2\,dt$. Since $V_t$ has zero quadratic variation, the product rule gives
\[
\frac{d(\widetilde S_t^{\;\beta}V_t)}{\widetilde S_t^{\;\beta}V_t}
=\beta\sigma_{\mathrm{inst}}\,dW_t.
\]
This equals the SDE for $\widetilde L_t$. Both processes start at 1, so pathwise uniqueness gives \eqref{eq:factor}.
\end{proof}

\begin{corollary}[Spot inversion]
For $\beta\neq0$,
\begin{equation}\label{eq:inversion}
S_t = F_t^S\left(\frac{L_t}{F_t^L\,V_t}\right)^{\!1/\beta}.
\end{equation}
\end{corollary}

\begin{definition}[Induced LETF instantaneous variance]\label{def:induced}
Once the variance-drag state $z$ is specified, the instantaneous variance of $L_t$ at level $l$ is
\begin{equation}\label{eq:induced-LV}
b^{LV}(t,l,z)
= \beta^2\,\sigma^2\!\left(t,\;F_t^S\left(\frac{l}{F_t^L\,z}\right)^{\!1/\beta}\right).
\end{equation}
\end{definition}

%==============================================================================
\section{First-Order Result under Local Volatility}\label{sec:first-LV}
%==============================================================================

\subsection{Base Model and Perturbation}

We expand around a flat Black reference model. Choose a constant variance $a_0>0$; a natural practical choice is the target at-the-money-forward (ATMF) implied variance $\sigma_S^2(T,F_T^S)$. Introduce the formal perturbation family
\begin{equation}
\sigma_\varepsilon^2(t,s)=a_0+\varepsilon a_1(t,s),
\qquad 0\leq\varepsilon\leq1,
\end{equation}
where $\varepsilon=0$ is the flat base and $\varepsilon=1$ is the target surface. The induced LETF instantaneous variance has the corresponding expansion
\begin{equation}
 b_\varepsilon(t,l,z)=b_0+\varepsilon b_1(t,l,z),
 \qquad
 b_0=\beta^2a_0,
\end{equation}
where $b_1$ is $\beta^2a_1$ evaluated at the spot recovered from $(l,z)$ by \eqref{eq:inversion}. The base LETF volatility is $\sigma_0=|\beta|\sqrt{a_0}$. In what follows, implied volatilities refer to the $\varepsilon$-model and the error is understood as $\varepsilon\to0$.

The full derivation, comprising the gamma-weighted implied-variance representation, the first-order cancellation and the Brownian-bridge strike map, is collected in Appendix~\ref{app:first-order-proof}.
\begin{theorem}[First-order strike remapping under LV]\label{thm:first-LV}
Fix $T$, $K$, and $\beta\neq0$. Under a regular perturbation for which the gamma-weighted expectations in Appendix~\ref{app:first-order-proof} are differentiable at $\varepsilon=0$,
\begin{equation}\label{eq:main-first}
\hat{\sigma}_L^2(T,K)=\beta^2\sigma_S^2(T,K^*)+O(\varepsilon^2),
\end{equation}
where
\begin{equation}
K^*=F_T^S\left(\frac{K}{F_T^L V_T^*}\right)^{1/\beta},
\qquad
V_T^*=\exp\!\left\{-\frac{\beta(\beta-1)}{2}a_0T\right\}.
\end{equation}
\end{theorem}

\begin{remark}[Scope of the remainder]
The $O(\varepsilon^2)$ statement is a formal, pointwise expansion at fixed $(T,K)$. No uniform wing estimate is claimed: the gamma-weight denominator becomes small at extreme strikes, so additional bounds would be required for a uniform result.
\end{remark}

%==============================================================================
\section{Setup for the Conditional Method}\label{sec:conditional-setup}
%==============================================================================

We condition on the terminal underlying return and approximate the remaining law of $I_T$. Section~\ref{sec:conditional-gamma} fits its conditional mean and variance with a Gamma distribution. Section~\ref{sec:shifted-gamma} adds a non-negative shift when the measured conditional skewness permits one.

Throughout, we work under Dupire local volatility \cite{dupire1994,gatheral2006} and reuse the notation of Section~\ref{sec:setup}. It is convenient to write the log-forward dynamics as
\begin{equation}\label{eq:X-dynamics}
dX_t=-\frac12 a(t,X_t)\,dt+\sqrt{a(t,X_t)}\,dW_t,
\qquad
a(t,x)=\sigma^2(t,F_t^S e^x),
\end{equation}
where $X_t=\log\widetilde S_t$ and $R_S=X_T$. The exact factorisation of Proposition~\ref{prop:factor} then gives the LETF log-return
\begin{equation}\label{eq:RL-identity}
R_L=\log \widetilde L_T
=\beta R_S-\gamma I_T,
\qquad
\gamma=\frac{\beta(\beta-1)}{2},
\qquad
I_T=\int_0^T a(t,X_t)\,dt,
\end{equation}
with $\E[e^{R_L}]=1$. At a fixed terminal state $X_T=x$, LETF pricing depends on the conditional law of $I_T$ and retains the strike dependence carried by $x$.

%==============================================================================
\section{Exact Conditional Representation}
%==============================================================================

Let $p_T(x)$ denote the $\Q$-density of $X_T=R_S$. For an LETF strike $K$, define forward log-moneyness
\begin{equation}\label{eq:k-def}
k=\log\left(\frac{K}{F_T^L}\right).
\end{equation}
For the exact model no additional normaliser is required.

\begin{definition}[Conditional integrated-variance moments]\label{def:conditional-moments}
For $x\in\R$, define
\begin{equation}\label{eq:m-v-def}
m(x)=\E[I_T\mid X_T=x],
\quad
v(x)=\Var(I_T\mid X_T=x),
\quad
\mu_3(x)=\E\!\left[\big(I_T-m(x)\big)^3\;\middle|\;X_T=x\right],
\end{equation}
and, where $v(x)>0$, the conditional coefficient of variation and conditional skewness
\begin{equation}\label{eq:cv-skew-def}
\mathrm{cv}(x)=\frac{\sqrt{v(x)}}{m(x)},
\qquad
\lambda_3(x)=\frac{\mu_3(x)}{v(x)^{3/2}}.
\end{equation}
\end{definition}

The Gamma fit uses $m$ and $v$, while its shifted extension also uses $\mu_3$. Section~\ref{sec:moment-pdes} derives all three quantities from one forward hierarchy.

Conditioning the LETF payoff on $X_T$ gives
\begin{equation}\label{eq:exact-conditional-price}
C_L(K,T)
=D_T\int_{\R}
\E\left[\left(F_T^L e^{\beta x-\gamma I_T}-K\right)^+\mid X_T=x\right]
p_T(x)\,dx,
\end{equation}
where $D_T$ is the discount factor.

\begin{remark}[Source of strike dependence]
An unconditional cumulant expansion approximates the distribution of $R_L$ directly. It can match ATM level, skew, and curvature locally, but it weakens the link between strike $K$ and terminal underlying state $x$. Formula \eqref{eq:exact-conditional-price} keeps that link: each strike selects different regions of the terminal density through the function $\beta x-\gamma I_T$.
\end{remark}

%==============================================================================
\section{Conditional-Gamma Closure}\label{sec:conditional-gamma}
%==============================================================================

The realised variance is a non-negative additive functional,
\begin{equation}
I_T=\int_0^T a(t,X_t)\,dt\geq0.
\end{equation}
Even after conditioning on $X_T=x$, its law need not be symmetric and is often right-skewed. A Gaussian closure would place artificial mass on $I_T<0$; because the LETF payoff contains $e^{-\gamma I_T}$, that mass can create disproportionately large terminal LETF values when $\gamma>0$. We approximate the conditional law by a Gamma distribution that matches the conditional mean and variance obtained from the moment PDEs. The actual conditional skewness $\lambda_3(x)$ is itself computable, from the same hierarchy extended to $M_3$, and Section~\ref{sec:shifted-gamma} uses it to relax the fit where the Gamma shape is too rigid.

For $m(x)>0$ and $v(x)>0$, define the shape and scale
\begin{equation}\label{eq:gamma-parameters}
\alpha(x)=\frac{m(x)^2}{v(x)},
\qquad
\vartheta(x)=\frac{v(x)}{m(x)}.
\end{equation}
The moment-matched closure is
\begin{equation}\label{eq:conditional-gamma}
I_T\mid X_T=x\ \approx\ G_x,
\qquad
G_x\sim\operatorname{Gamma}\big(\alpha(x),\vartheta(x)\big),
\end{equation}
where the second parameter is the scale, so that
\begin{equation}
f_{G_x}(i)=\frac{i^{\alpha(x)-1}e^{-i/\vartheta(x)}}
{\Gamma(\alpha(x))\vartheta(x)^{\alpha(x)}}\one_{\{i>0\}}.
\end{equation}
This construction gives $\E[G_x]=m(x)$ and $\Var(G_x)=v(x)$. We refer to the resulting scheme---the moment PDEs of Section~\ref{sec:moment-pdes} followed by the Gamma fit---as the \emph{conditional-Gamma method}. Let
\begin{equation}\label{eq:regularized-gamma}
\mathcal P(a,z)=\frac{1}{\Gamma(a)}\int_0^z t^{a-1}e^{-t}\,dt
\end{equation}
denote the regularised lower incomplete gamma function.

\subsection{Motivation for the Gamma closure}

A Gamma law respects the pathwise constraint $I_T\geq0$ and can represent the right skew commonly associated with integrated variance. It also gives a parsimonious continuous approximation to an accumulation of non-negative, though dependent, contributions. Gamma-based representations are well established for integrated CIR variance in the Heston model \cite{glasserman_kim2011}, while the exact conditional law used in affine-model simulation is already non-trivial \cite{broadie_kaya2006}. Finally, its Laplace transform and truncated Laplace transform are explicit, so the conditional LETF call retains a closed form despite the payoff kink.

The closure also has the correct small-dispersion behaviour. Its conditional coefficient of variation and skewness are
\begin{equation}\label{eq:gamma-cv-skew}
\frac{\sqrt{\Var(G_x)}}{\E[G_x]}=\frac{1}{\sqrt{\alpha(x)}},
\qquad
\operatorname{Skew}(G_x)=\frac{2}{\sqrt{\alpha(x)}}
=2\frac{\sqrt{v(x)}}{m(x)}.
\end{equation}
As $v(x)/m(x)^2\to0$, $\alpha(x)\to\infty$ and the Gamma law concentrates at $m(x)$, with a locally Gaussian central region but without introducing negative variance. This remains a two-moment closure: it imposes the skewness relation $\lambda_3=2\,\mathrm{cv}$ in \eqref{eq:gamma-cv-skew} and places the lower edge of the fitted law at zero. Section~\ref{sec:shifted-gamma} uses a third moment to relax both restrictions where a non-negative shift is admissible. Appendix~\ref{app:gamma-mixture} gives an unrestricted three-moment alternative, and the joint-law equation remains the closure-free reference.

\subsection{Closed-form conditional call value}

We first consider the standard LETF cases $\beta>1$ and $\beta<0$, for which
\begin{equation}
\gamma=\frac{\beta(\beta-1)}{2}>0.
\end{equation}
Let $k=\log(K/F_T^L)$ and define the conditional exercise boundary in realised variance,
\begin{equation}\label{eq:gamma-boundary}
h(x)=\frac{\beta x-k}{\gamma}.
\end{equation}
Conditional on $X_T=x$, the LETF call is in the money precisely when $G_x<h(x)$. The two truncated Gamma quantities required for pricing are
\begin{align}
\Q(G_x<h\mid X_T=x)
&=\mathcal P\!\left(\alpha(x),\frac{h}{\vartheta(x)}\right),\label{eq:gamma-cdf}\\
\E\!\left[e^{-\gamma G_x}\one_{\{G_x<h\}}\mid X_T=x\right]
&=\big(1+\gamma\vartheta(x)\big)^{-\alpha(x)}
\mathcal P\!\left(
\alpha(x),\frac{(1+\gamma\vartheta(x))h}{\vartheta(x)}
\right).\label{eq:gamma-truncated-laplace}
\end{align}

\begin{theorem}[Conditional-Gamma LETF price]\label{thm:conditional-gamma-price}
Under approximation \eqref{eq:conditional-gamma}, with $\gamma>0$, define
\begin{align}\label{eq:gamma-positive-call}
\mathcal A_{\Gamma}(x;K)
={}&F_T^Le^{\beta x}\big(1+\gamma\vartheta(x)\big)^{-\alpha(x)}
\mathcal P\!\left(
\alpha(x),\frac{(1+\gamma\vartheta(x))h(x)}{\vartheta(x)}
\right)\notag\\
&-K\mathcal P\!\left(
\alpha(x),\frac{h(x)}{\vartheta(x)}
\right).
\end{align}
\begin{equation}\label{eq:gamma-conditional-call}
\mathcal C_{\Gamma}(x;K)=
\begin{cases}
\mathcal A_{\Gamma}(x;K),&h(x)>0,\\[3pt]
0,&h(x)\leq0.
\end{cases}
\end{equation}
Then the LETF call price is the one-dimensional integral
\begin{equation}\label{eq:conditional-gamma-price}
C_L^{\Gamma}(K,T)=D_T\int_{\R}\mathcal C_{\Gamma}(x;K)p_T(x)\,dx.
\end{equation}
If $v(x)=0$, the integrand is interpreted by continuity as
\begin{equation}\label{eq:gamma-deterministic-limit}
\mathcal C_{\Gamma}(x;K)
=\left(F_T^Le^{\beta x-\gamma m(x)}-K\right)^+.
\end{equation}
\end{theorem}

\begin{proof}
For $\gamma>0$, the payoff is positive exactly on $G_x<h(x)$. Hence
\begin{align}
\E\!\left[\left(F_T^Le^{\beta x-\gamma G_x}-K\right)^+\mid X_T=x\right]
={}&F_T^Le^{\beta x}
\E\!\left[e^{-\gamma G_x}\one_{\{G_x<h(x)\}}\mid X_T=x\right]\notag\\
&-K\Q(G_x<h(x)\mid X_T=x).
\end{align}
Substitution of \eqref{eq:gamma-cdf} and \eqref{eq:gamma-truncated-laplace} gives \eqref{eq:gamma-positive-call}, and the case $h(x)\leq0$ carries no payoff, which is \eqref{eq:gamma-conditional-call}; integrating against $p_T(x)$ proves \eqref{eq:conditional-gamma-price}.
\end{proof}

\begin{remark}[De-leveraged case]
For $0<\beta<1$, one has $\gamma<0$ and the exercise region is an upper Gamma tail rather than a lower tail. Let
\begin{equation}
h_+(x)=\max\{h(x),0\},
\qquad
\mathcal Q(a,z)=1-\mathcal P(a,z).
\end{equation}
Provided $1+\gamma\vartheta(x)>0$, the conditional call value is
\begin{align}
\mathcal C_{\Gamma}(x;K)
={}&F_T^Le^{\beta x}\big(1+\gamma\vartheta(x)\big)^{-\alpha(x)}
\mathcal Q\!\left(
\alpha(x),\frac{(1+\gamma\vartheta(x))h_+(x)}{\vartheta(x)}
\right)\notag\\
&-K\mathcal Q\!\left(
\alpha(x),\frac{h_+(x)}{\vartheta(x)}
\right).
\end{align}
The moment condition is necessary because $e^{-\gamma G_x}$ grows exponentially when $\gamma<0$. In addition, the outer integral over $x$ and the corresponding normalising moment must be finite. The remaining implementation and shifted extensions are stated for $\gamma>0$; extending every formula to de-leveraged products is straightforward only after these stronger integrability conditions have been verified.
\end{remark}

\begin{remark}[Computational form]
The Gamma closure needs only the first two conditional moments, the same inputs as a Gaussian closure. Unlike a Gaussian closure it keeps the conditional law non-negative, and it handles the payoff kink analytically. The variance-drag correction therefore enters at the price level before implied-volatility inversion, and the only remaining numerical step is quadrature over $x$.
\end{remark}

%==============================================================================
\section{Conditional Shifted-Gamma Closure}\label{sec:shifted-gamma}
%==============================================================================

The Gamma closure reproduces $m(x)$ and $v(x)$ exactly, but by \eqref{eq:gamma-cv-skew} it then forces
\begin{equation}\label{eq:two-moment-skew-constraint}
\lambda_3(x)=2\,\mathrm{cv}(x),
\end{equation}
so the conditional skewness is not free: it is whatever the first two moments dictate. There is no reason for the true conditional law of $I_T$ given $X_T=x$ to obey \eqref{eq:two-moment-skew-constraint}, and the discrepancy is measurable, since $\lambda_3(x)$ follows from the same triangular hierarchy at the cost of one further one-dimensional solve.

An unshifted Gamma assigns positive density arbitrarily close to the origin. If the local variance is bounded below by $a_{\min}>0$ on the model domain, realised variance instead satisfies
\begin{equation}\label{eq:hard-floor}
I_T=\int_0^T a(t,X_t)\,dt\ \geq\ a_{\min}T>0
\end{equation}
on every path. Each conditional law then has the same positive lower bound, suggesting a floor plus a positive fluctuation rather than a distribution starting at zero.

An unshifted Gamma cannot represent such a lower bound independently of its dispersion: increasing $\alpha(x)$ reduces mass near zero but also fixes the variance through $\Var=m^2/\alpha$.

\begin{figure}[p]
\centering
\panel{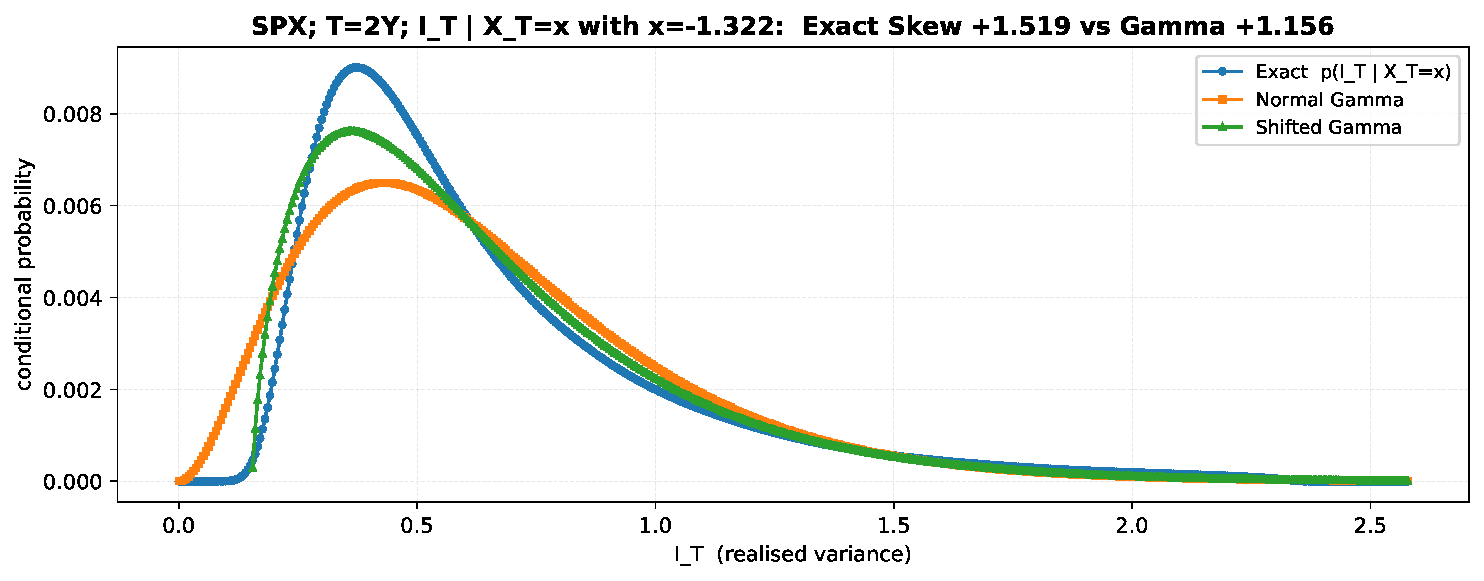}
\caption{Shifted Gamma can capture the lower bound}
\label{fig:shifted-gamma}
\end{figure}

We therefore add a deterministic non-negative shift $\zeta(x)$ to the Gamma variable. The shift locates the fitted lower edge, while the Gamma component describes variation above it (see Figure~\ref{fig:shifted-gamma} for example). This additional parameter frees the skewness while retaining non-negative support, an explicit truncated Laplace transform, and a closed-form conditional call value.

\subsection{Three-moment fit and admissibility}

Fix $x$ with $m(x)>0$, $v(x)>0$ and $\lambda_3(x)>0$, and approximate
\begin{equation}\label{eq:shifted-gamma-law}
I_T\mid X_T=x\ \approx\ \zeta(x)+G^s_x,
\qquad
G^s_x\sim\operatorname{Gamma}\big(\alpha_s(x),\vartheta_s(x)\big),
\end{equation}
with $\zeta(x)\geq0$ the floor and $G^s_x$ the fluctuation above it. Matching the three conditional moments of Definition~\ref{def:conditional-moments} gives
\begin{align}\label{eq:shifted-gamma-parameters}
\alpha_s(x)&=\frac{4}{\lambda_3(x)^2},\notag\\
\vartheta_s(x)&=\sqrt{\frac{v(x)}{\alpha_s(x)}}
=\frac{\lambda_3(x)\sqrt{v(x)}}{2},\notag\\
\zeta(x)&=m(x)-\alpha_s(x)\vartheta_s(x)
=m(x)-\frac{2\sqrt{v(x)}}{\lambda_3(x)}.
\end{align}
Indeed the shifted law has mean $\zeta+\alpha_s\vartheta_s=m$, variance $\alpha_s\vartheta_s^2=v$, and third central moment $2\alpha_s\vartheta_s^3=\mu_3$; a shift changes no central moment, so it is available to absorb exactly the mismatch between $\lambda_3$ and the Gamma shape.

The fit is not always usable. To retain the pathwise support $I_T\geq0$ the shift must be non-negative, which by \eqref{eq:shifted-gamma-parameters} means
\begin{equation}\label{eq:shifted-gamma-admissibility}
\zeta(x)\geq0
\quad\Longleftrightarrow\quad
\lambda_3(x)\geq2\,\mathrm{cv}(x).
\end{equation}
Equality in \eqref{eq:shifted-gamma-admissibility} returns the unshifted closure of Section~\ref{sec:conditional-gamma}. The shifted family therefore represents only conditional skewness at or above $2\,\mathrm{cv}(x)$. When $\lambda_3(x)<2\,\mathrm{cv}(x)$, no non-negatively shifted Gamma law matches all three moments. The Gamma-mixture closure of Appendix~\ref{app:gamma-mixture} removes this restriction; by Remark~\ref{rem:mixture-skew-range}, it covers the full skewness range compatible with $I_T\geq0$.

\begin{definition}[Admissible set and fallback]\label{def:shifted-gamma-fallback}
Let
\begin{equation}\label{eq:shifted-gamma-admissible-set}
\mathcal A=\big\{x:\ M_0(T,x)\ \text{resolved},\ m(x)>0,\ v(x)>0,\ \lambda_3(x)\geq2\,\mathrm{cv}(x)\big\}.
\end{equation}
The conditional shifted-Gamma closure uses \eqref{eq:shifted-gamma-law} on $\mathcal A$ and the Gamma closure \eqref{eq:conditional-gamma} on its complement. Since the two families agree at $\lambda_3=2\,\mathrm{cv}$, this hybrid is continuous across the boundary of $\mathcal A$.
\end{definition}

Useful diagnostics are the fraction of resolved nodes in $\mathcal A$ and the gap $\lambda_3(x)-2\,\mathrm{cv}(x)$ on that set. They measure, respectively, the scope of the shifted fit and the departure from the two-moment Gamma skewness.

\subsection{Closed-form conditional call value}

For $\gamma>0$, the drag conditional on $X_T=x$ is $\gamma\zeta(x)+\gamma G^s_x$, so the deterministic part of the shift simply rescales the conditional forward and the exercise boundary moves to
\begin{equation}\label{eq:shifted-gamma-boundary}
h_s(x)=\frac{\beta x-\gamma\zeta(x)-k}{\gamma}
=h(x)-\zeta(x),
\end{equation}
with $h(x)$ from \eqref{eq:gamma-boundary}. Conditional exercise occurs precisely when $G^s_x<h_s(x)$.

\begin{theorem}[Conditional shifted-Gamma LETF price]\label{thm:shifted-gamma-price}
Under approximation \eqref{eq:shifted-gamma-law} with $\gamma>0$ and $\zeta(x)\geq0$, set $\mathcal C_{s\Gamma}(x;K)=0$ when $h_s(x)\leq0$ and
\begin{align}\label{eq:shifted-gamma-call}
\mathcal C_{s\Gamma}(x;K)
={}&F_T^Le^{\beta x-\gamma\zeta(x)}
\big(1+\gamma\vartheta_s(x)\big)^{-\alpha_s(x)}
\mathcal P\!\left(
\alpha_s(x),
\frac{(1+\gamma\vartheta_s(x))h_s(x)}{\vartheta_s(x)}
\right)\notag\\
&-K\mathcal P\!\left(
\alpha_s(x),\frac{h_s(x)}{\vartheta_s(x)}
\right)
\end{align}
when $h_s(x)>0$. Then the LETF call price is the one-dimensional integral
\begin{equation}\label{eq:shifted-gamma-price}
C_L^{s\Gamma}(K,T)
=D_T\int_{\R}\mathcal C_{s\Gamma}(x;K)\,p_T(x)\,dx.
\end{equation}
\end{theorem}

\begin{proof}
Conditional on $X_T=x$ the payoff is $\big(F_T^Le^{\beta x-\gamma\zeta(x)}e^{-\gamma G^s_x}-K\big)^+$, which is the payoff appearing in the proof of Theorem~\ref{thm:conditional-gamma-price} with $F_T^Le^{\beta x}$ replaced by $F_T^Le^{\beta x-\gamma\zeta(x)}$ and $G_x$ by $G^s_x$. The exercise region is $\{G^s_x<h_s(x)\}$ by \eqref{eq:shifted-gamma-boundary}, and \eqref{eq:gamma-cdf}--\eqref{eq:gamma-truncated-laplace} apply verbatim with parameters $(\alpha_s,\vartheta_s)$ and boundary $h_s$. Integrating against $p_T$ gives \eqref{eq:shifted-gamma-price}.
\end{proof}

\begin{remark}[Degenerate limits]
If $\zeta(x)=0$, then $\alpha_s=\alpha$, $\vartheta_s=\vartheta$, $h_s=h$, and \eqref{eq:shifted-gamma-call} is \eqref{eq:gamma-positive-call}. If $v(x)=0$ the conditional law degenerates and the integrand is interpreted by continuity as in \eqref{eq:gamma-deterministic-limit}, independently of the shift.
\end{remark}

\begin{remark}[Cost of the extra moment]\label{rem:shifted-gamma-cost}
At each maturity, relative to Section~\ref{sec:conditional-gamma}, the method adds the forward solve for $M_3$ on that maturity's finite-difference grid and the closed-form parameter map \eqref{eq:shifted-gamma-parameters} per quadrature node. Pricing still uses the same two incomplete-gamma evaluations per node, and the outer quadrature over $x$, strike loop and martingale normalisation are unchanged.

On a fixed maturity grid, all the $M_n$ obey the same forward operator and differ only in the source $n\,a\,M_{n-1}$. At each time step, the local variance, tridiagonal operator and factorisation are computed once and reused for every $n$. Because the system is triangular, each additional moment requires only one right-hand-side assembly and one back-substitution, both $O(n_x)$, plus one grid vector in memory. The terminal density, conditional quadrature and implied-volatility inversion are also shared with the two-moment method. Constructing a full surface requires the additional $M_3$ solve separately at each maturity. Table~\ref{tab:timings} reports the resulting incremental cost.
\end{remark}

%==============================================================================
\section{Computing the Conditional Moments by Forward Moment PDEs}\label{sec:moment-pdes}
%==============================================================================

The remaining task is to compute the conditional moments of Definition~\ref{def:conditional-moments} under Dupire local volatility. They are terminal-conditioned functionals of the augmented process $(X_t,I_t)$ whose joint density solves the Fokker--Planck equation~\eqref{eq:forward-xi-pde} below. They can be obtained without a distributional closure and without solving the full two-dimensional joint density: it suffices to propagate a small triangular system of one-dimensional forward PDEs for unnormalised partial moments. The Gamma method stops at $n=2$; the shifted fit carries one more equation.

\subsection{Partial-moment functions}

Define the unnormalised partial-moment measures by
\begin{equation}\label{eq:partial-moment-measures}
\mu_n(t,A)=\E\!\left[I_t^n\one_{\{X_t\in A\}}\right],
\qquad n=0,1,2,3,
\end{equation}
for Borel sets $A\subset\R$. Assume these measures admit densities $M_n(t,x)$ with respect to $dx$. If the joint law of $(X_t,I_t)$ has a density $p(t,x,i)$, this definition reduces to
\begin{equation}\label{eq:partial-moments-def}
M_n(t,x)=\int_{\R_+}i^np(t,x,i)\,di.
\end{equation}
The measure definition is more general: it remains meaningful, for example, when constant variance makes $I_t$ deterministic and the joint law singular in the $i$ direction. Here $M_0(t,x)=p_t(x)$ is the marginal density of $X_t$, while $M_1,M_2,M_3$ are the $I_t$-weighted densities. Writing the conditional raw moments as
\begin{equation}\label{eq:raw-moments-from-M}
r_n(x)=\frac{M_n(T,x)}{M_0(T,x)}=\E[I_T^{\,n}\mid X_T=x],
\qquad n=1,2,3,
\end{equation}
the quantities of Definition~\ref{def:conditional-moments} are
\begin{equation}\label{eq:moments-from-M}
m(x)=r_1(x),
\qquad
v(x)=r_2(x)-r_1(x)^2,
\qquad
\mu_3(x)=r_3(x)-3r_1(x)r_2(x)+2r_1(x)^3 .
\end{equation}

\subsection{The triangular moment system}

The system can be derived without assuming a joint density. It is the moment hierarchy associated with the Feynman--Kac transform of the additive functional $I_t$ \cite{kac1949}. Let $\varphi\in C_c^2(\R)$ be a test function.

With $dI_t=a(t,X_t)\,dt$ and $d\langle X\rangle_t=a(t,X_t)\,dt$ from \eqref{eq:X-dynamics},
\begin{align}\label{eq:weak-ito}
d\big(I_t^n\varphi(X_t)\big)
={}&I_t^n\,a(t,X_t)\left(
-\tfrac12\varphi'(X_t)+\tfrac12\varphi''(X_t)
\right)dt\notag\\
&+n\,I_t^{n-1}a(t,X_t)\varphi(X_t)\,dt
+I_t^n\varphi'(X_t)\sqrt{a(t,X_t)}\,dW_t.
\end{align}
Taking expectations removes the martingale term, and each remaining expectation has the form $\E[I_t^{\,j}g(X_t)]=\int_\R g\,M_j(t,\cdot)\,dx$ by \eqref{eq:partial-moment-measures}, since $a$ depends on the path only through $X_t$. Hence
\begin{equation}\label{eq:weak-integral-form}
\int_{\R}\varphi\,\partial_tM_n\,dx
=\int_{\R}a\left(-\tfrac12\varphi'+\tfrac12\varphi''\right)M_n\,dx
+n\int_{\R}a\,\varphi\,M_{n-1}\,dx.
\end{equation}
Integrating by parts, with no boundary terms because $\varphi$ has compact support,
\begin{equation}\label{eq:ibp}
\int_{\R}a\varphi'M_n\,dx=-\int_{\R}\varphi\,\partial_x\!\big(aM_n\big)dx,
\qquad
\int_{\R}a\varphi''M_n\,dx=\int_{\R}\varphi\,\partial_{xx}\!\big(aM_n\big)dx,
\end{equation}
so that \eqref{eq:weak-integral-form} becomes
\begin{equation}\label{eq:weak-residual}
\int_{\R}\varphi\left[
\partial_tM_n
-\tfrac12\partial_x\!\big(aM_n\big)
-\tfrac12\partial_{xx}\!\big(aM_n\big)
-n\,a\,M_{n-1}
\right]dx=0 .
\end{equation}
Since \eqref{eq:weak-residual} holds for every $\varphi\in C_c^2(\R)$, the bracket vanishes as a distribution, giving the triangular system
\begin{equation}\label{eq:moment-pde}
\partial_t M_n
=\tfrac12\partial_x\!\big(a(t,x)M_n\big)
+\tfrac12\partial_{xx}\!\big(a(t,x)M_n\big)
+n\,a(t,x)\,M_{n-1},
\qquad n=0,1,2,3,
\end{equation}
with the convention $M_{-1}\equiv0$. All four equations share the same one-dimensional Kolmogorov-forward operator, and $M_{n-1}$ acts as a known source for $M_n$, so one solves $M_0$ first, then $M_1$ with source $a M_0$, then $M_2$ with source $2a M_1$, and finally $M_3$ with source $3a M_2$ for the shifted fit.

\begin{definition}[Initial conditions]\label{def:moment-ic}
Since $X_0=0$ after forward normalisation and $I_0=0$ deterministically, the initial joint law is the point mass $\delta_0(dx)\delta_0(di)$. Its zeroth partial moment is $\delta(x)$ and all positive-order partial moments vanish, so
\begin{equation}\label{eq:moment-ic}
M_0(0,x)=\delta(x),
\qquad
M_n(0,x)=0\ \ \text{for } n\ge1.
\end{equation}
Thus $M_0$ carries the usual Kolmogorov-forward point mass at $x=0$, while $M_1$, $M_2$ and $M_3$ start identically zero and accumulate mass through the source terms $n\,a\,M_{n-1}$. In a finite-difference implementation, $M_0(0,\cdot)$ is initialised as a conservative discrete approximation of $\delta(x)$ and $M_n\equiv0$ for $n\ge1$. If the solve is instead started at a small $t_0>0$, replace $\delta$ by the short-time Gaussian for $X_{t_0}$ and use the locally constant-variance initialisation
\begin{equation}
M_n(t_0,x)\approx\big(a(0,0)t_0\big)^nM_0(t_0,x),
\qquad n=1,2,3,
\end{equation}
which preserves a non-negative conditional variance to leading order.
\end{definition}

\begin{remark}[Exactness and cost]
Subject to the stated regularity and boundary conditions, the moment system \eqref{eq:moment-pde} delivers the conditional moments $m(x)$, $v(x)$ and $\mu_3(x)$ without a distributional closure wherever $M_0(T,x)>0$. Each maturity is solved on its own finite-difference grid, selected for that tenor. On a fixed maturity grid, the equations for $M_0$, $M_1$, $M_2$ and $M_3$ share the same tridiagonal operator: local-variance evaluation, matrix assembly and factorisation are performed once per time step, while each higher moment adds only a right-hand-side assembly and back-substitution. The two-moment closure therefore requires three one-dimensional solves and the three-moment closure four, with the marginal density $M_0=p_T$ reused in the pricing integrals. Constructing a full surface requires a separate set of moment solves for each maturity; approximation enters only when the resulting moments are used to close the law of $I_T\mid X_T=x$. See Remark~\ref{rem:shifted-gamma-cost} for the incremental cost of the fourth solve.
\end{remark}

%==============================================================================
\section{Approximate Martingale Normalisation}
%==============================================================================

The exact variable $Y=\beta X_T-\gamma I_T$ satisfies $\E[e^Y]=1$ under the standing true-martingale assumption. Neither conditional closure is guaranteed to preserve that identity. A direct scalar normalisation enforces a consistent LETF forward at the approximation level, provided the displayed exponential moment is finite.

\subsection{Direct normalisation}

If a numerical scheme produces samples or quadrature weights for the raw return
\begin{equation}
Y=\beta X_T-\gamma I_T,
\end{equation}
replace it by
\begin{equation}\label{eq:direct-normalisation}
R_L^{\mathrm{app}}=Y-\log\E_{\mathrm{app}}[e^Y].
\end{equation}
Then use $R_L^{\mathrm{app}}$ in the payoff. Under the conditional-Gamma closure with $\gamma>0$, the required exponential moment is explicit:
\begin{equation}\label{eq:gamma-normaliser}
c_T^{\Gamma}
=\log\E_{\Gamma}[e^Y]
=\log\int_{\R}e^{\beta x}
\big(1+\gamma\vartheta(x)\big)^{-\alpha(x)}p_T(x)\,dx.
\end{equation}
At points where $v(x)=0$, the factor $(1+\gamma\vartheta)^{-\alpha}$ is interpreted by continuity as $e^{-\gamma m(x)}$.
The normalised conditional price is obtained from \eqref{eq:gamma-conditional-call} by replacing
\begin{equation}\label{eq:gamma-normalised-substitution}
F_T^Le^{\beta x}\ \longmapsto\ F_T^Le^{\beta x-c_T^{\Gamma}},
\qquad
h(x)\ \longmapsto\ h_c(x)=\frac{\beta x-c_T^{\Gamma}-k}{\gamma}.
\end{equation}
Equivalently, if $C_{L,\mathrm{raw}}^{\Gamma}$ denotes the unnormalised pricing formula, then
\begin{equation}
C_{L,\mathrm{norm}}^{\Gamma}(K,T)
=e^{-c_T^{\Gamma}}C_{L,\mathrm{raw}}^{\Gamma}
\big(Ke^{c_T^{\Gamma}},T\big).
\end{equation}
The same scalar must be used for every strike at a given maturity.

\subsection{Normalisation of the shifted-Gamma closure}

The shifted closure requires the same treatment with the shift carried through the exponential moment. Under \eqref{eq:shifted-gamma-law},
\begin{equation}\label{eq:shifted-gamma-normaliser}
c_T^{s\Gamma}
=\log\int_{\R}e^{\beta x-\gamma\zeta(x)}
\big(1+\gamma\vartheta_s(x)\big)^{-\alpha_s(x)}p_T(x)\,dx,
\end{equation}
after which $e^{\beta x-\gamma\zeta(x)}$ is replaced by $e^{\beta x-\gamma\zeta(x)-c_T^{s\Gamma}}$ and $h_s(x)$ by $\big(\beta x-\gamma\zeta(x)-c_T^{s\Gamma}-k\big)/\gamma$ in \eqref{eq:shifted-gamma-call}. At nodes falling back to the two-moment closure by Definition~\ref{def:shifted-gamma-fallback}, the integrand of \eqref{eq:shifted-gamma-normaliser} is the corresponding term of \eqref{eq:gamma-normaliser}; the normaliser is a single scalar computed once for the hybrid, not one scalar per family.

%==============================================================================
\section{Forward Joint-Law PDE Reference}\label{sec:joint-law}
%==============================================================================

The conditional closures are assessed against a method that retains the full joint law. Under Dupire LV, use the augmented Markov process $(X_t,I_t)$:
\begin{equation}\label{eq:augmented-process}
dX_t=-\frac12a(t,X_t)dt+\sqrt{a(t,X_t)}dW_t,
\qquad
dI_t=a(t,X_t)dt,
\qquad I_0=0.
\end{equation}
A single forward solve propagates the joint law for all LETF strikes:
\begin{equation}
\Pi_t(dx,di)=\Q(X_t\in dx,I_t\in di).
\end{equation}
When this law admits a density, write $\Pi_t(dx,di)=p(t,x,i)dx\,di$; otherwise $p$ below is understood as a generalised density and the equation in the weak sense. The Fokker--Planck equation corresponding to \eqref{eq:augmented-process} is
\begin{equation}\label{eq:forward-xi-pde}
\partial_t p
=\frac12\partial_x\!\big(a(t,x)p\big)
-\partial_i\!\big(a(t,x)p\big)
+\frac12\partial_{xx}\!\big(a(t,x)p\big),
\end{equation}
with initial condition
\begin{equation}
p(0,x,i)=\delta(x)\delta(i).
\end{equation}
The equation is two-dimensional but degenerate: it diffuses in $x$ and transports in $i$. Since $I_t$ is monotone increasing, the $i$ direction can be handled as an upwind/transport direction in finite-difference implementations.

Once $\Pi_T$ is available, the entire LETF smile at maturity $T$ is obtained by integrating the payoff against the same law. In density notation,
\begin{equation}
C_L(K,T)
=D_T\iint_{\mathbb{R}\times\mathbb{R}_+}
\left(F_T^Le^{\beta x-\gamma i}-K\right)^+p(T,x,i)\,dx\,di.
\end{equation}
If a numerical scheme introduces a small martingale error, apply a scalar normalisation consistent with \eqref{eq:direct-normalisation} by replacing $\beta x-\gamma i$ with $\beta x-\gamma i-c_T^{\mathrm{num}}$, where
\begin{equation}
c_T^{\mathrm{num}}
=\log\iint e^{\beta x-\gamma i}p(T,x,i)\,dx\,di.
\end{equation}
This keeps the LETF forward consistent before implied-volatility inversion.

\begin{remark}[Dimensionality]
The forward equation is two-dimensional in $(x,i)$, whereas the conditional-Gamma pricing quadrature in \eqref{eq:conditional-gamma-price} is one-dimensional after the moment solves. Its advantage as a reference is that one joint-law solve produces all LETF strikes at a fixed maturity without an assumed Gamma closure.
\end{remark}

%==============================================================================
\section{Implementation Algorithm}\label{sec:algorithm}
%==============================================================================

For each maturity $T$:
\begin{enumerate}
\item Build or read the Dupire local variance surface $a(t,x)$ from the underlying market surface.
\item Obtain the terminal underlying density $p_T(x)$ for the conditional approximation, either from the Dupire forward equation or from market call prices by Breeden--Litzenberger differentiation \cite{breeden_litzenberger1978}. If $C_S(K,T)$ is the underlying call price and $f_{S_T}$ is the density of $S_T$, then
\begin{equation}\label{eq:bl-density}
f_{S_T}(K)=\frac{1}{D_T}\frac{\partial^2 C_S}{\partial K^2}(K,T),
\qquad
p_T(x)=F_T^Se^x f_{S_T}(F_T^Se^x).
\end{equation}
\item Choose an integration grid in terminal log-moneyness $x$ covering the relevant LETF strikes.
\item Solve the triangular forward moment system \eqref{eq:moment-pde} with initial conditions \eqref{eq:moment-ic} for $n=0,1,2$, and additionally $n=3$ if the three-moment closure is wanted, then form $m(x)$, $v(x)$ and, where applicable, $\mu_3(x)$ from \eqref{eq:moments-from-M} wherever $M_0(T,x)$ is numerically resolved.
\item Where $m(x),v(x)>0$, form the two-moment Gamma parameters
\[
\alpha(x)=\frac{m(x)^2}{v(x)},
\qquad
\vartheta(x)=\frac{v(x)}{m(x)}.
\]
Use the deterministic limit \eqref{eq:gamma-deterministic-limit} where $v(x)$ is numerically negligible. Do not form moment ratios where $M_0(T,x)$ is below a density cutoff; materially negative values of $v(x)$ indicate inadequate resolution rather than a valid Gamma input.
\item For the three-moment closure, form $\lambda_3(x)$ from \eqref{eq:cv-skew-def}, test the admissibility condition \eqref{eq:shifted-gamma-admissibility}, and on the admissible set \eqref{eq:shifted-gamma-admissible-set} form $\alpha_s(x)$, $\vartheta_s(x)$ and $\zeta(x)$ from \eqref{eq:shifted-gamma-parameters}. Elsewhere fall back to the two-moment parameters as in Definition~\ref{def:shifted-gamma-fallback}. Record the admissible fraction and the gap $\lambda_3(x)-2\,\mathrm{cv}(x)$ as diagnostics.
\item If the approximation does not preserve $\E[e^Y]=1$, compute \eqref{eq:gamma-normaliser}, or \eqref{eq:shifted-gamma-normaliser} for the shifted closure, and apply the corresponding substitutions consistently to every strike.
\item Evaluate the conditional price integral \eqref{eq:conditional-gamma-price}, or its shifted counterpart \eqref{eq:shifted-gamma-price}, for each LETF strike, using stable library implementations of the regularised incomplete gamma function.
\item Invert the corrected prices to Black implied volatilities.
\item Verify probability conservation and non-negativity of $M_0$, and check the unconditional identities $\int M_n(T,x)\,dx=\E[I_T^n]$ against an independent calculation or Monte Carlo where available.
\item Check monotonicity and convexity of prices in $K$. If smoothing is required, use an arbitrage-aware total-variance parameterisation instead of raw volatility; see Gatheral \cite{gatheral2006}.
\end{enumerate}

For a closure-free reference at the same maturity, replace steps 2--8 by a solve of the forward joint-law equation \eqref{eq:forward-xi-pde}, then evaluate the two-dimensional payoff integral for each LETF strike on the grid.

%==============================================================================
\section{Numerical Validation}\label{sec:numerics}
%==============================================================================

We first test the forward joint-law calculation by grid refinement and by an
independent Monte Carlo simulation of the LETF dynamics.  The converged joint
law is then used to measure the errors of the strike map, the Gamma fit and its
shifted extension. Finally, the strike map and the Gamma fit are compared with
quoted bid and ask volatilities of a liquid listed LETF.

%------------------------------------------------------------------------------
\subsection{Test configuration}
%------------------------------------------------------------------------------

All tests use Dupire local-volatility surfaces calibrated to two liquid index
markets, S\&P~500 (SPX) and Nikkei~225 (NKY), on the valuation date
22~June~2026.  For each market we test four leverage ratios,
$$
\beta\in\{-2,-1,2,3\},
$$
covering two inverse and two long ratios of the kind used by listed LETFs.  In
all four cases
$$
\gamma=\frac{\beta(\beta-1)}{2}>0,
$$
so the tests are inside the standard long/inverse LETF regime treated in
Sections~\ref{sec:conditional-gamma}--\ref{sec:joint-law}.

Results are reported in Black implied volatility, the market convention, which
also normalises the strike-dependent scale of the option premium. LETF strikes
are expressed in LETF forward log-moneyness,
$$
k=\log(K/F_T^L),
$$
so that $k=0$ is at the money for every leverage ratio.  In the convergence,
approximation and closure-comparison figures the four subfigures are the
leverage ratios $\beta\in\{-2,-1,2,3\}$, and within each subfigure the four
panels are the maturities $T\in\{3\text{M},6\text{M},1\text{Y},2\text{Y}\}$.
The empirical figure of Section~\ref{sec:numerics-empirical} uses listed
expiries instead.

Four pricing methods are compared: the first-order strike remapping
of Theorem~\ref{thm:first-LV}, the conditional-Gamma method of
Theorem~\ref{thm:conditional-gamma-price}, the conditional shifted-Gamma method
of Theorem~\ref{thm:shifted-gamma-price}, and the forward joint-law solve of
Section~\ref{sec:joint-law}, which serves as the reference once it has been
validated in Section~\ref{sec:numerics-convergence}.  The first three are
assessed against that reference in two stages: Section~\ref{sec:numerics-approximation}
compares the first-order remapping and the two-moment conditional-Gamma method,
and Section~\ref{sec:numerics-shifted} then isolates what the third conditional
moment adds. Section~\ref{sec:numerics-empirical} compares the strike
remapping, the conditional-Gamma method and the reference with market quotes.

%------------------------------------------------------------------------------
\subsection{Validation of the forward joint-law solve}
\label{sec:numerics-convergence}
%------------------------------------------------------------------------------

\subsubsection{Grid design}

The joint-law equation~\eqref{eq:forward-xi-pde} is solved on a tensor grid in
the terminal underlying log-return $x$ and the realised integrated variance $i$.
The two directions play different numerical roles.  The $x$ direction resolves
the terminal underlying density and carries the diffusion, so it is the more
demanding of the two and is given the finer mesh; the $i$ direction is a pure
transport direction and only has to resolve the variance drag entering the
payoff through $\exp(-\gamma i)$. The third numerical parameter is the truncation of the $x$ domain, expressed as
a multiple $n_\sigma$ of the standard deviation of the terminal underlying
log-return.

Table~\ref{tab:grids} lists the four configurations.  The sequence is nested:
each successive configuration refines both directions and widens the truncation,
so the plots measure the aggregate discretisation error of the scheme rather
than the isolated effect of one parameter.

\begin{table}[htbp]
\centering
\caption{Grid configurations for the forward joint-law solve.  Here $n_x$ is the
number of nodes in log-moneyness, $n_i$ the number of nodes in integrated
variance, and $n_\sigma$ the $x$-domain truncation in standard deviations of the
terminal underlying log-return.  The time step is held fixed across
configurations.}
\label{tab:grids}
\begin{tabular}{lrrrr}
\toprule
Configuration & $n_x$ & $n_i$ & Steps per year & $n_\sigma$ \\
\midrule
Coarse   & 1600 & 800  & 250 & 8.0 \\
Baseline & 2000 & 1000 & 250 & 10.0 \\
Mid      & 2400 & 1200 & 250 & 12.0 \\
Fine     & 3000 & 1500 & 250 & 15.0 \\
\bottomrule
\end{tabular}
\end{table}

The baseline configuration is chosen as the reference setting.  The coarse
configuration is included to make the direction and size of the discretisation
error visible, and the mid and fine configurations to show that refining beyond
the baseline moves the smile only marginally.

\subsubsection{Monte Carlo benchmark}

The Monte Carlo benchmark evolves the LETF as its own process.  Under Dupire
local volatility the two forward-normalised processes are driven by the same
Brownian motion,
\begin{equation}\label{eq:mc-sde}
d\widetilde S_t=\sqrt{a(t,X_t)}\,\widetilde S_t\,dW_t,
\qquad
d\widetilde L_t=\beta\sqrt{a(t,X_t)}\,\widetilde L_t\,dW_t,
\qquad
\widetilde S_0=\widetilde L_0=1,
\end{equation}
with $X_t=\log\widetilde S_t$.  Writing $Y_t=\log\widetilde L_t$ and applying a
log-Euler scheme on a uniform time grid of step $\Delta$, one increment is
\begin{align}\label{eq:mc-scheme}
X_{t+\Delta}&=X_t-\tfrac12a(t,X_t)\Delta
+\sqrt{a(t,X_t)}\,\sqrt{\Delta}\,Z,\notag\\
Y_{t+\Delta}&=Y_t-\tfrac12\beta^2a(t,X_t)\Delta
+\beta\sqrt{a(t,X_t)}\,\sqrt{\Delta}\,Z,
\qquad Z\sim\mathcal N(0,1),
\end{align}
with the same draw $Z$ in both lines.  The realised variance $I_T$ is never
accumulated and $Y_T$ is never reconstructed from $\beta X_T-\gamma I_T$, so the
simulation is independent of the identity \eqref{eq:RL-identity} and of the
forward equation derived from it.

Implied volatilities are obtained by pricing sample call options.
For each strike the sample call price is
\begin{equation}\label{eq:mc-price}
\hat C_L(K,T)=\frac{D_T}{N}\sum_{j=1}^N
\left(F_T^Le^{Y_T^{(j)}}-K\right)^+
\end{equation}
The implied volatility $\hat\sigma_L(T,K)$ is the Black volatility reproducing
$\hat C_L(K,T)$ at forward $F_T^L$.

Agreement between this simulation and the joint-law solve jointly tests the
pathwise LETF identity, its forward equation and the numerical scheme.

\subsubsection{Convergence results}

\begin{figure}[p]
\centering
\begin{subfigure}[b]{\panelwidth}
\centering
\panel{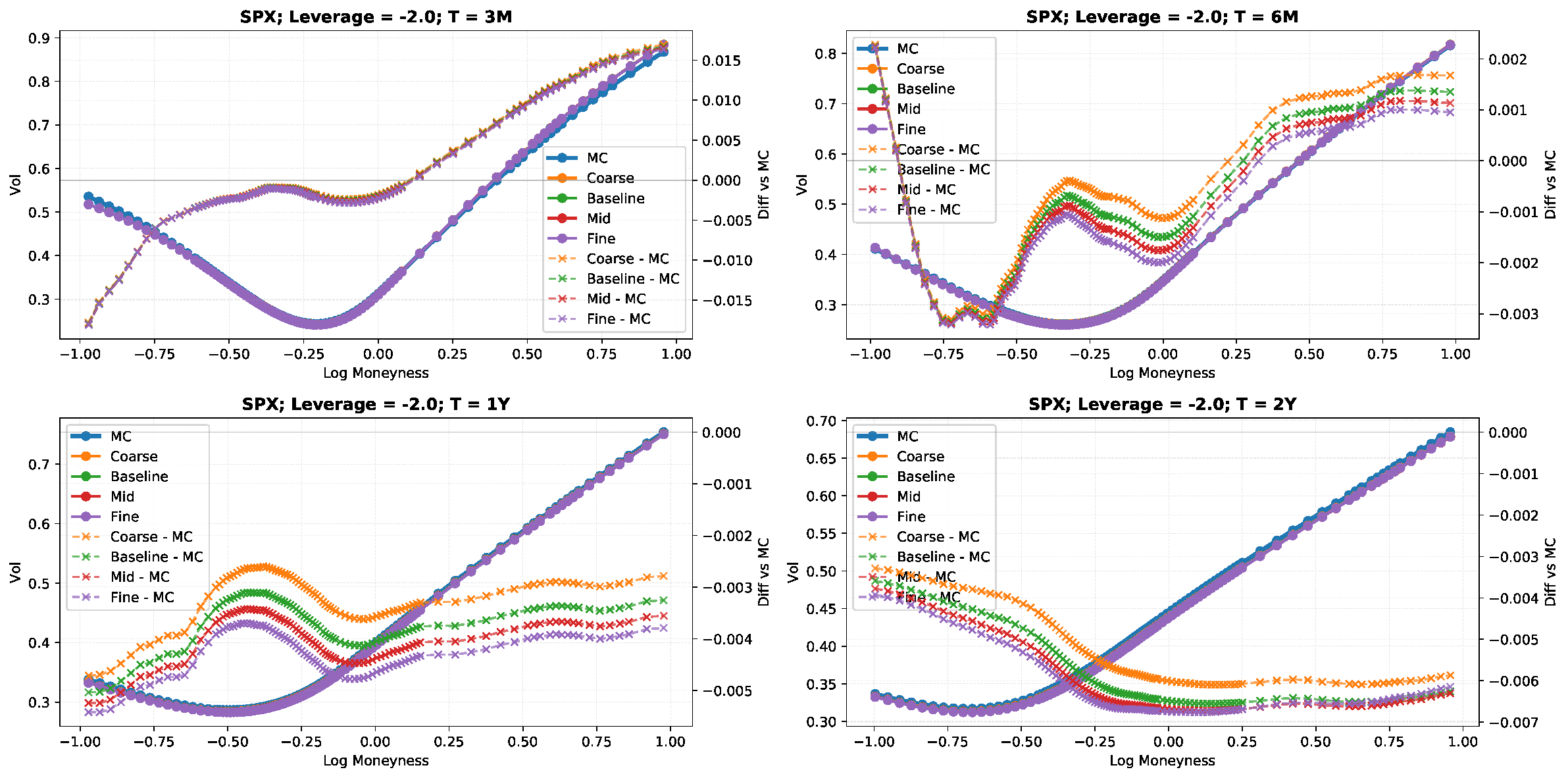}
\caption{$\beta=-2$}
\end{subfigure}

\vfill

\begin{subfigure}[b]{\panelwidth}
\centering
\panel{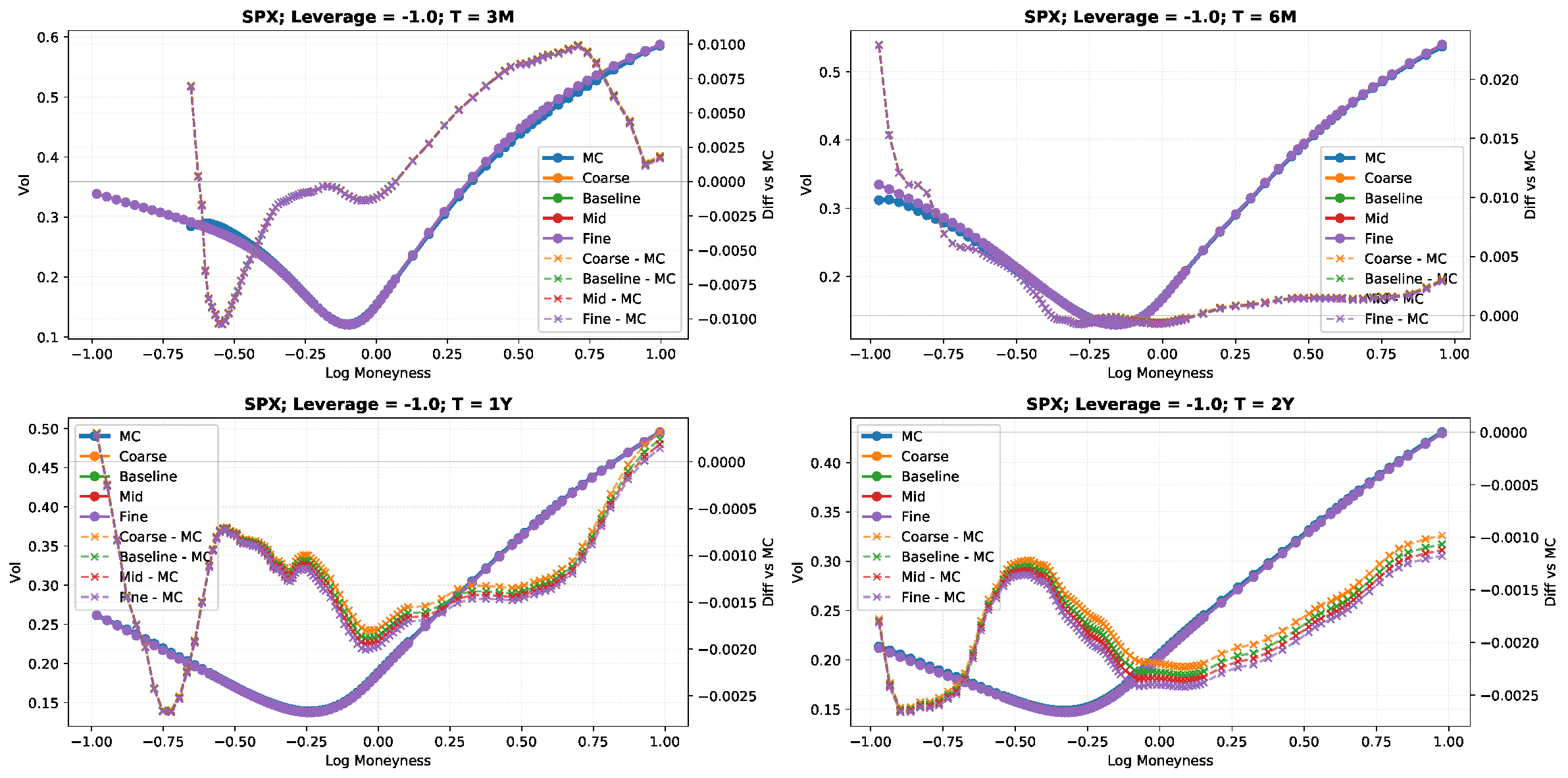}
\caption{$\beta=-1$}
\end{subfigure}

\vfill

\begin{subfigure}[b]{\panelwidth}
\centering
\panel{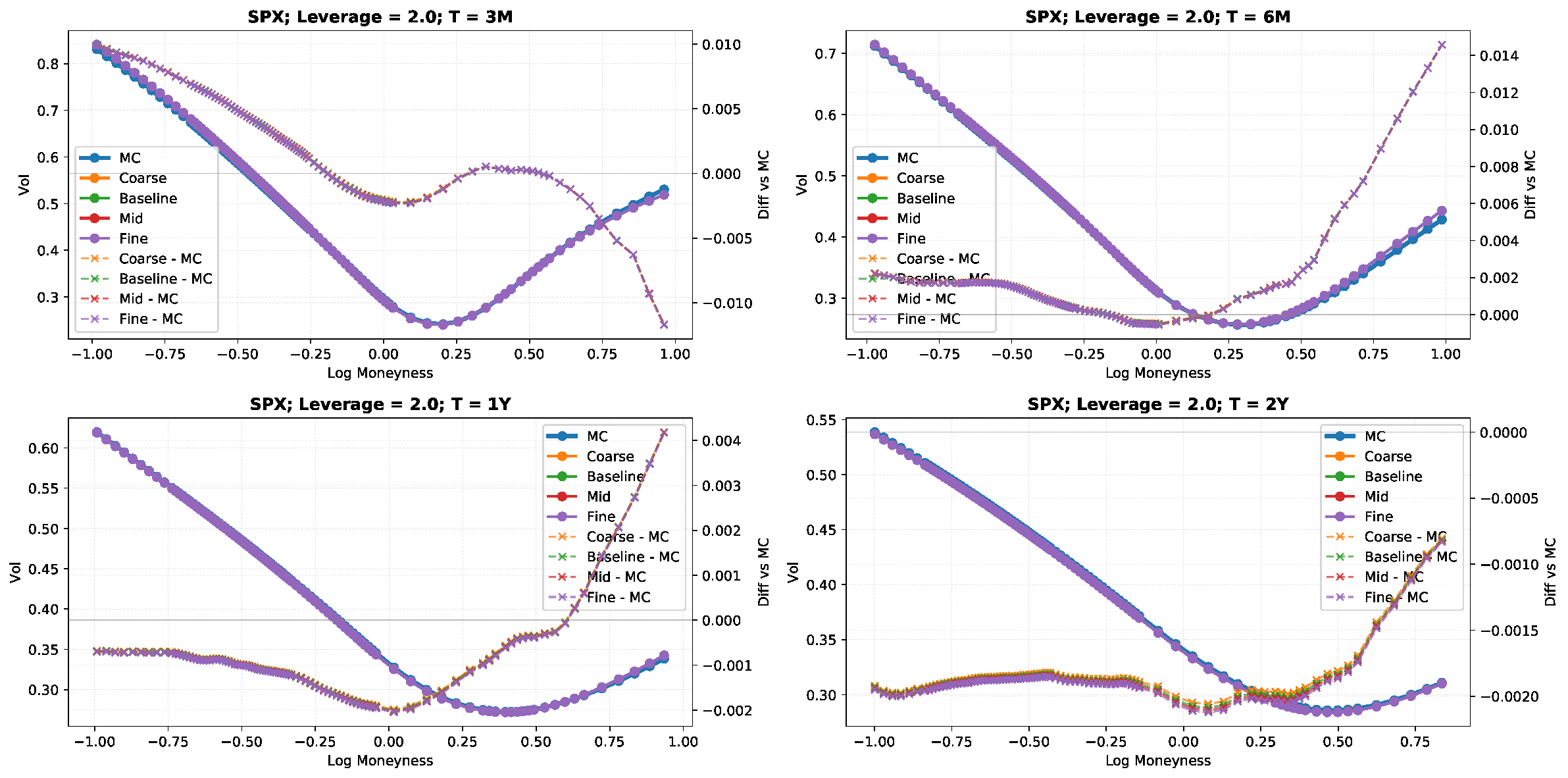}
\caption{$\beta=2$}
\end{subfigure}

\vfill

\begin{subfigure}[b]{\panelwidth}
\centering
\panel{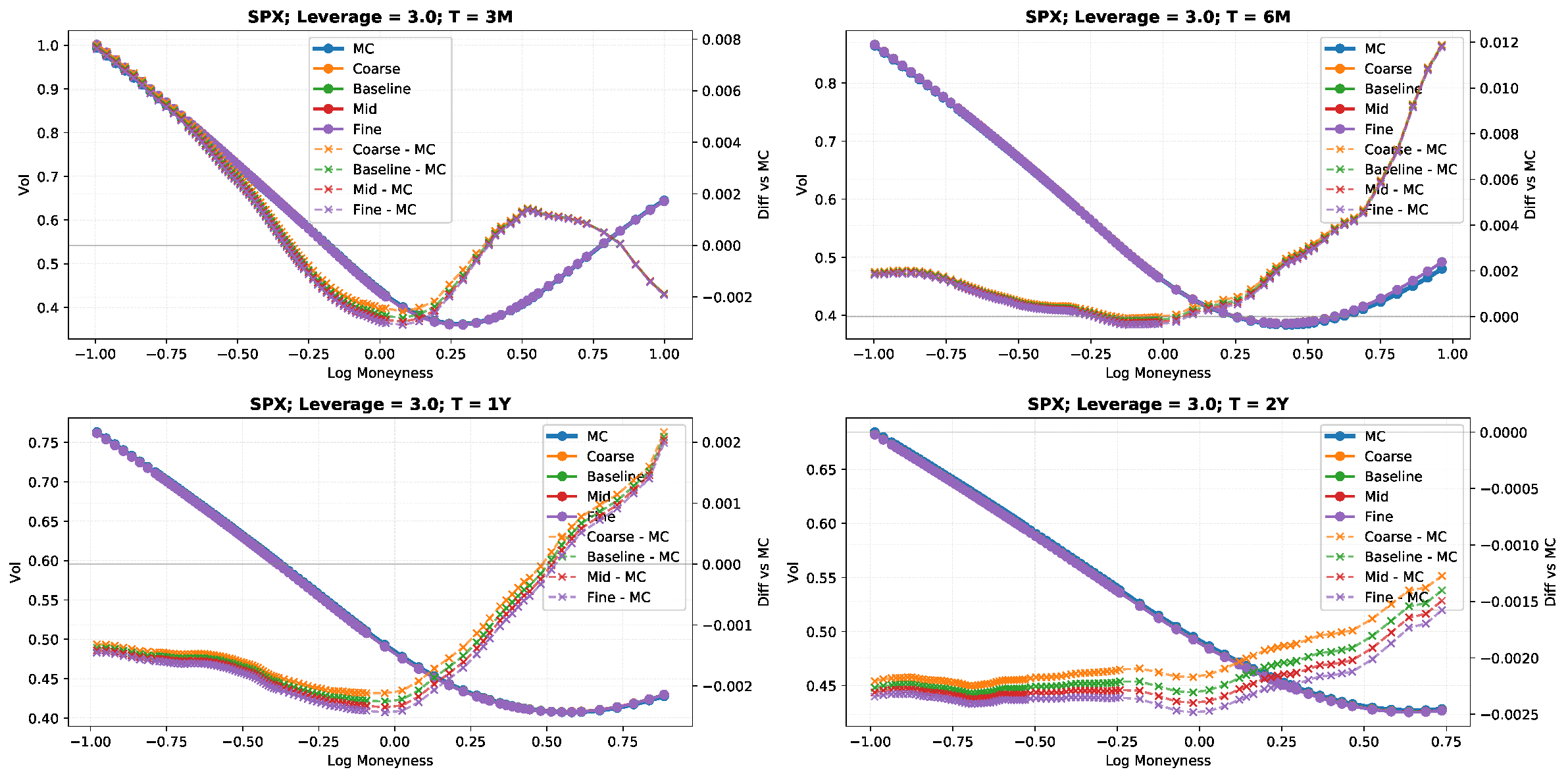}
\caption{$\beta=3$}
\end{subfigure}

\caption{Grid convergence of the forward joint-law solve against the Monte
Carlo benchmark, SPX, 22~June~2026.}
\label{fig:conv-spx}
\end{figure}

\begin{figure}[p]
\centering
\begin{subfigure}[b]{\panelwidth}
\centering
\panel{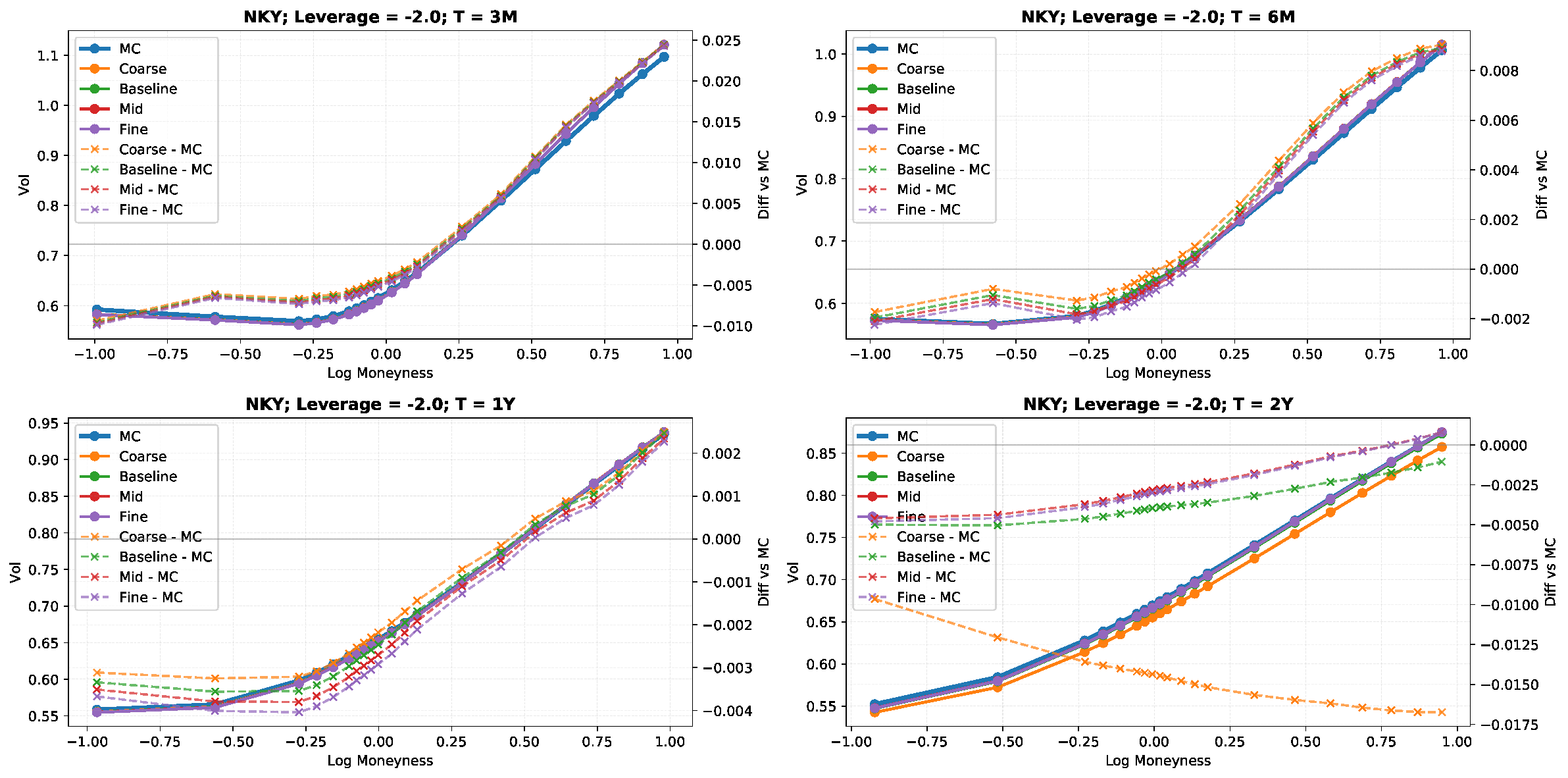}
\caption{$\beta=-2$}
\end{subfigure}

\vfill

\begin{subfigure}[b]{\panelwidth}
\centering
\panel{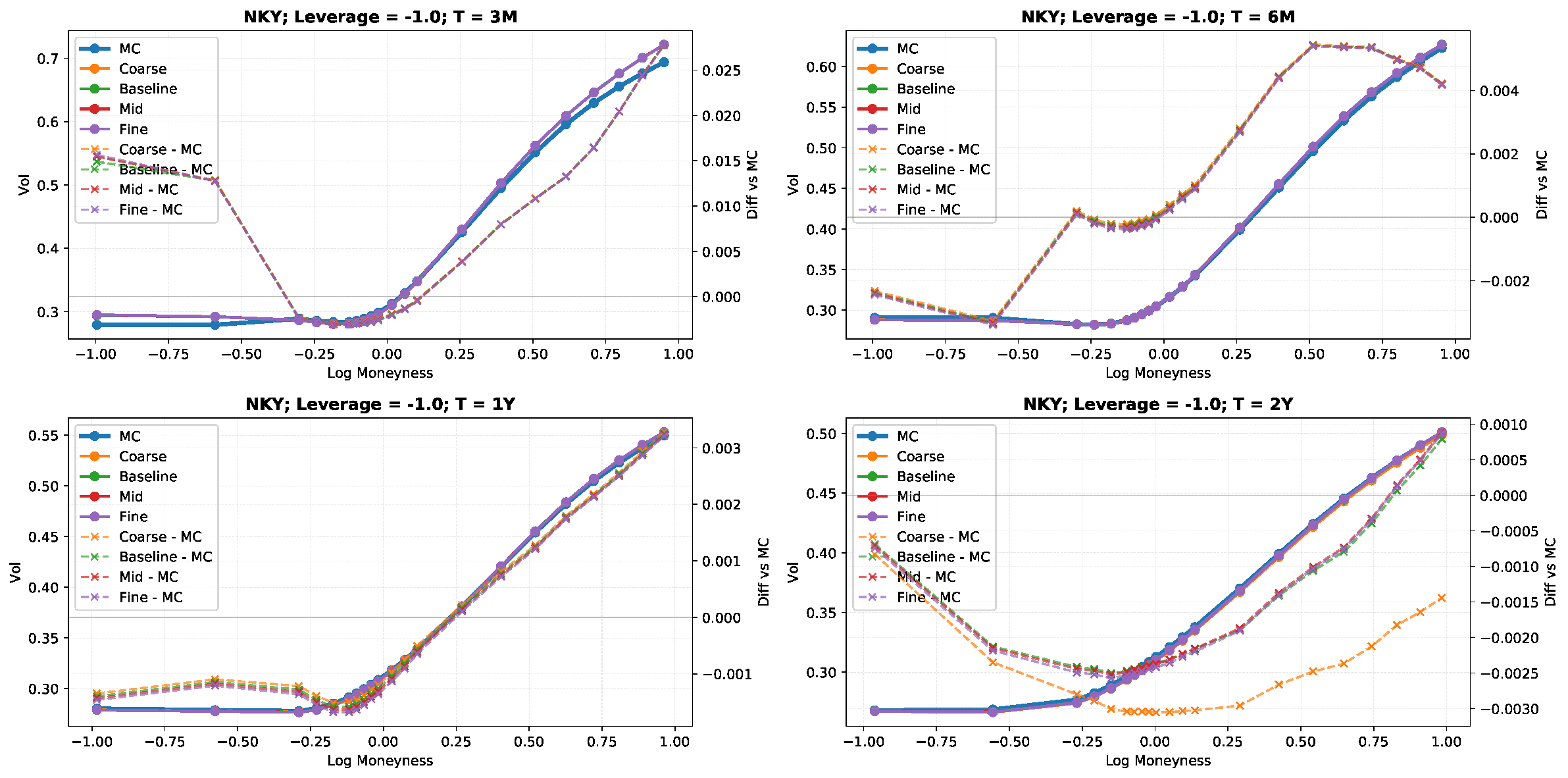}
\caption{$\beta=-1$}
\end{subfigure}

\vfill

\begin{subfigure}[b]{\panelwidth}
\centering
\panel{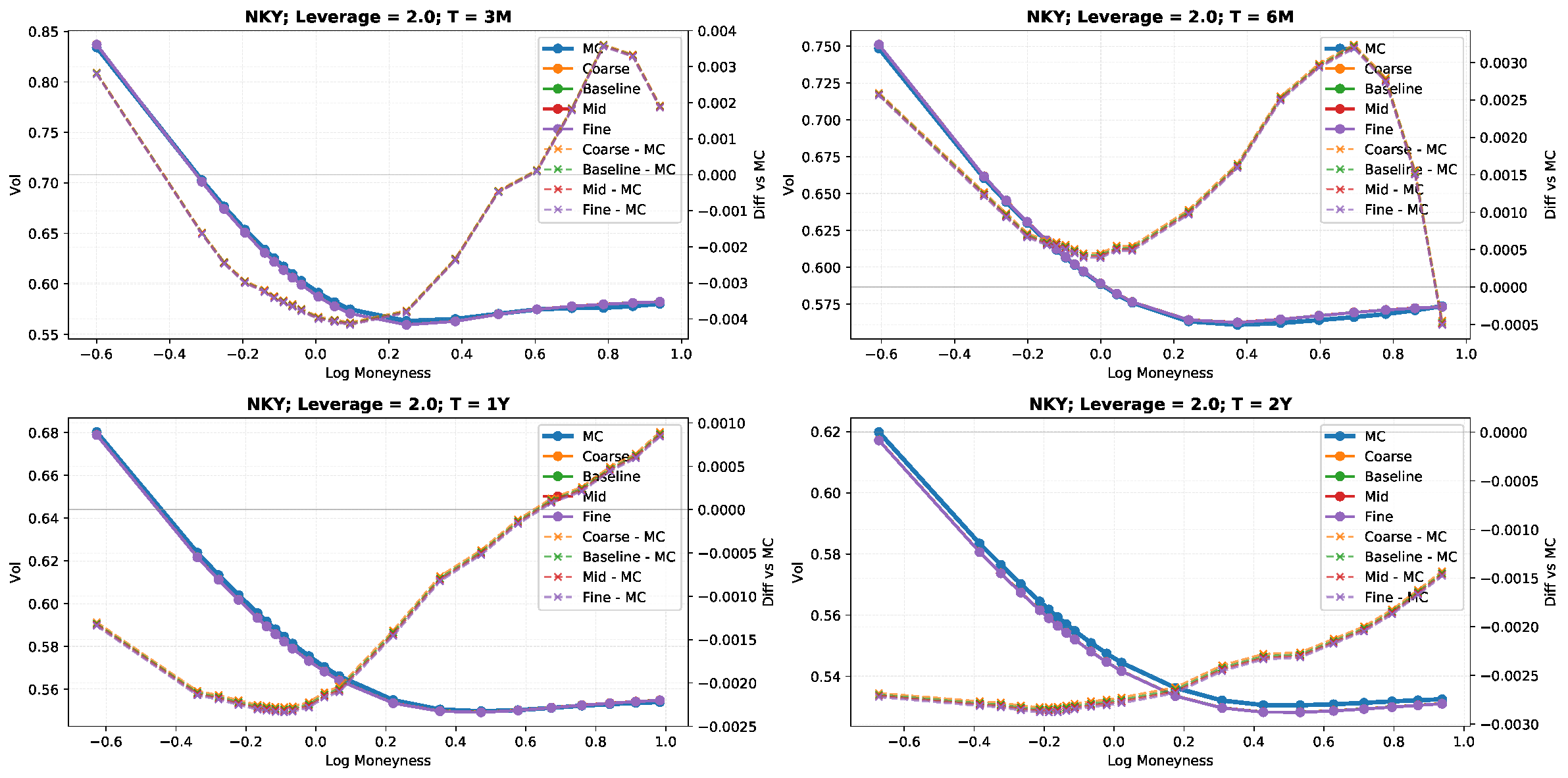}
\caption{$\beta=2$}
\end{subfigure}

\vfill

\begin{subfigure}[b]{\panelwidth}
\centering
\panel{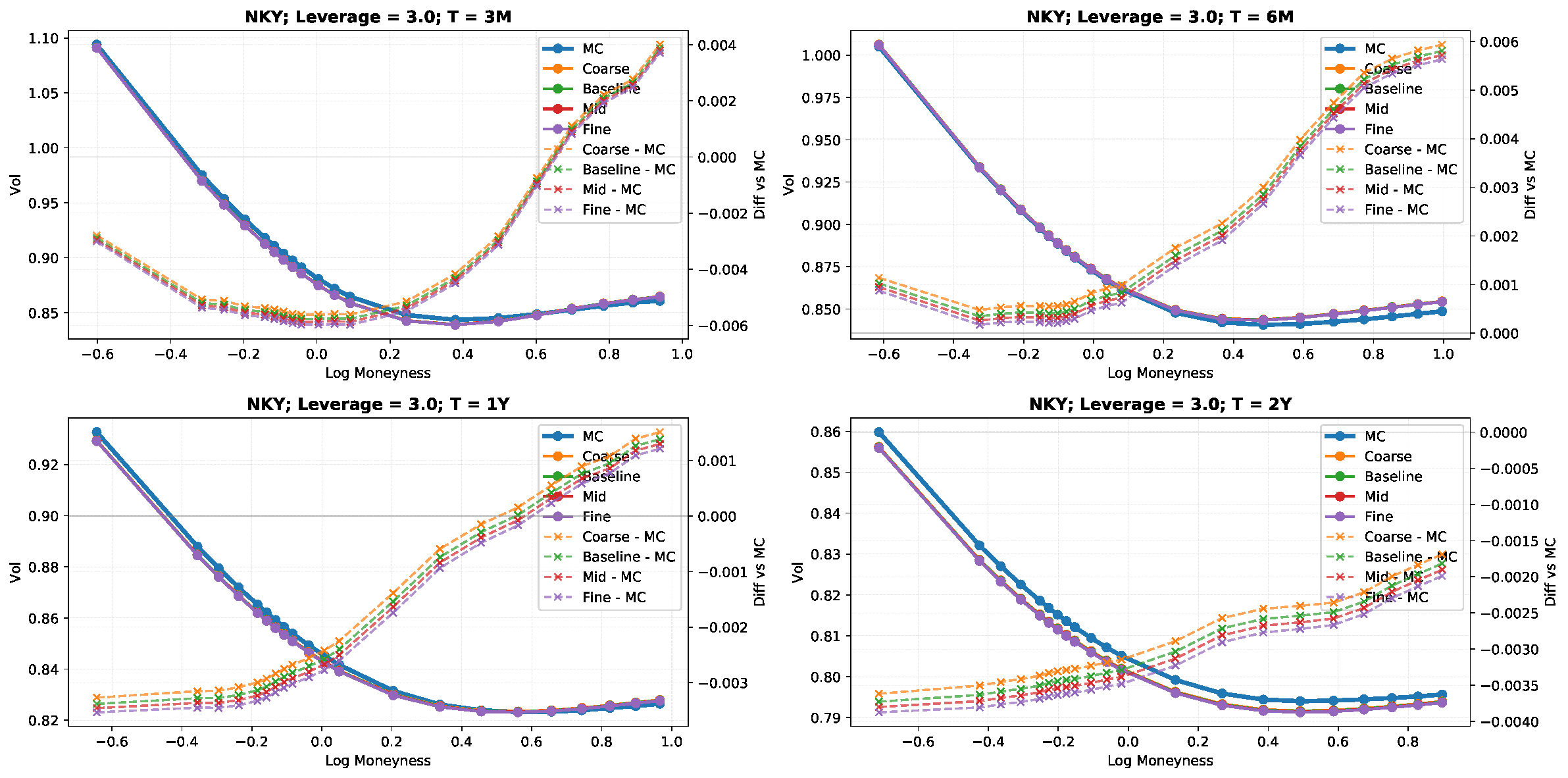}
\caption{$\beta=3$}
\end{subfigure}

\caption{Grid convergence of the forward joint-law solve against the Monte
Carlo benchmark, NKY, 22~June~2026.}
\label{fig:conv-nky}
\end{figure}

Figures~\ref{fig:conv-spx} and~\ref{fig:conv-nky} show the convergence study for
SPX and NKY.  Each panel plots the LETF implied-volatility smile for the four
configurations of Table~\ref{tab:grids} together with the Monte Carlo benchmark
and its sampling band.

The curves move in an ordered way under refinement, and the change from
baseline to mid to fine is far smaller than the change from coarse to baseline.
The baseline smile also lies inside, or very close to, the Monte Carlo band over
the quoted strike range. Thus further grid refinement has little effect, and the
joint-law solve is consistent with the independent simulation.

Residual differences are largest in the wings, and grow with the leverage
magnitude.  A fixed LETF wing maps back to a more remote region of the
underlying distribution when $|\beta|$ is large, and that region carries less
probability mass, fewer effective grid points, and the largest simulation error.
These wing discrepancies come from discretisation and sampling error, not from
the joint-law method itself.

%------------------------------------------------------------------------------
\subsection{Accuracy of the approximations}
\label{sec:numerics-approximation}
%------------------------------------------------------------------------------

Since the baseline joint-law solve is converged, it is used as the reference for
the approximation study.  Three curves are compared at each strike.

\begin{enumerate}
\item \textbf{First order.}  The strike remapping of
Theorem~\ref{thm:first-LV}, obtained in closed form by evaluating the underlying
implied-volatility surface at the remapped strike $K^*$ and scaling by $|\beta|$.
It is labelled as ``1stOrder'' in the figures.

\item \textbf{Conditional-Gamma.}  The price of
Theorem~\ref{thm:conditional-gamma-price}, with $m(x)$ and $v(x)$ from the
triangular moment system~\eqref{eq:moment-pde} and the martingale normalisation
\eqref{eq:gamma-normaliser} applied before implied-volatility inversion.
It is labelled as ``2ndOrder'' in the figures.

\item \textbf{Reference.}  The closure-free forward joint-law price on the
baseline grid.
It is labelled as ``2dPDE'' in the figures.
\end{enumerate}

Cost increases in the same order: no PDE solve, three one-dimensional forward
solves on a shared grid plus a one-dimensional quadrature per strike, and a
two-dimensional degenerate forward solve in $(x,i)$.  The comparison measures
the accuracy retained by the one-dimensional conditional-Gamma method relative
to the two-dimensional reference.

\subsubsection{Approximation accuracy}

\begin{figure}[p]
\centering
\begin{subfigure}[b]{\panelwidth}
\centering
\panel{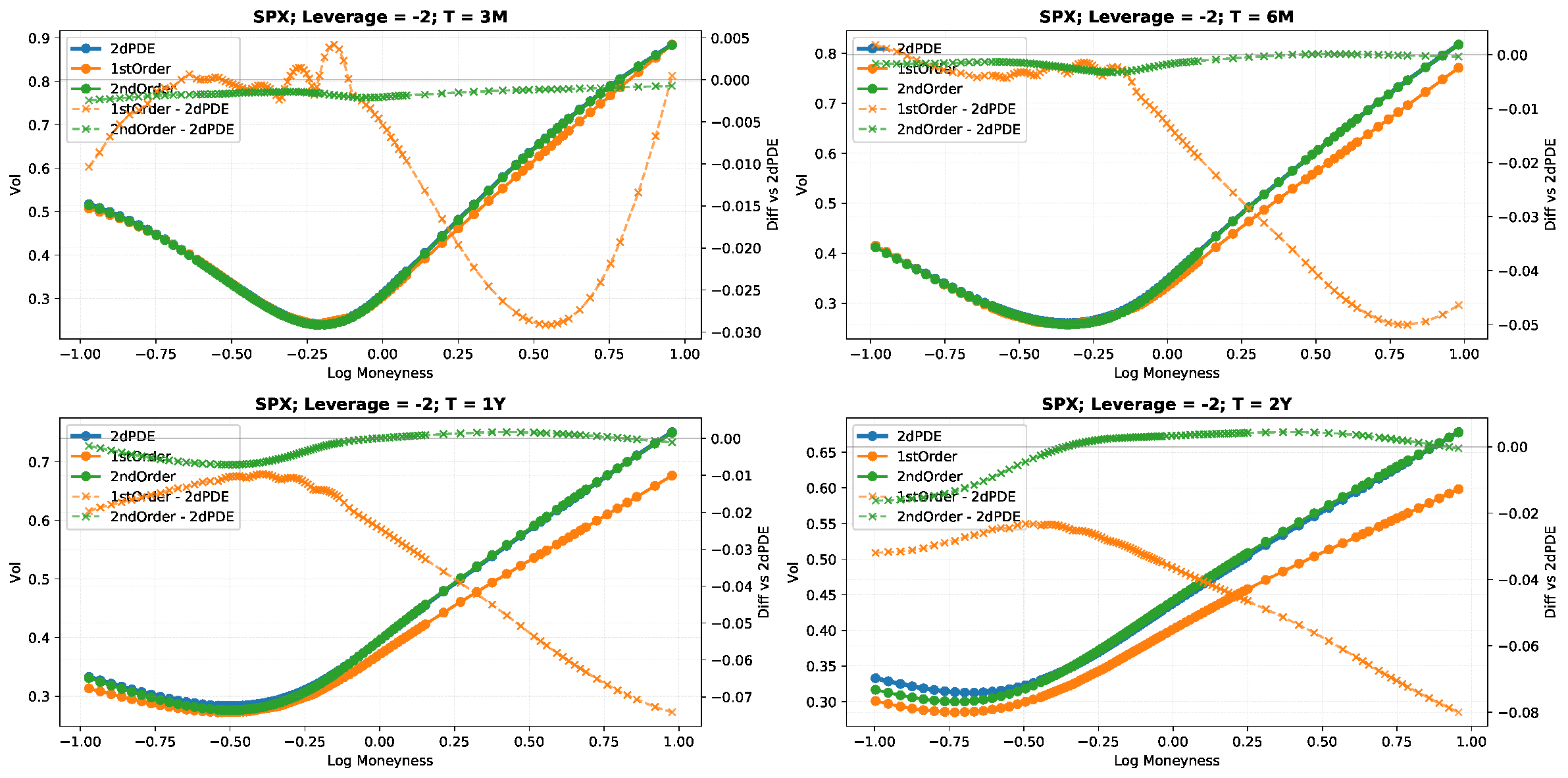}
\caption{$\beta=-2$}
\end{subfigure}

\vfill

\begin{subfigure}[b]{\panelwidth}
\centering
\panel{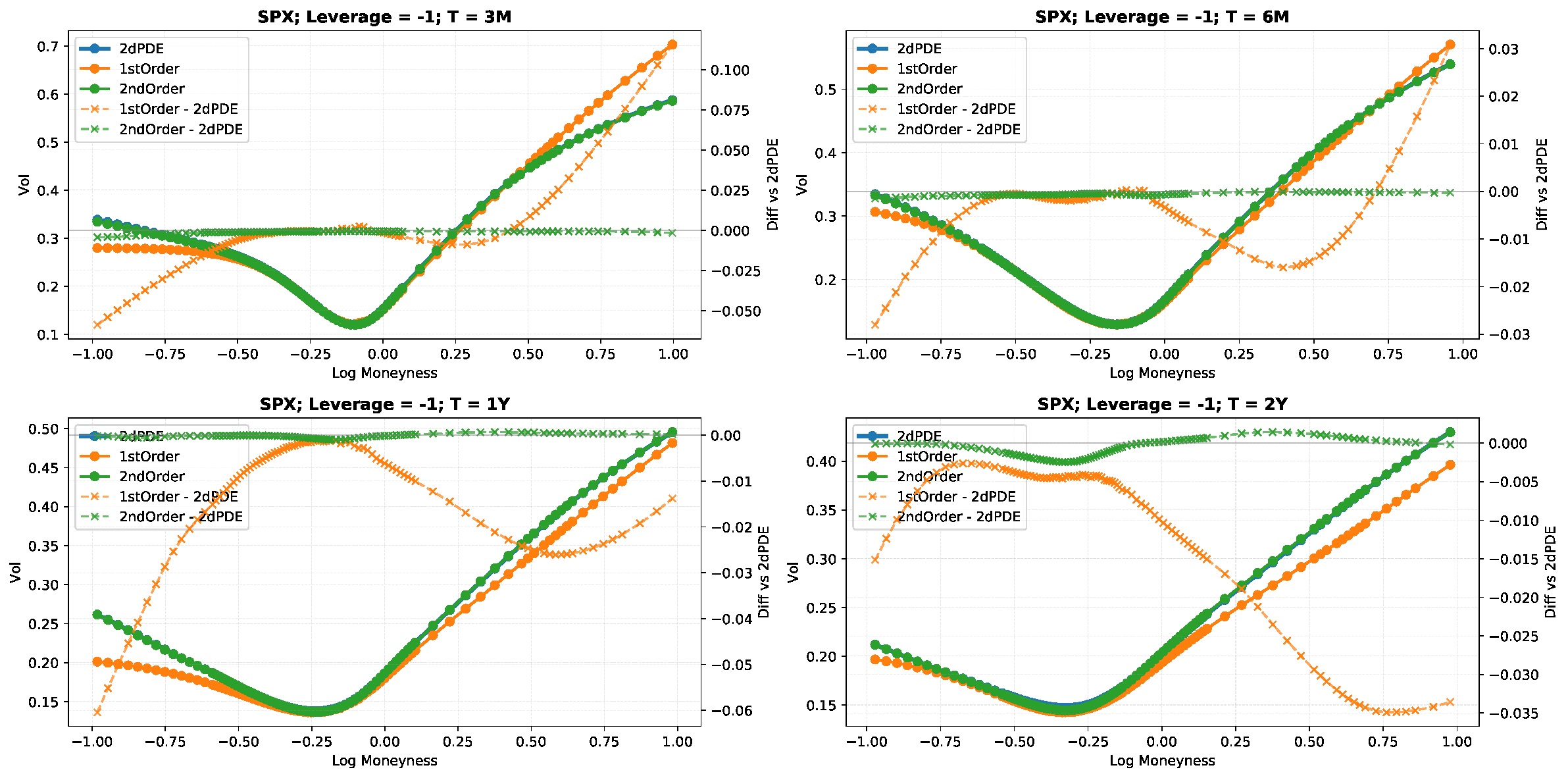}
\caption{$\beta=-1$}
\end{subfigure}

\vfill

\begin{subfigure}[b]{\panelwidth}
\centering
\panel{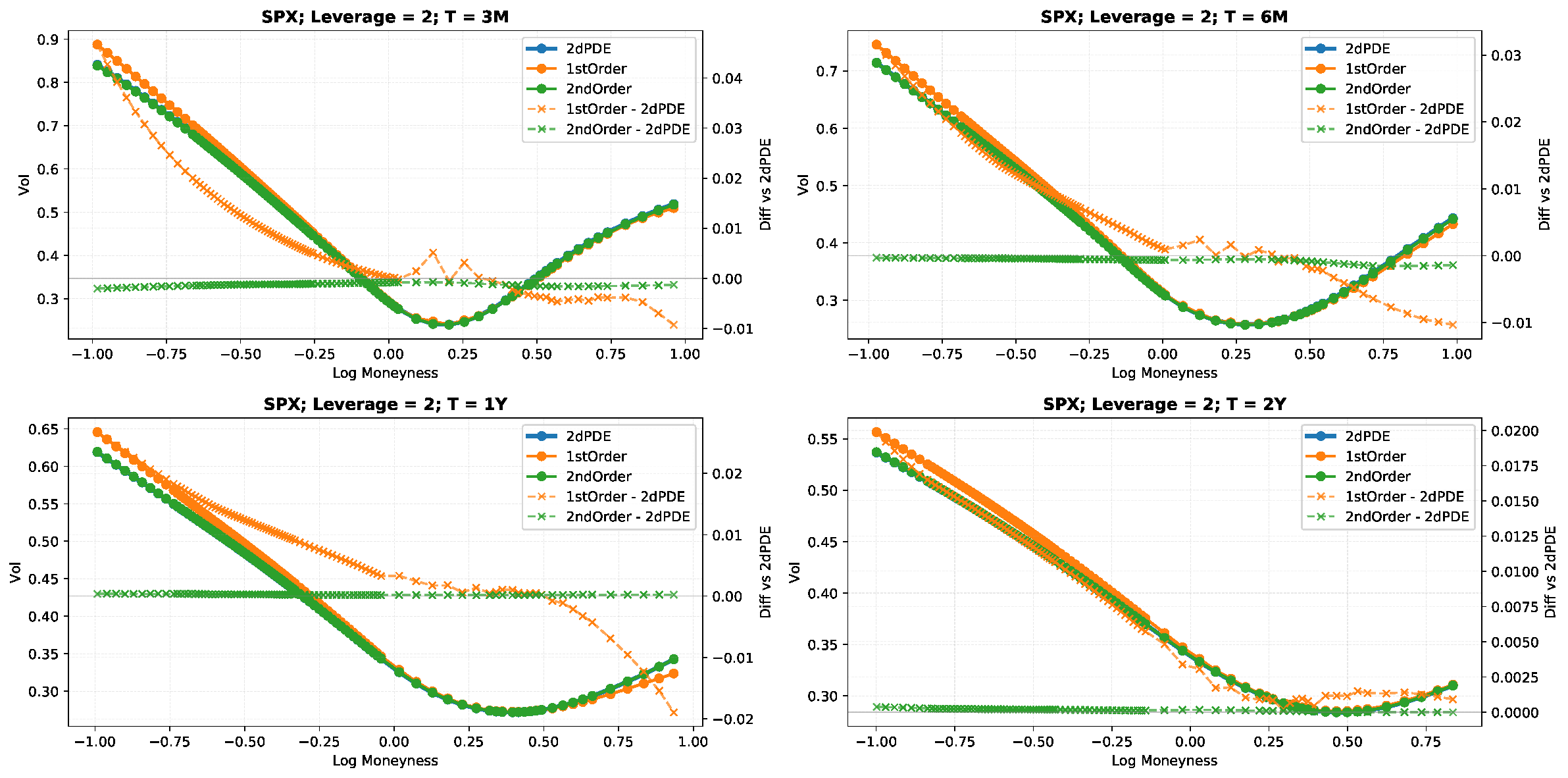}
\caption{$\beta=2$}
\end{subfigure}

\vfill

\begin{subfigure}[b]{\panelwidth}
\centering
\panel{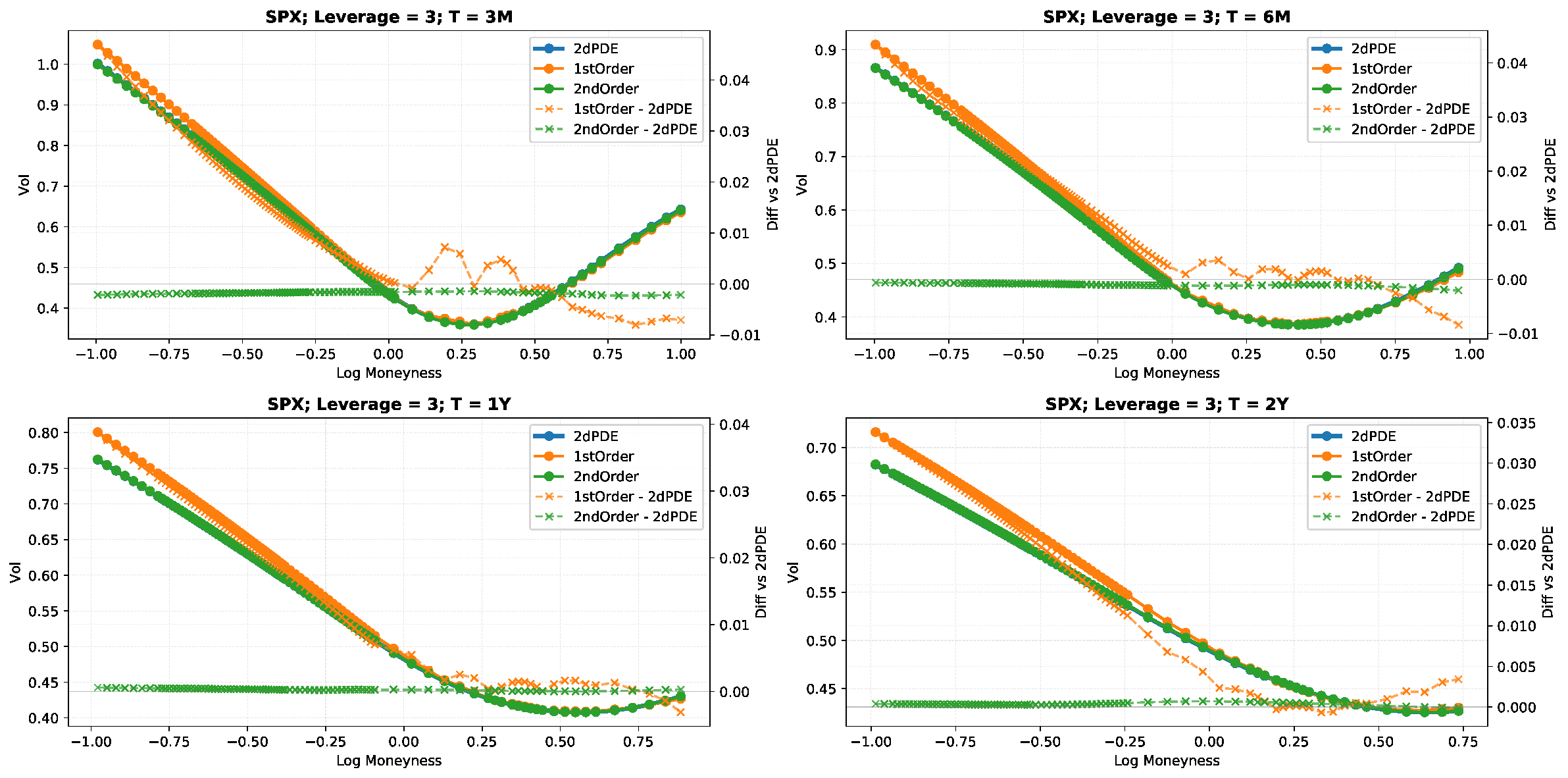}
\caption{$\beta=3$}
\end{subfigure}

\caption{First-order strike remapping and conditional-Gamma method against the
baseline forward joint-law reference, SPX, 22~June~2026.}
\label{fig:approx-spx}
\end{figure}

\begin{figure}[p]
\centering
\begin{subfigure}[b]{\panelwidth}
\centering
\panel{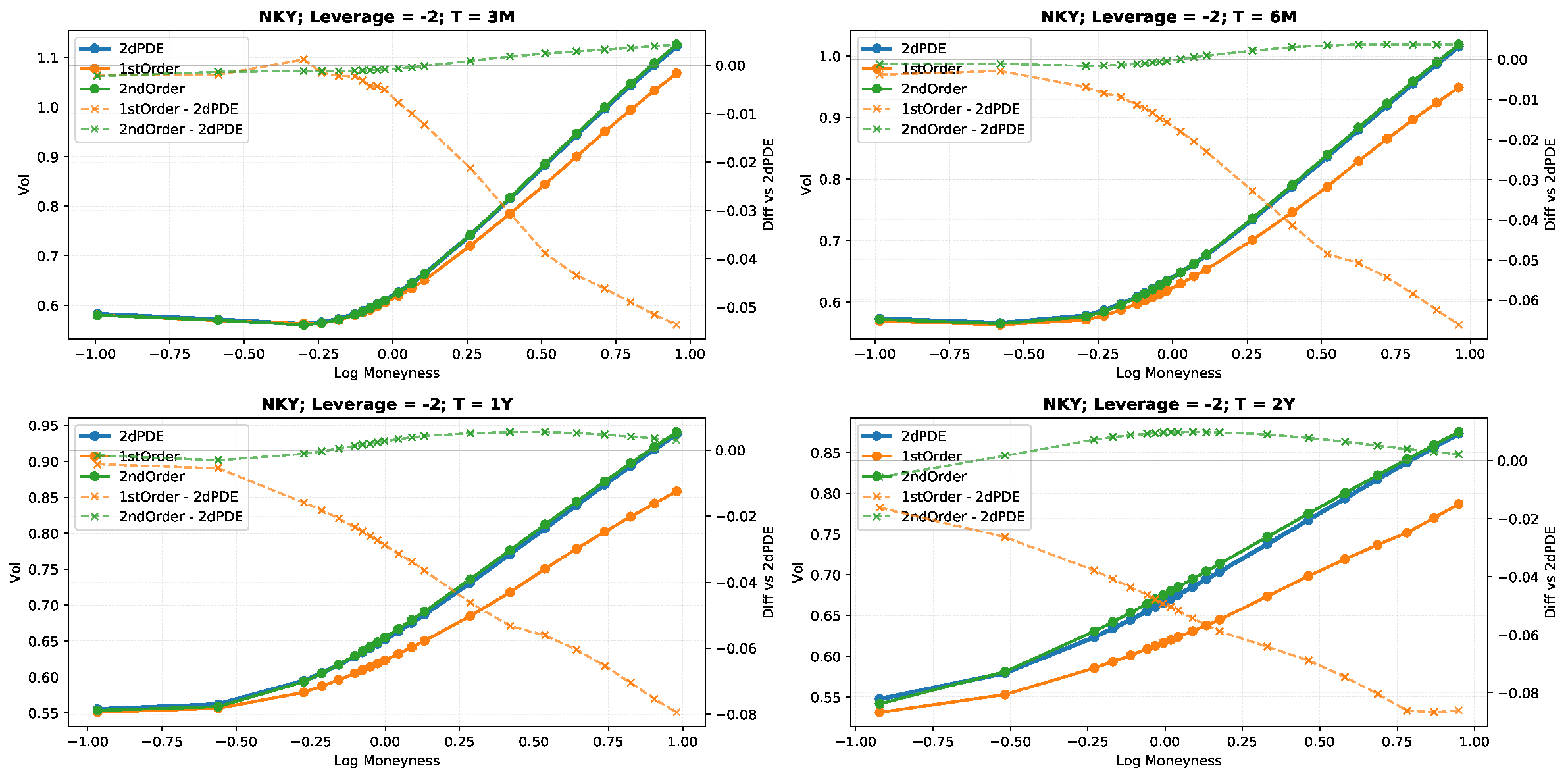}
\caption{$\beta=-2$}
\end{subfigure}

\vfill

\begin{subfigure}[b]{\panelwidth}
\centering
\panel{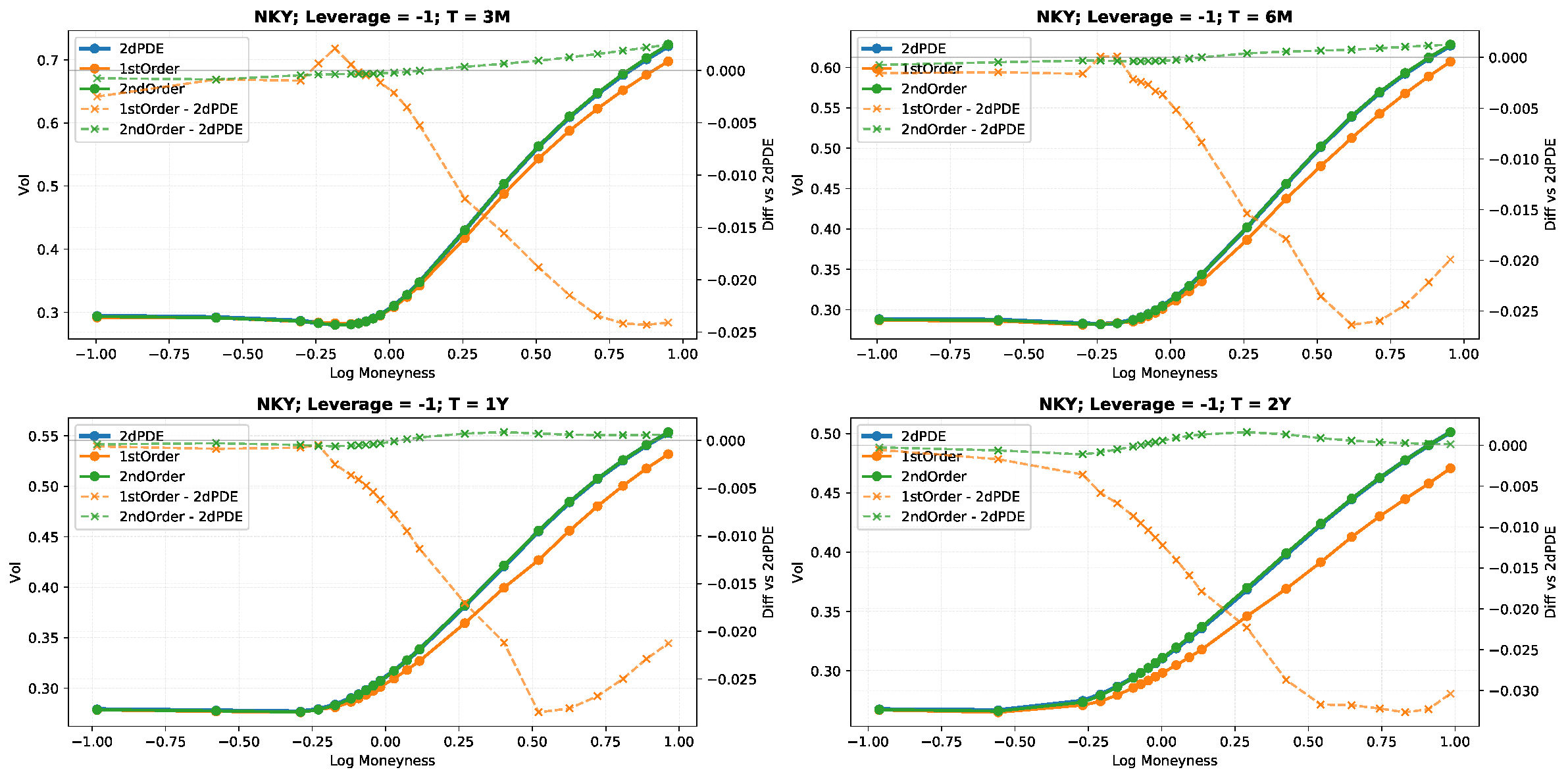}
\caption{$\beta=-1$}
\end{subfigure}

\vfill

\begin{subfigure}[b]{\panelwidth}
\centering
\panel{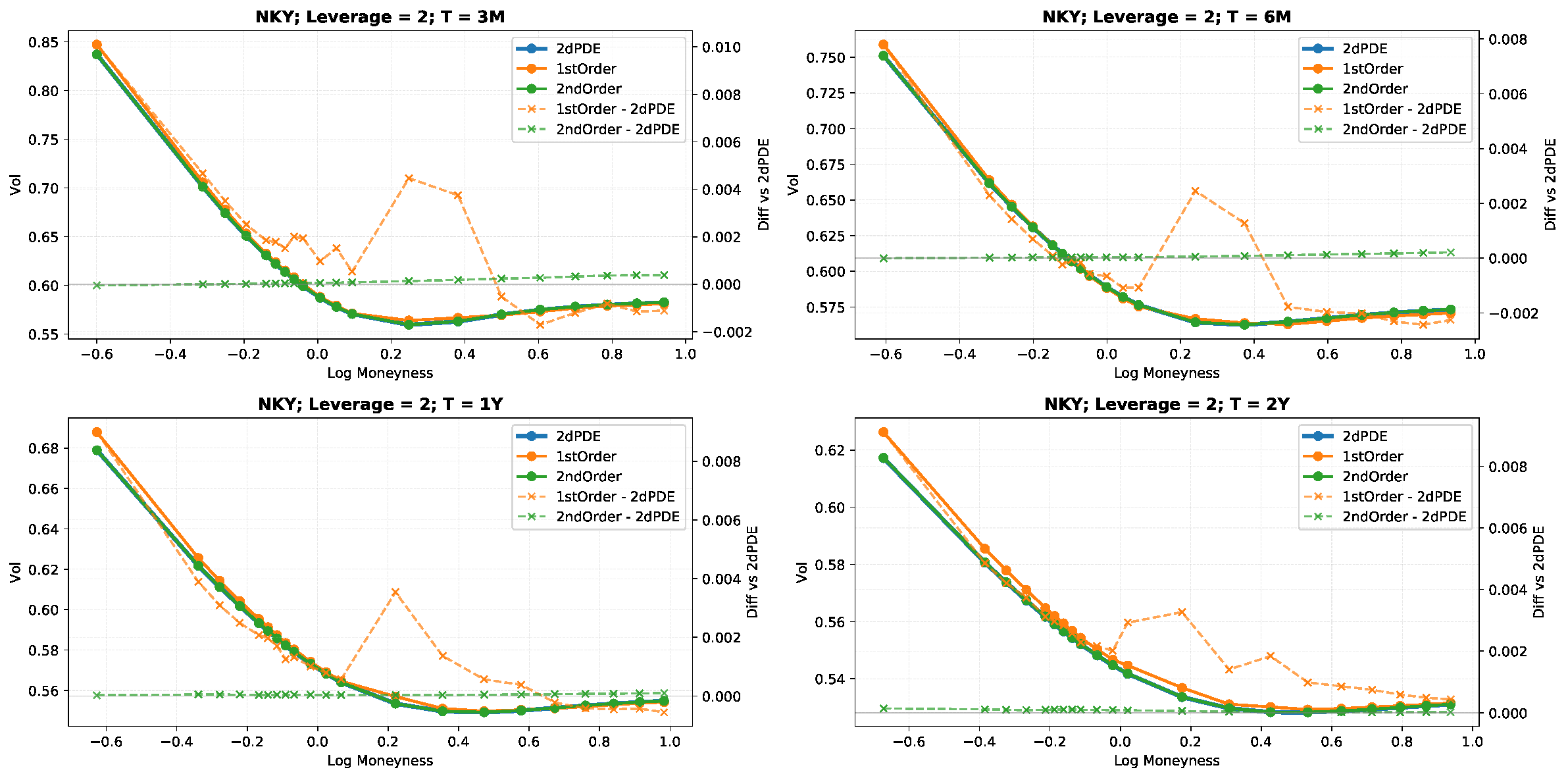}
\caption{$\beta=2$}
\end{subfigure}

\vfill

\begin{subfigure}[b]{\panelwidth}
\centering
\panel{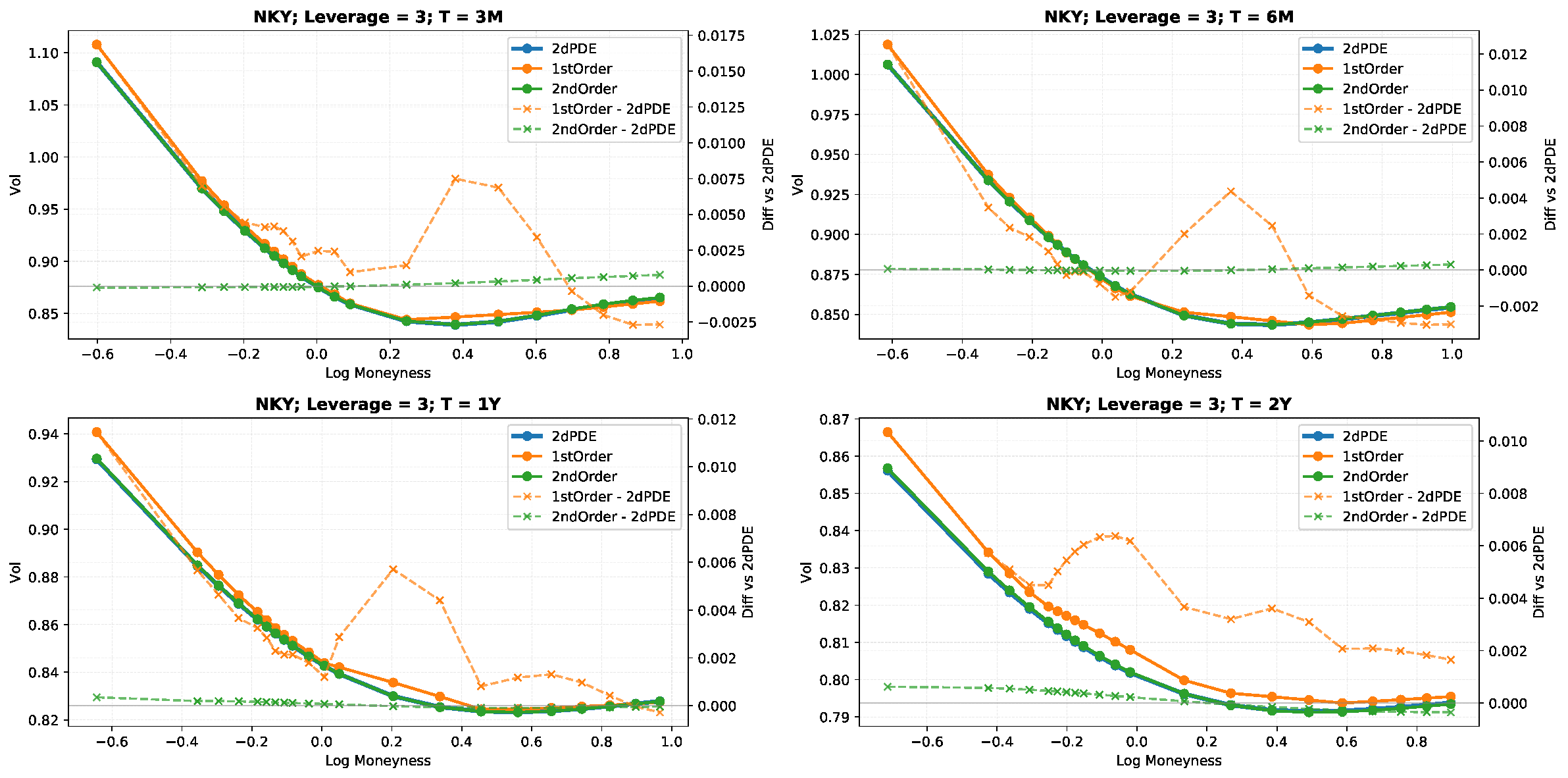}
\caption{$\beta=3$}
\end{subfigure}

\caption{First-order strike remapping and conditional-Gamma method against the
baseline forward joint-law reference, NKY, 22~June~2026.}
\label{fig:approx-nky}
\end{figure}

Figures~\ref{fig:approx-spx} and~\ref{fig:approx-nky} compare the two
approximations against the joint-law reference.

The first-order curve captures the level and slope of the reference smile near
the money, since it evaluates the underlying smile at a leverage-adjusted strike
and inherits the dominant shape of the vanilla surface. It drifts
away from the reference in the wings because it replaces the random variance
drag $I_T$ by the deterministic base-model quantity entering $V_T^*$, capturing
an average drag but neither the conditional dispersion of $I_T$ nor its
dependence on the terminal underlying return.

The conditional-Gamma curve stays close to the reference across the quoted
strike range, and the improvement is largest exactly where the first-order error
is largest.  The correction is therefore strike-dependent rather than a parallel
shift of the first-order smile. Conditioning on $X_T=x$ associates each strike
with the relevant region of the terminal underlying distribution, while $v(x)$
supplies the convexity correction from $e^{-\gamma I_T}$.

The remaining conditional-Gamma error is larger at $\gamma=3$ ($\beta\in\{3,-2\}$)
than at $\gamma=1$ ($\beta\in\{2,-1\}$) because the payoff is
more sensitive to dispersion in realised variance when the drag coefficient is
larger.  It is also larger for NKY than for SPX at matched leverage. A likely
cause, not isolated here, is the steeper NKY skew: local variance then varies
more across paths that end at the same $x$, which raises the conditional
dispersion of $I_T$.  In every tested case the error is smaller than the
first-order error shown in the same panel.

%------------------------------------------------------------------------------
\subsection{Two-moment against three-moment closure}
\label{sec:numerics-shifted}
%------------------------------------------------------------------------------

The conditional-Gamma error remaining in
Figures~\ref{fig:approx-spx} and~\ref{fig:approx-nky} has two possible sources:
the conditional moments themselves, which are exact up to discretisation, and
the shape of the closure, which is not.  Only the second is addressable without
leaving the one-dimensional cost regime, and the constraint
\eqref{eq:two-moment-skew-constraint} identifies it precisely: the two-moment
closure cannot choose its conditional skewness.  This subsection measures how
much of the residual error that constraint accounts for, by replacing the
two-moment closure with the conditional shifted-Gamma method of
Theorem~\ref{thm:shifted-gamma-price}.

\begin{figure}[p]
\centering
\begin{subfigure}[b]{\panelwidth}
\centering
\panel{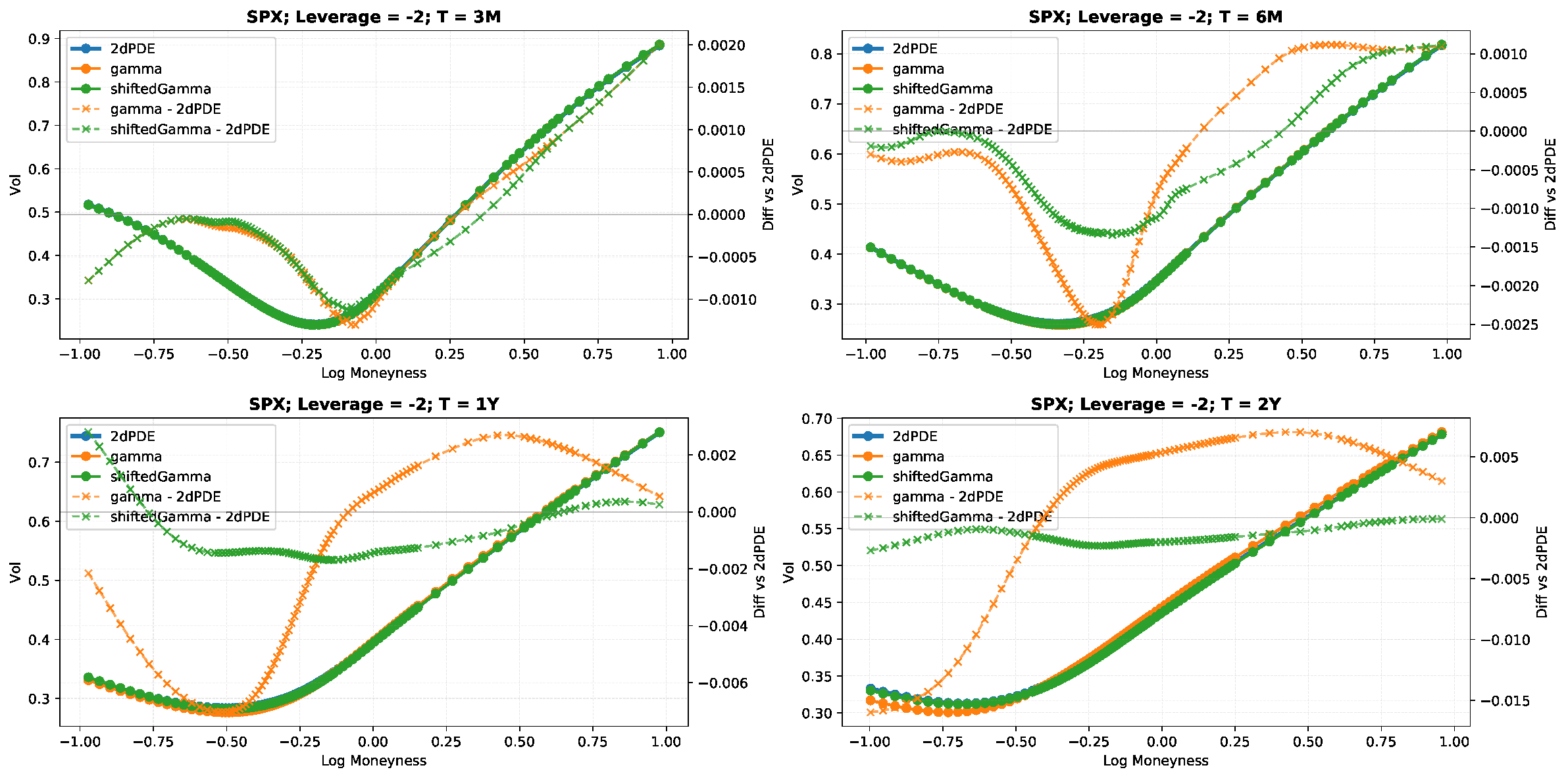}
\caption{$\beta=-2$}
\end{subfigure}

\vfill

\begin{subfigure}[b]{\panelwidth}
\centering
\panel{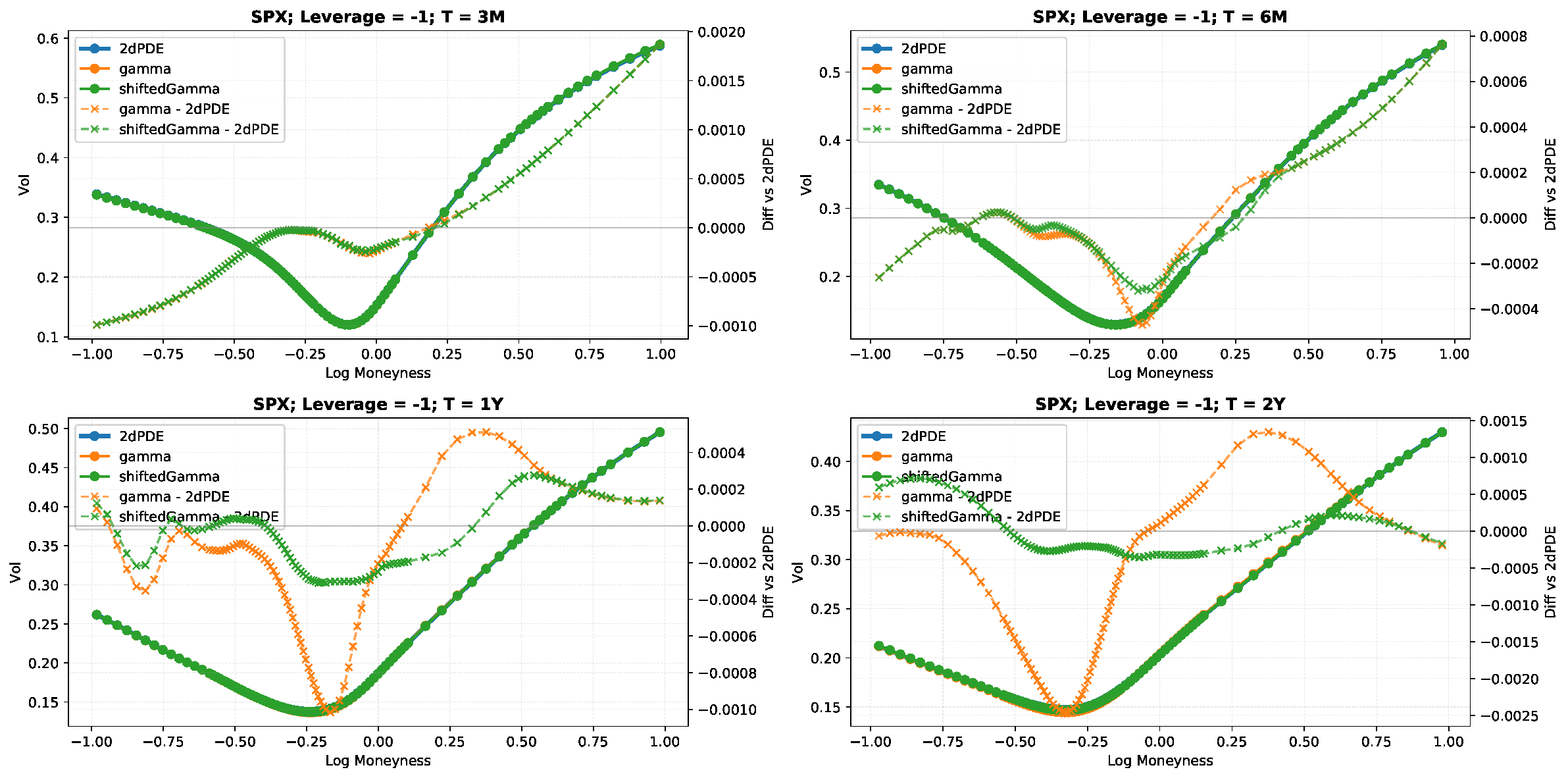}
\caption{$\beta=-1$}
\end{subfigure}

\vfill

\begin{subfigure}[b]{\panelwidth}
\centering
\panel{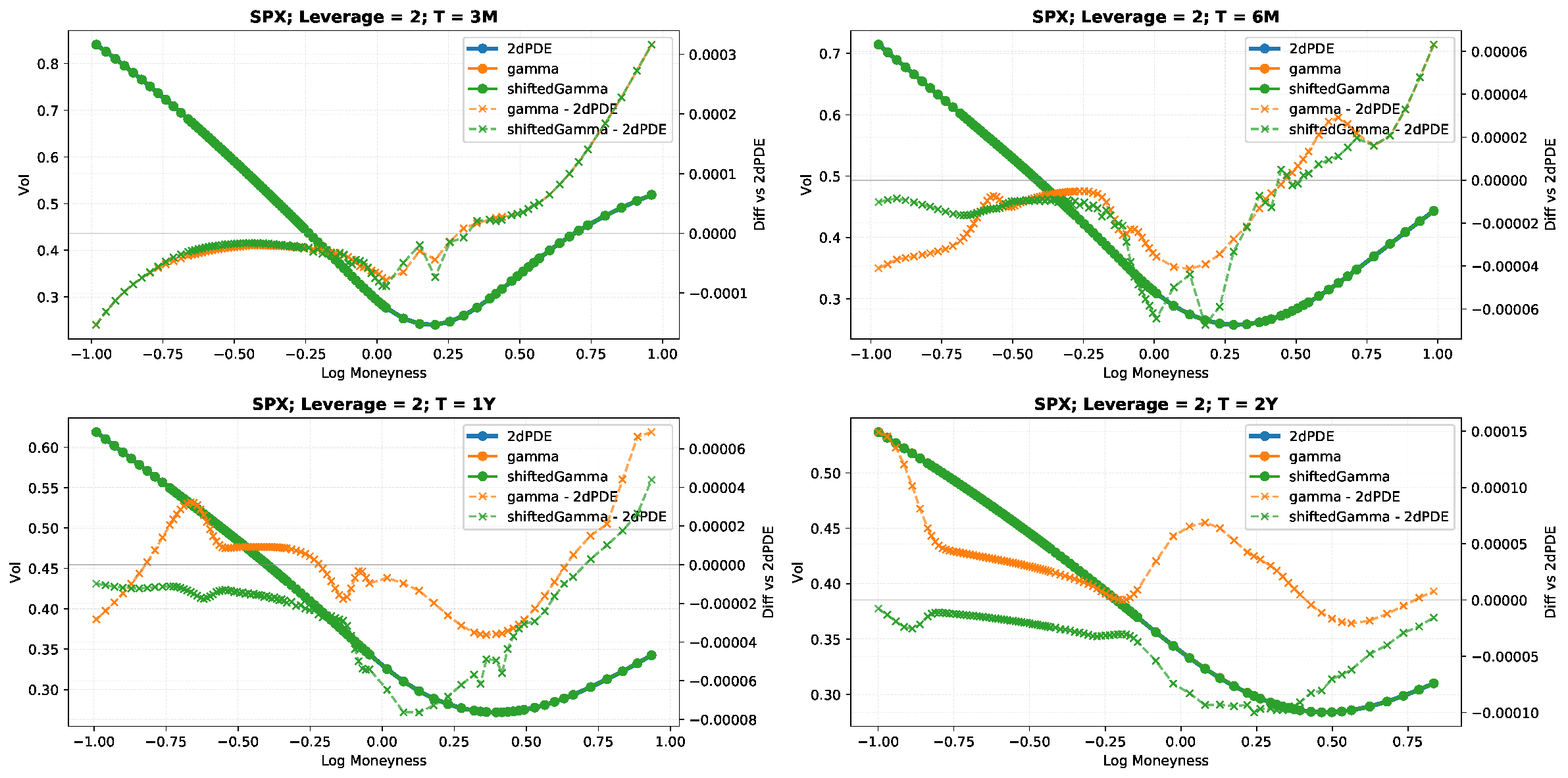}
\caption{$\beta=2$}
\end{subfigure}

\vfill

\begin{subfigure}[b]{\panelwidth}
\centering
\panel{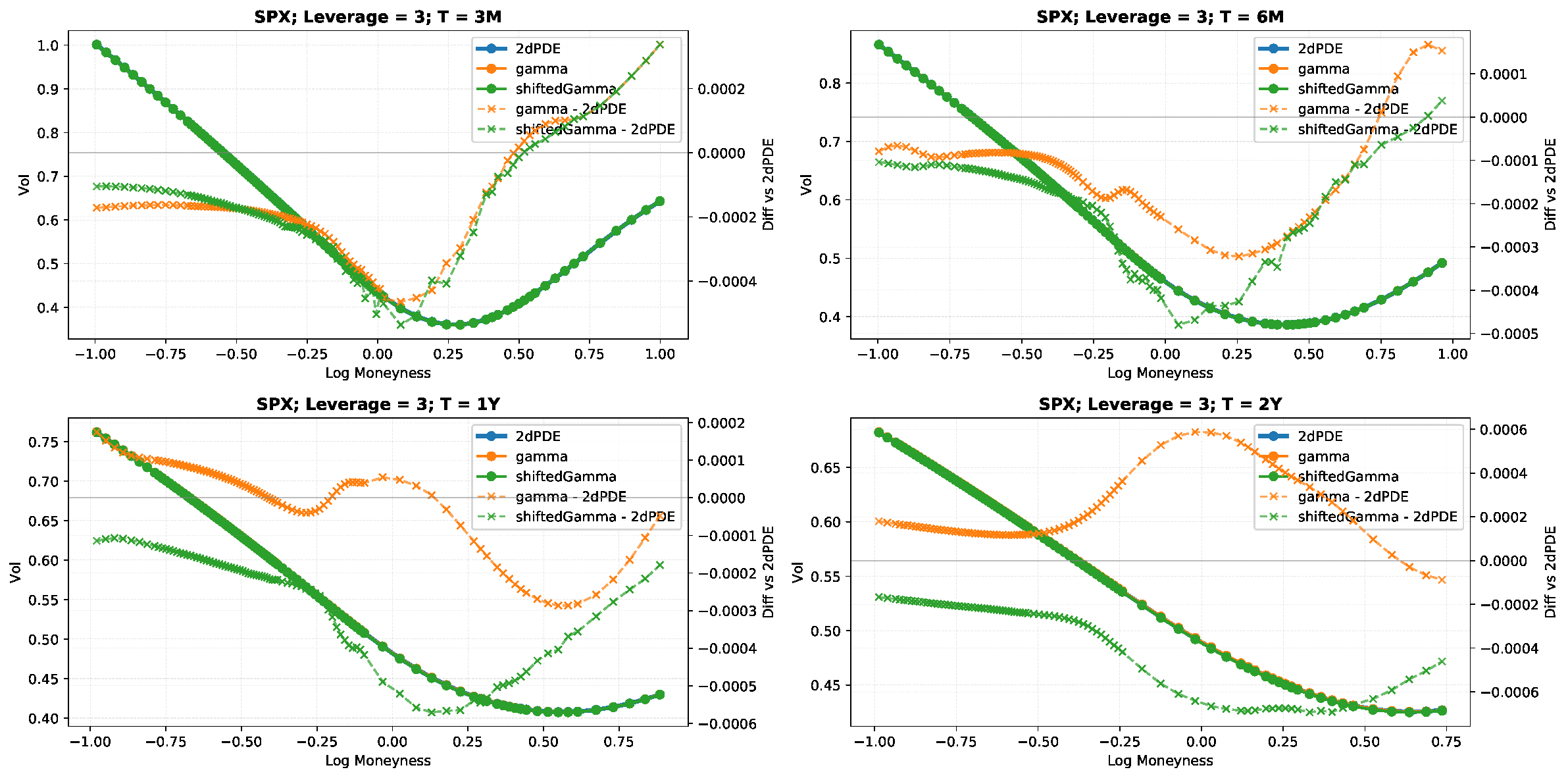}
\caption{$\beta=3$}
\end{subfigure}

\caption{Two-moment conditional-Gamma and three-moment conditional
shifted-Gamma methods against the baseline forward joint-law reference, SPX,
22~June~2026.}
\label{fig:shifted-spx}
\end{figure}

\begin{figure}[p]
\centering
\begin{subfigure}[b]{\panelwidth}
\centering
\panel{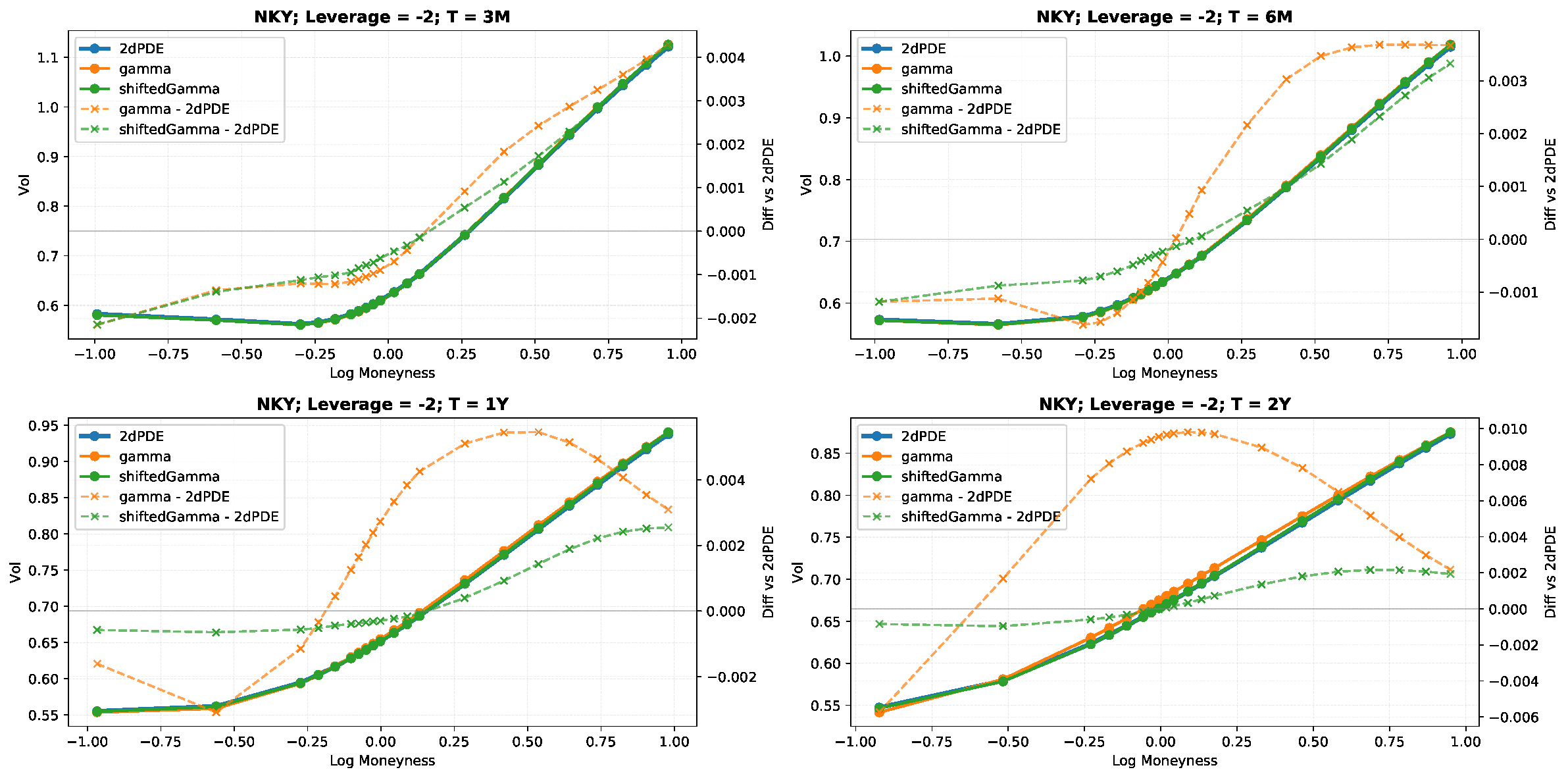}
\caption{$\beta=-2$}
\end{subfigure}

\vfill

\begin{subfigure}[b]{\panelwidth}
\centering
\panel{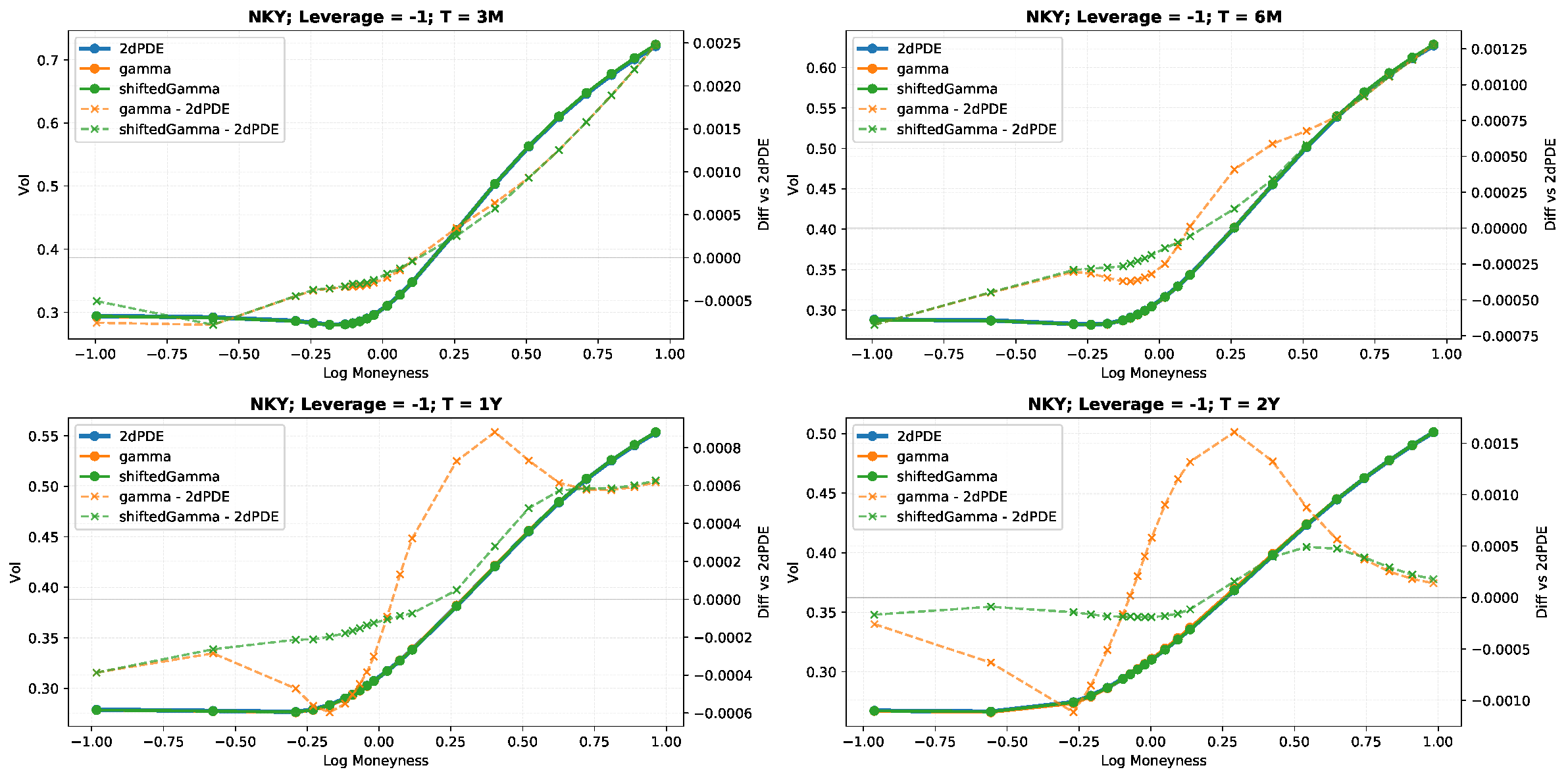}
\caption{$\beta=-1$}
\end{subfigure}

\vfill

\begin{subfigure}[b]{\panelwidth}
\centering
\panel{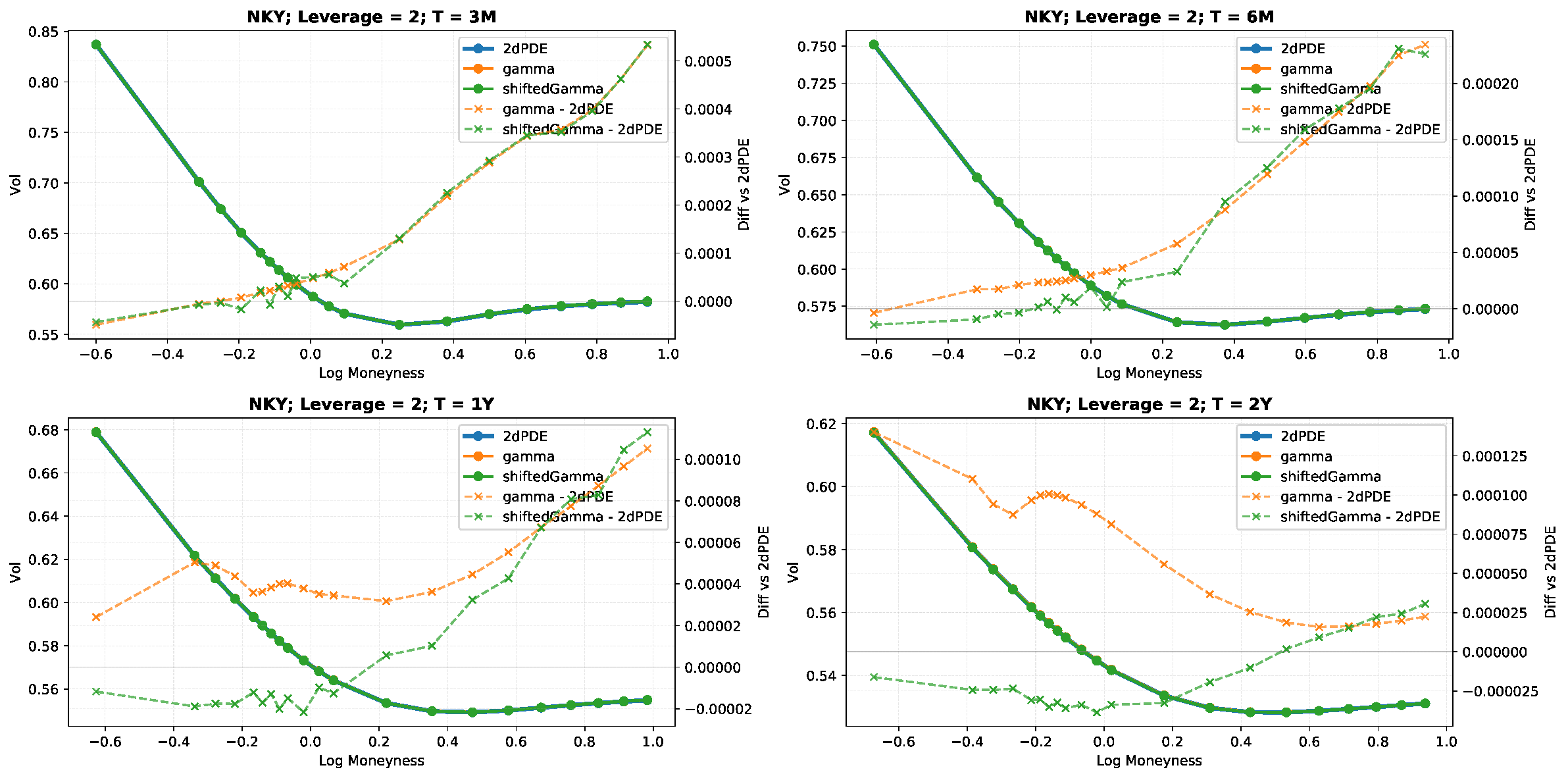}
\caption{$\beta=2$}
\end{subfigure}

\vfill

\begin{subfigure}[b]{\panelwidth}
\centering
\panel{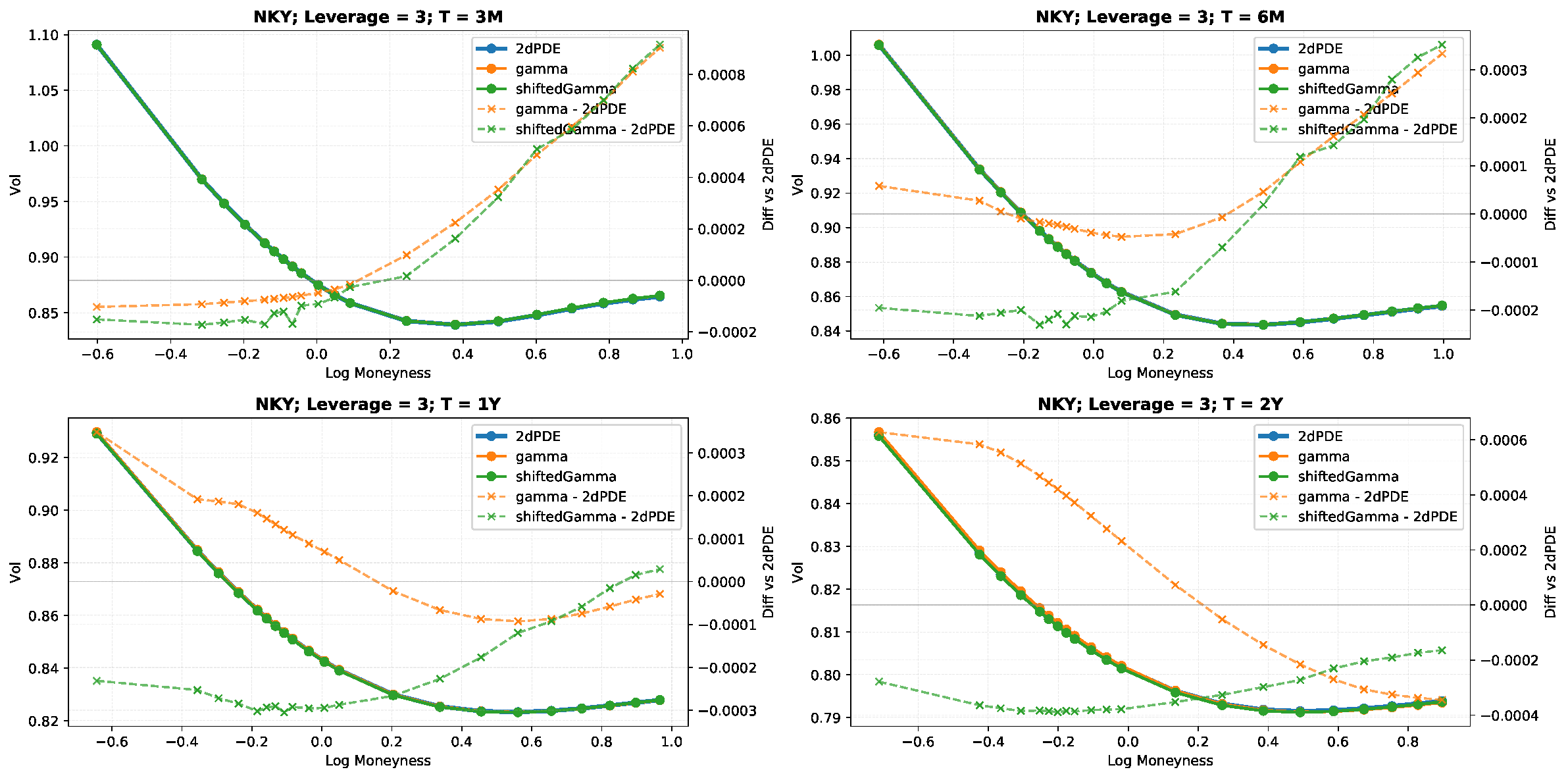}
\caption{$\beta=3$}
\end{subfigure}

\caption{Two-moment conditional-Gamma and three-moment conditional
shifted-Gamma methods against the baseline forward joint-law reference, NKY,
22~June~2026.}
\label{fig:shifted-nky}
\end{figure}

Figures~\ref{fig:shifted-spx} and~\ref{fig:shifted-nky} show the outcome.  Each
panel plots the two conditional smiles and the reference, labelled
``gamma'', ``shiftedGamma'' and ``2dPDE'', together with the two differences
against the reference on a second axis.

\subsubsection{Comparison results}

The two-moment error is small for the long ratios and at short maturities, but
grows with maturity for the inverse ratios. In those cases the third moment
removes most of the residual error, with a larger reduction in the peak error
than in the typical error. At $3$M the two closures are almost
indistinguishable; their difference widens as $T$ grows.

For $\beta\in\{2,3\}$ both closures are accurate to a few hundredths of a
volatility point, and their difference curves overlap over the central strikes.
At SPX $\beta=3$ the three-moment closure has marginally larger errors at the
longer maturities. These differences are comparable to the discretisation error
of the reference and the outer quadrature, so they do not support a robust
ranking of the closures in those cases.

The split follows the sign of $\beta$, not the drag coefficient alone.
The cases $\beta=-2$ and $\beta=3$ share the same $\gamma=3$, yet the
inverse ratio is much the harder for the two-moment closure.  The reason is
where the LETF strikes sit in the underlying distribution: for $\beta<0$ the map
$x\mapsto\beta x$ sends the LETF call payoff to the region of low underlying
returns, where a skewed local-volatility surface makes the conditional law of
$I_T$ both more dispersed and more asymmetric, so the constraint
\eqref{eq:two-moment-skew-constraint} binds hardest.  This is consistent with
the reading of Section~\ref{sec:shifted-gamma}: the floor $\zeta(x)$ is largest,
and the shape restriction of the unshifted fit costliest, where conditional
realised variance is large and its law far from the Gamma shape that two
moments impose.

%------------------------------------------------------------------------------
\subsection{Empirical test against listed LETF quotes}
\label{sec:numerics-empirical}
%------------------------------------------------------------------------------

Sections~\ref{sec:numerics-convergence}--\ref{sec:numerics-shifted}
are internal consistency tests: every curve is produced from the same
calibrated local-volatility surface, and the reference is a numerical solution
of the model rather than a traded price. We now ask whether the LETF smile
implied by the underlying surface agrees with an LETF option market quoted
independently of it.

\subsubsection{Data and procedure}

We use the Nasdaq-100 ETF (QQQ) as the underlying and its three-times leveraged
listed product (TQQQ), so that $\beta=3$ and $\gamma=3$, the largest drag
coefficient considered in Section~\ref{sec:numerics-approximation}. Three
market dates are tested, 20~March~2026, 22~June~2026 and 25~September~2026.
They are spaced roughly one quarter apart.

For each date four listed expiries are retained: the quoted maturities nearest
to 1M, 3M and 6M, together with a long-dated expiry near one year.
The last is dictated by the listed expiry calendar
and is 10M on 20~March~2026, 18M on 22~June~2026 and 16M on
25~September~2026.

Listed QQQ and TQQQ options are American-style, so both sets of quotes are
converted to European implied volatilities: QQQ before the local-volatility
calibration and TQQQ before comparison. For each expiry the LETF
forward $F_T^L$ is backed out from put--call parity at the strike where the
call and put prices are closest. This market-implied forward absorbs the
fund's fees and financing; daily rebalancing and tracking error remain outside
the model.

The curve labels in Figure~\ref{fig:empirical-tqqq} follow
Section~\ref{sec:numerics-approximation}: ``2dPDE'' is the joint-law reference
of Section~\ref{sec:joint-law}, ``2ndOrder'' the conditional-Gamma method of
Theorem~\ref{thm:conditional-gamma-price} and ``1stOrder'' the strike remapping
of Theorem~\ref{thm:first-LV}. These are plotted against the TQQQ bid and ask
volatilities, all as functions of LETF forward log-moneyness.

\subsubsection{Results}

\begin{figure}[p]
\centering
\begin{subfigure}[b]{\emppanelwidth}
\centering
\emppanel{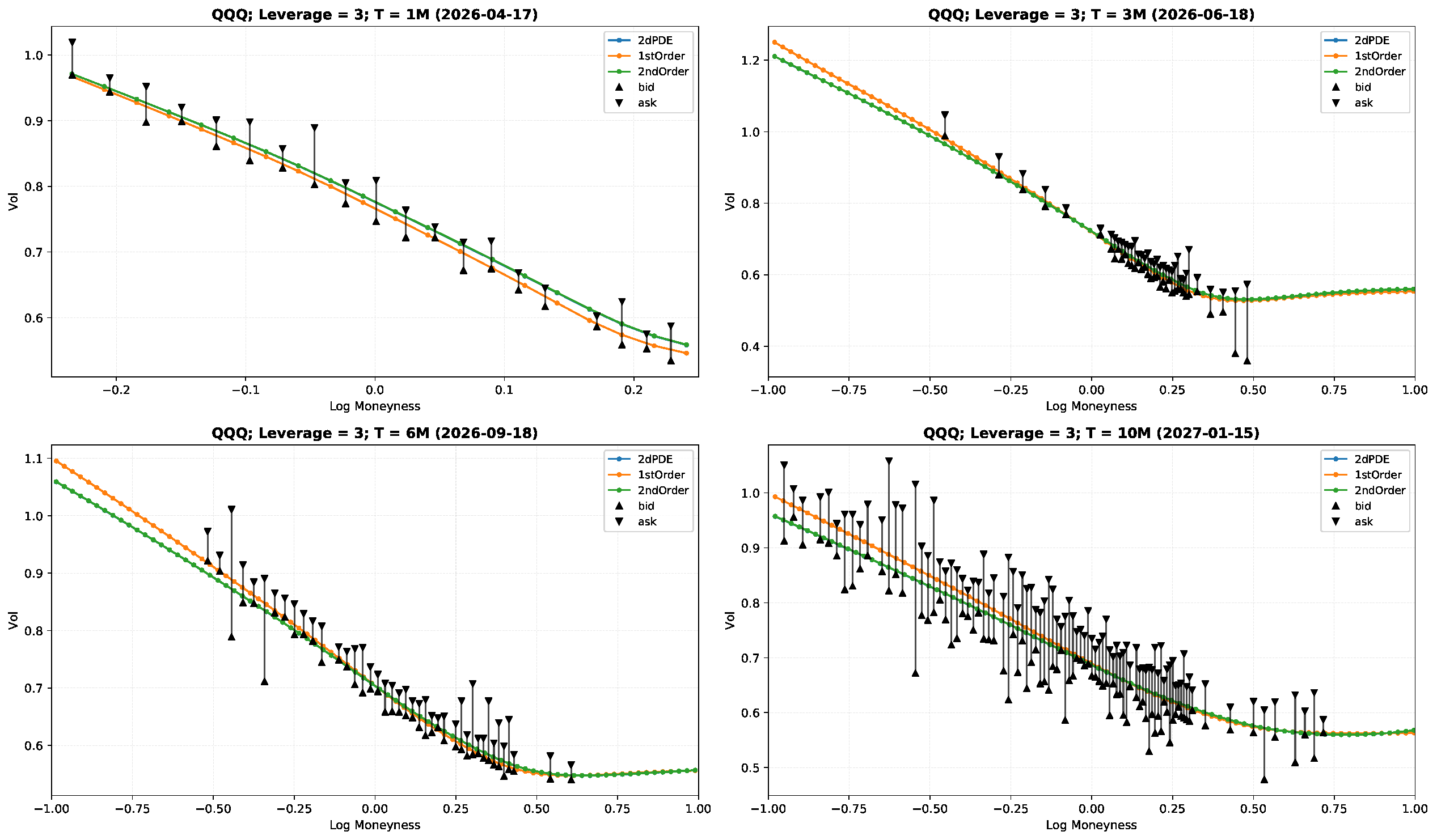}
\caption{20~March~2026; tenors 1M, 3M, 6M, 10M}
\end{subfigure}

\vfill

\begin{subfigure}[b]{\emppanelwidth}
\centering
\emppanel{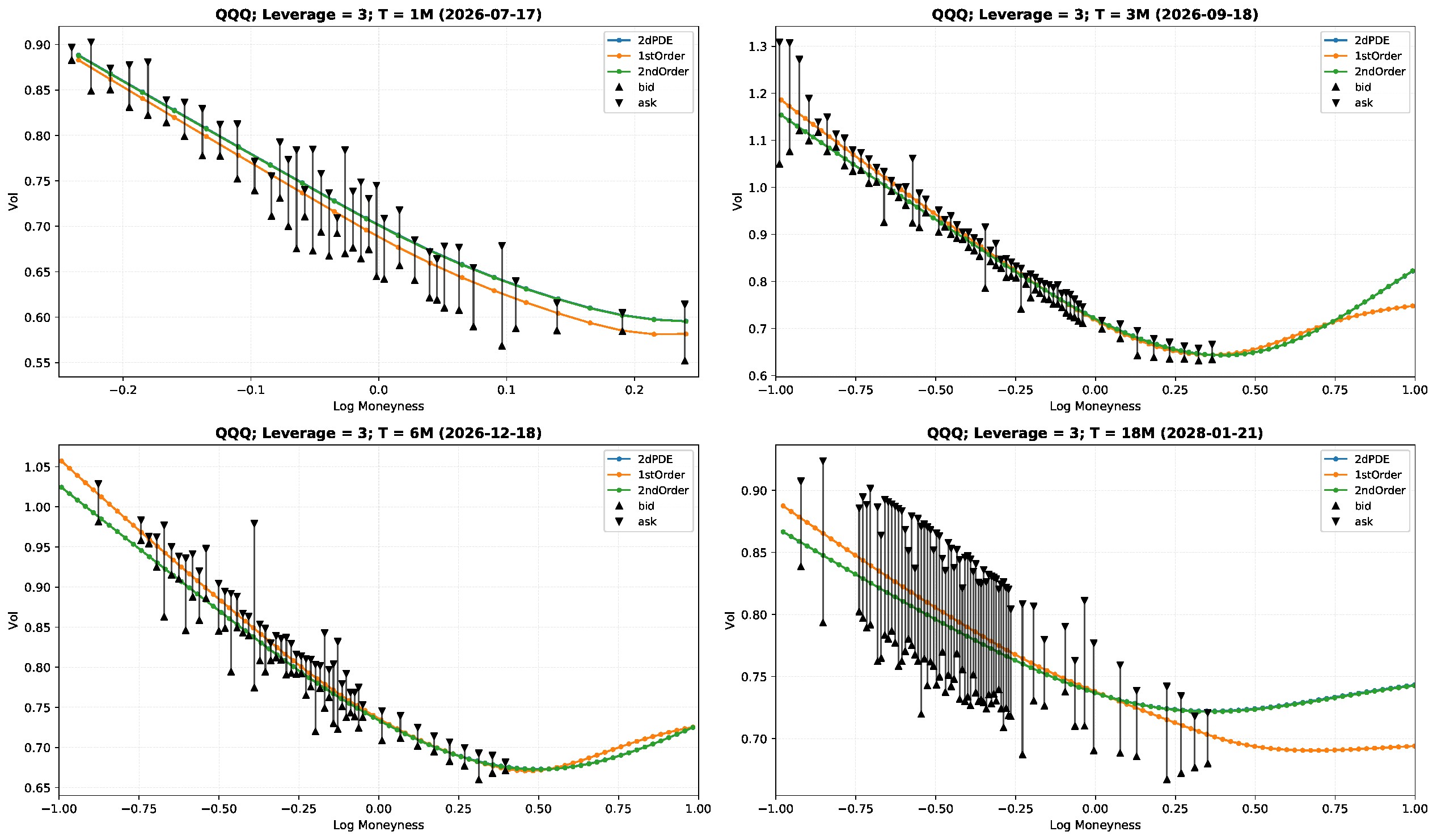}
\caption{22~June~2026; tenors 1M, 3M, 6M, 18M}
\end{subfigure}

\vfill

\begin{subfigure}[b]{\emppanelwidth}
\centering
\emppanel{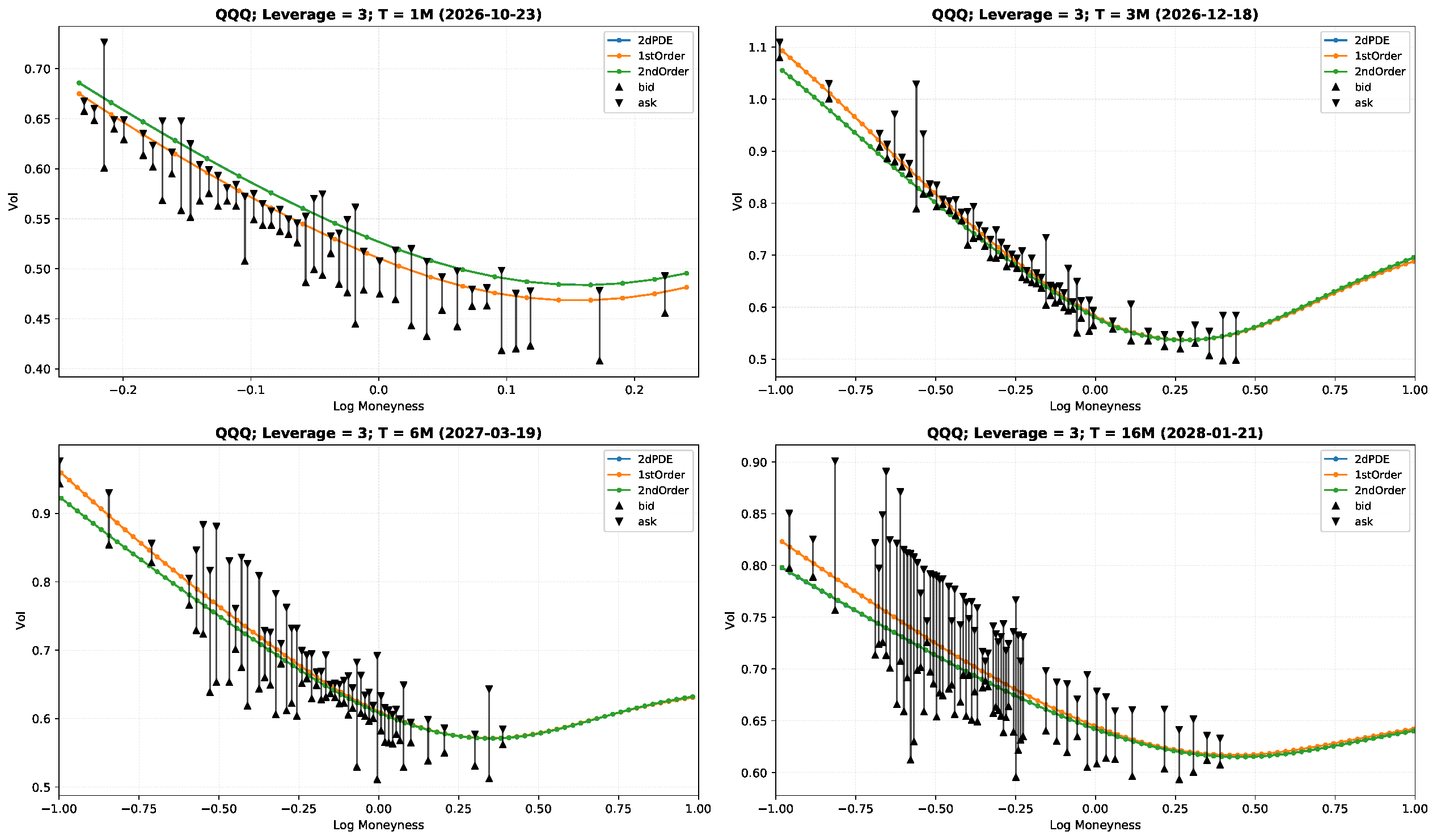}
\caption{25~September~2026; tenors 1M, 3M, 6M, 16M}
\end{subfigure}

\caption{Forward joint-law solve (``2dPDE''), conditional-Gamma method
(``2ndOrder'') and first-order strike remapping (``1stOrder''),
against TQQQ market bid and ask volatilities.}
\label{fig:empirical-tqqq}
\end{figure}

At strikes with a two-sided quote in Figure~\ref{fig:empirical-tqqq}, the
joint-law and conditional-Gamma curves overlap over essentially the whole quoted
range, with a separation orders of magnitude smaller than the spread. At market
resolution, the approximation error measured in
Section~\ref{sec:numerics-approximation} is therefore invisible.

The quotes also follow all three curves. Each method lies inside the band at
the large majority of strikes, and the exits are small. They occur mainly at
the shortest tenor, where the quoted band is narrowest. On 25~September~2026 at
1M, for instance, the joint-law and Gamma curves sit slightly above the ask
across most strikes.

%------------------------------------------------------------------------------
\subsection{Computational cost}
\label{sec:numerics-cost}
%------------------------------------------------------------------------------

Table~\ref{tab:timings} reports the wall-clock time to produce a full LETF
smile at the tested maturity for one $(\text{index},\beta)$ pair.  Each reported
timing includes the complete pricing chain for its method.  For the conditional
methods, this comprises the forward moment solves, Gamma parameterisation,
martingale normalisation and conditional quadrature over all strikes; for the
reference, it comprises the two-dimensional joint-law solve on the baseline
grid of Table~\ref{tab:grids} and the payoff integration.

\begin{table}[htbp]
\centering
\caption{Wall-clock time in seconds to produce a full LETF smile; the
shifted-Gamma column includes the extra $M_3$ solve.  Implementation in
Python~3.7.9, single-threaded.}
\label{tab:timings}
\begin{tabular}{llrrr}
\toprule
Index & $\beta$ & Cond.\ Gamma (s) & Shifted Gamma (s) & Joint law (s) \\
\midrule
NKY & $-2$ & 0.85 & 1.06 & 200.97 \\
NKY & $-1$ & 0.88 & 1.09 & 196.82 \\
NKY & $2$  & 0.85 & 1.06 & 196.61 \\
NKY & $3$  & 0.89 & 1.07 & 197.38 \\
SPX & $-2$ & 1.30 & 1.54 & 218.69 \\
SPX & $-1$ & 1.42 & 1.53 & 223.94 \\
SPX & $2$  & 1.23 & 1.49 & 223.19 \\
SPX & $3$  & 1.28 & 1.43 & 215.30 \\
\bottomrule
\end{tabular}
\end{table}

Both conditional methods run on the order of one second: $0.85$ to
$1.42$~seconds for the two-moment closure against $1.06$ to $1.54$ for the
three-moment closure. The reference takes about $197$ to $224$~seconds. This
cost difference reflects the two-dimensional density propagated by the
reference, compared with one-dimensional solves and quadrature for the
conditional methods.

The third-moment extension adds approximately $0.2$ seconds, consistent with
the shared-operator structure described in Remark~\ref{rem:shifted-gamma-cost}. Cost is
also nearly independent of leverage, the spread across $\beta\in\{-2,-1,2,3\}$
being a few percent within each index and method, since $\beta$ enters only
through the payoff mapping and the drag coefficient $\gamma$ and not through any
grid or the number of solves.  The joint-law cost varies little between the two
indices for the same reason, its grid being fixed by Table~\ref{tab:grids}
rather than by the calibrated surface.

%------------------------------------------------------------------------------
\subsection{Discussion}
%------------------------------------------------------------------------------

Grid refinement changes the baseline joint-law smiles only marginally, and the
remaining differences from Monte Carlo are concentrated in the wings.  Against
this benchmark, the Gamma fit is consistently more accurate than the strike map
for both markets and all leverage ratios tested.

The shifted fit has little visible effect for the long products.  For the
inverse products, its benefit grows with maturity and removes most of the
residual Gamma error beyond the first few months.  Computing the additional
moment adds about 0.2 seconds to runs of order one second.  On listed TQQQ
options the conditional-Gamma and joint-law smiles are indistinguishable at
market resolution and lie inside the bid--ask at most strikes.  The conditional
calculations are fast enough for repeated smile construction, whereas the
joint-law solve, at about 200 seconds, is better reserved for validation and
stress tests.  The strike map remains useful as a quick diagnostic or initial
guess.

%==============================================================================
\section{Conclusion}\label{sec:conclusion}
%==============================================================================

Options on leveraged ETFs are priced and hedged off the LETF volatility
surface, and path-dependent variance drag makes that surface hard to build
consistently with the underlying smile. Starting from the exact identity
$R_L=\beta R_S-\gamma I_T$, we compared a first-order strike remapping, a
conditional-Gamma method and a two-dimensional joint-law benchmark.

The conditional-Gamma method is the main contribution. Conditioning on the
terminal underlying return keeps the state dependence of variance drag that a
single remapped strike discards. The conditional mean and variance of $I_T$
come from one triangular system of one-dimensional forward PDEs, and the
moment-matched Gamma law keeps realised variance non-negative while pricing
through incomplete gamma functions. The method thus supplies the
strike-dependent convexity correction that the first-order map misses, without
a two-dimensional solve.

The joint-law benchmark was itself validated by grid refinement and by an
independent Monte Carlo simulation of the LETF dynamics. Against it, the
conditional-Gamma smile stays close at every leverage tested on SPX and NKY,
including the wings where the strike remapping drifts. The conditional-Gamma
method is also 158 to 236 times faster than the benchmark, producing a full
smile in about one second, and on TQQQ it lies inside the quoted bid--ask at
most strikes. A shifted-Gamma variant that also matches the conditional third
moment gives its clearest gains for inverse products at longer maturities, and
the Gamma-mixture extension of Appendix~\ref{app:gamma-mixture} covers the
full range of conditional skewness compatible with non-negative realised
variance. The strike remapping remains a useful closed-form check near the
money.

\section*{Disclaimer}
{\small
Opinions and estimates constitute our judgement as of the date of this Material, are for informational purposes only and are subject to change without notice. This Material is not the product of J.P. Morgan's Research Department and therefore, has not been prepared in accordance with legal requirements to promote the independence of research, including but not limited to, the prohibition on the dealing ahead of the dissemination of investment research. This Material is not intended as research, a recommendation, advice, offer or solicitation for the purchase or sale of any financial product or service, or to be used in any way for evaluating the merits of participating in any transaction. It is not a research report and is not intended as such. Past performance is not indicative of future results. Please consult your own advisors regarding legal, tax, accounting or any other aspects including suitability implications for your particular circumstances. J.P. Morgan disclaims any responsibility or liability whatsoever for the quality, accuracy or completeness of the information herein, and for any reliance on, or use of this material in any way.
\par}

\appendix

%==============================================================================
\section{Derivation of the First-Order Strike Remapping}\label{app:first-order-proof}
%==============================================================================

This appendix derives Theorem~\ref{thm:first-LV} by expanding around the flat Black reference and reducing the first-order correction to a Brownian-bridge average of local variance.

The argument combines two standard results on the mapping from local to implied volatility. Bergomi's \emph{Stochastic Volatility Modeling} \cite[Ch.~2]{bergomi_smiledynamics} writes implied variance exactly as a ratio of time-integrated, discounted dollar-gamma-weighted expectations of local variance and expands it to first order around a flat surface, which is where the cancellation of the weighting-kernel perturbation comes from. Gatheral's \emph{The Volatility Surface} \cite[Ch.~3]{gatheral2006} writes implied variance as a time average of per-time gamma-weighted expectations of local variance and approximates it by the local variance along the most probable path from spot to strike. The two forms agree at first order because the expected dollar gamma under the flat model does not depend on $t$ (see the denominator calculation below), and the resulting Brownian bridge is centred on Gatheral's most probable path. Our contribution here is the application to the LETF: the leverage power transformation $l\mapsto F_t^S(l/(F_t^L V_t^*))^{1/\beta}$ together with the variance-drag factor $V_t^*$ carries the underlying-smile bridge into the LETF bridge and produces the strike remap $K\mapsto K^*$.

\subsection{Bergomi's Exact Implied Variance Representation}

Let $\Gamma_{\hat\sigma_L}(t,l)$ be the gamma of the Black price with volatility $\hat\sigma_L(T,K)$, strike $K$, maturity $T$, and the same deterministic discount and carry as $L$. The delta-hedging P\&L identity gives the exact representation \cite[Ch.~2]{bergomi_smiledynamics}
\begin{equation}\label{eq:bergomi-exact}
\hat{\sigma}_L^2(T,K)
=\frac{\E\!\left[\displaystyle\int_0^T
D_tL_t^2\Gamma_{\hat\sigma_L}(t,L_t)b(t,L_t,V_t)\,dt\right]}
{\E\!\left[\displaystyle\int_0^T
D_tL_t^2\Gamma_{\hat\sigma_L}(t,L_t)\,dt\right]},
\end{equation}
where $b(t,L_t,V_t)$ is the instantaneous LETF variance from \eqref{eq:induced-LV}.

\subsection{First-Order Expansion}

\begin{proposition}[Key cancellation]\label{prop:cancel}
For the perturbation family of Section~\ref{sec:first-LV}, changes in the path law and in the Black gamma affect the implied variance only at second order when multiplied by the explicit $O(\varepsilon)$ variance perturbation. Consequently,
\begin{equation}\label{eq:first-order-correction}
\hat\sigma_L^2(T,K)
=b_0+\frac{\varepsilon}{\mathcal D_0}
\E_0\!\left[\int_0^T
D_tL_t^2\Gamma_{\sigma_0}(t,L_t)b_1(t,L_t,V_t^*)\,dt\right]
+O(\varepsilon^2),
\end{equation}
where
\begin{equation}\label{eq:D0-definition}
\mathcal D_0=\E_0\!\left[\int_0^T
D_tL_t^2\Gamma_{\sigma_0}(t,L_t)\,dt\right]
\end{equation}
and $\E_0$ denotes expectation under the flat base model.
\end{proposition}

\begin{proof}
Subtract $b_0$ from both sides of \eqref{eq:bergomi-exact}. Since $b_\varepsilon-b_0=\varepsilon b_1$, one obtains the exact identity
\begin{equation}
\hat\sigma_L^2-b_0
=\varepsilon\,
\frac{\E_\varepsilon\!\left[\int_0^T
D_tL_t^2\Gamma_{\hat\sigma_L}(t,L_t)b_1(t,L_t,V_t)\,dt\right]}
{\E_\varepsilon\!\left[\int_0^T
D_tL_t^2\Gamma_{\hat\sigma_L}(t,L_t)\,dt\right]}.
\end{equation}
Under a regular perturbation, replacing the path law, gamma, and denominator by their $\varepsilon=0$ values changes this ratio by $O(\varepsilon)$, so the implied variance changes by $O(\varepsilon^2)$. Under the flat base model, integrated variance is $a_0t$ and the drag state is deterministic:
\begin{equation}\label{eq:Vstar}
V_t^* = \exp\!\left\{-\frac{\beta(\beta-1)}{2}\,a_0\,t\right\}.
\end{equation}
Substitution gives \eqref{eq:first-order-correction}.
\end{proof}

\subsection{Denominator: Constant-Expectation Property}

Under base GBM, $D_tL_t^2\Gamma_{\sigma_0}(t,L_t)$ has constant unconditional expectation:
\begin{equation}
\mathcal D_0
=T L_0^2\Gamma_{\sigma_0}(0,L_0)
=T F_T^L D_T\frac{1}{\sqrt{2\pi\sigma_0^2T}}
\exp\!\left\{-\frac{(x_K-\tfrac12\sigma_0^2T)^2}{2\sigma_0^2T}\right\}.
\end{equation}
Consequently, at first order the ratio of time integrals in \eqref{eq:first-order-correction} equals the time average of per-time gamma-weighted averages of $b_1$, which is the form given by Gatheral \cite[Ch.~3]{gatheral2006}.

\subsection{Numerator: Brownian Bridge Representation}

Define the log-deviation $x = \log(L_t/F_t^L)$. Under base GBM:
\[
x \sim \mathcal{N}\!\left(-\tfrac{1}{2}\sigma_0^2 t,\;\sigma_0^2 t\right), \quad \text{density } A_t(x).
\]
In terms of $x$, the dollar gamma is
\[
L_t^2\Gamma_{\sigma_0}(t,L_t)
=F_T^Le^x\frac{D_T}{D_t}B_t(x),
\]
where
\[
B_t(x) = \frac{1}{\sqrt{2\pi\sigma_0^2(T-t)}}\exp\!\left\{-\frac{(x - x_K + \tfrac{1}{2}\sigma_0^2(T-t))^2}{2\sigma_0^2(T-t)}\right\}.
\]

\begin{lemma}[Exponential tilting]
\begin{equation}
e^xA_t(x)=\widetilde A_t(x)
=\frac{1}{\sqrt{2\pi\sigma_0^2t}}
\exp\!\left\{-\frac{(x-\tfrac12\sigma_0^2t)^2}{2\sigma_0^2t}\right\}.
\end{equation}
The exponential tilting shifts the mean from $-\tfrac{1}{2}\sigma_0^2 t$ to $+\tfrac{1}{2}\sigma_0^2 t$ with unit prefactor (by the MGF of the Gaussian).
\end{lemma}

\begin{lemma}[Product of Gaussians gives Brownian bridge kernel]
The product $\widetilde A_t(x)B_t(x)$ equals $C\phi_{\bar v}(x-\bar x)$, where
\begin{align}
\bar{v} &= \sigma_0^2\,\frac{t(T-t)}{T}, \label{eq:bridge-var}\\
\bar{x} &= \frac{t}{T}\,x_K, \label{eq:bridge-mean}\\
C &= \frac{1}{\sqrt{2\pi\sigma_0^2 T}}\exp\!\left\{-\frac{(x_K - \tfrac{1}{2}\sigma_0^2 T)^2}{2\sigma_0^2 T}\right\}. \label{eq:bridge-C}
\end{align}
This is the distribution of a Brownian bridge (diffusion $\sigma_0$) at time $t$, conditioned on starting at $0$ and reaching $x_K$ at time $T$.
\end{lemma}

\begin{proof}
Standard product-of-Gaussians identity. With $\mu_1 = \tfrac{1}{2}\sigma_0^2 t$, $v_1 = \sigma_0^2 t$, $\mu_2 = x_K - \tfrac{1}{2}\sigma_0^2(T-t)$, $v_2 = \sigma_0^2(T-t)$:
\[
\bar{v} = \left(\frac{1}{v_1}+\frac{1}{v_2}\right)^{-1} = \sigma_0^2\frac{t(T-t)}{T},
\]
\[
\bar{x} = \bar{v}\left(\frac{\mu_1}{v_1}+\frac{\mu_2}{v_2}\right) = \sigma_0^2\frac{t(T-t)}{T}\cdot\frac{x_K}{\sigma_0^2(T-t)} = \frac{t}{T}x_K.
\]
The prefactor $C$ is the ``overlap integral'': $\mu_1 - \mu_2 = \tfrac{1}{2}\sigma_0^2 T - x_K$, $v_1+v_2 = \sigma_0^2 T$.
\end{proof}

\subsection{Combining: Bridge Average of Local Variance}

Since $\mathcal D_0=TF_T^LD_TC$, we obtain
\begin{equation}\label{eq:bridge-avg}
\hat\sigma_L^2(T,K)
=\frac{1}{T}\int_0^T\!\int_{-\infty}^{\infty}
\phi(y)b_\varepsilon\!\left(
t,F_t^L\exp\!\left\{\frac{t}{T}x_K
+\sigma_0\sqrt{\frac{t(T-t)}{T}}\,y\right\},V_t^*
\right)dy\,dt
+O(\varepsilon^2),
\end{equation}
where $y \sim \mathcal{N}(0,1)$ via the substitution $x = \bar{x} + \sqrt{\bar{v}}\,y$.

\subsection{Path Mapping to Underlying}

Substituting \eqref{eq:induced-LV} into \eqref{eq:bridge-avg} and using the spot inversion:

\begin{align}
F_t^S\left(\frac{l}{F_t^L V_t^*}\right)^{1/\beta} &= F_t^S\exp\!\left\{\frac{1}{\beta}\left(\frac{t}{T}x_K + \sigma_0\sqrt{\frac{t(T-t)}{T}}\,y\right) - \frac{1}{\beta}\log V_T^*\cdot\frac{t}{T}\right\} \nonumber\\
&=F_t^S\exp\!\left\{
\frac{t}{T}\frac{1}{\beta}\log\!\left(\frac{K}{F_T^LV_T^*}\right)
+\sqrt{a_0}\sqrt{\frac{t(T-t)}{T}}\,y
\right\},
\end{align}
where for $\beta<0$ we have replaced $y$ by $-y$, which leaves the symmetric Gaussian integral unchanged.

\subsection{The Remapped Strike}

\begin{definition}[Remapped strike]
\begin{equation}\label{eq:Kstar}
K^* = F_T^S\left(\frac{K}{F_T^L\,V_T^*}\right)^{1/\beta}, \qquad x_{K^*}^S = \frac{1}{\beta}\log\!\left(\frac{K}{F_T^L\,V_T^*}\right).
\end{equation}
\end{definition}

For compactness define the underlying bridge level
\begin{equation}
\mathcal B_t(y;q)=F_t^S\exp\!\left\{
\frac{t}{T}q
+\sqrt{a_0}\sqrt{\frac{t(T-t)}{T}}\,y
\right\}.
\end{equation}
With this notation, the formula becomes
\begin{equation}
\hat{\sigma}_L^2(T,K)
=\frac{\beta^2}{T}\int_0^T\!\int_{-\infty}^{\infty}
\phi(y)\,\sigma_\varepsilon^2\!\left(t,\mathcal B_t(y;x_{K^*}^S)\right)dy\,dt
+O(\varepsilon^2).
\end{equation}

\subsection{Recognition as Underlying Implied Variance}

Bergomi's first-order formula \cite[Ch.~2]{bergomi_smiledynamics} applied to the underlying at strike $K$ gives:
\begin{equation}
\sigma_S^2(T,K)
=\frac{1}{T}\int_0^T\!\int_{-\infty}^{\infty}
\phi(y)\,\sigma_\varepsilon^2\!\left(t,\mathcal B_t(y;x_K^S)\right)dy\,dt
+O(\varepsilon^2).
\end{equation}

Evaluating at $K = K^*$ completes the proof.

%==============================================================================
\section{Gamma-Mixture Extension}\label{app:gamma-mixture}
%==============================================================================

The two-moment Gamma closure matches $m(x)$ and $v(x)$ exactly but, by \eqref{eq:gamma-cv-skew}, forces the conditional skewness to equal twice the conditional coefficient of variation. Section~\ref{sec:shifted-gamma} relaxes that constraint with a shift, but only upwards: the admissibility condition \eqref{eq:shifted-gamma-admissibility} rules the shifted fit out wherever $\lambda_3(x)<2\,\mathrm{cv}(x)$. This appendix gives a closure that matches the same three moments for any conditional skewness compatible with $I_T\geq0$.

We use a mixture of Gamma laws with a common shape parameter. The components preserve non-negative support, so no analogue of \eqref{eq:shifted-gamma-admissibility} is required. Because the conditional call value is linear in the conditional law, pricing remains a weighted sum of the closed-form expressions in Theorem~\ref{thm:conditional-gamma-price}. The construction below matches three moments whenever they arise from a strictly positive conditional law and spans the skewness range compatible with $I_T\geq0$, including negative values.

\subsection{Conditional moments and notation}

The mixture uses the conditional raw moments $r_1,r_2,r_3$ of \eqref{eq:raw-moments-from-M} and the central quantities $m$, $v$, $\mu_3$, $\mathrm{cv}$, $\lambda_3$ of Definition~\ref{def:conditional-moments}, all supplied by the moment system \eqref{eq:moment-pde} carried to $n=3$. No further solve is required beyond the one the shifted closure already needs. Write
\begin{equation}\label{eq:baseline-gamma-shape}
\alpha_\Gamma(x)=\frac{m(x)^2}{v(x)}=\mathrm{cv}(x)^{-2}
\end{equation}
for the shape of the baseline two-moment closure, at which \eqref{eq:gamma-cv-skew} pins $\lambda_3=2\,\mathrm{cv}$. The gap between $\lambda_3(x)$ and $2\,\mathrm{cv}(x)$ measures the shape restriction imposed by that closure, and is what the mixture removes in either direction.

\subsection{Closure and pricing formula}

\begin{definition}[Gamma-mixture closure]\label{def:gamma-mixture}
Fix $N\geq1$. Approximate
\begin{equation}\label{eq:gamma-mixture-law}
I_T\mid X_T=x\ \approx\ \sum_{j=1}^{N}w_j(x)\,
\operatorname{Gamma}\big(\alpha(x),\vartheta_j(x)\big),
\qquad
w_j(x)>0,\quad\sum_{j=1}^{N}w_j(x)=1,
\end{equation}
meaning that the conditional density is the corresponding convex combination of Gamma densities, all carrying the common shape $\alpha(x)$ and differing only in scale.
\end{definition}

Write $\mathcal C_{\Gamma}(x;K;\alpha,\vartheta)$ for the conditional value \eqref{eq:gamma-conditional-call} with the Gamma parameters made explicit.

\begin{proposition}[Mixture price is a weighted average]\label{prop:gamma-mixture-price}
Under \eqref{eq:gamma-mixture-law} with $\gamma>0$,
\begin{equation}\label{eq:gamma-mixture-conditional-call}
\mathcal C_{\mathrm{mix}}(x;K)
=\sum_{j=1}^{N}w_j(x)\,
\mathcal C_{\Gamma}\big(x;K;\alpha(x),\vartheta_j(x)\big),
\end{equation}
and the LETF call price is the same one-dimensional integral
\begin{equation}\label{eq:gamma-mixture-price}
C_L^{\mathrm{mix}}(K,T)
=D_T\int_{\R}\mathcal C_{\mathrm{mix}}(x;K)\,p_T(x)\,dx.
\end{equation}
\end{proposition}

\begin{proof}
For $\gamma>0$ the conditional value is
$F_T^Le^{\beta x}\E[e^{-\gamma I_T}\one_{\{I_T<h(x)\}}\mid X_T=x]-K\,\Q(I_T<h(x)\mid X_T=x)$
with $h(x)$ from \eqref{eq:gamma-boundary}. Both the truncated Laplace transform and the distribution function are linear functionals of the conditional law, so under \eqref{eq:gamma-mixture-law} each equals the $w$-weighted sum of its Gamma-component values \eqref{eq:gamma-cdf}--\eqref{eq:gamma-truncated-laplace}. The exercise boundary $h(x)$ does not depend on the closure, so the same weights apply to both terms, giving \eqref{eq:gamma-mixture-conditional-call}.
\end{proof}

The mixture reuses the incomplete-gamma evaluations of Theorem~\ref{thm:conditional-gamma-price}, $N$ times per node. Martingale normalisation is unchanged in form,
\begin{equation}\label{eq:gamma-mixture-normaliser}
c_T^{\mathrm{mix}}
=\log\int_{\R}e^{\beta x}
\sum_{j=1}^{N}w_j(x)\big(1+\gamma\vartheta_j(x)\big)^{-\alpha(x)}
p_T(x)\,dx,
\end{equation}
followed by the substitutions \eqref{eq:gamma-normalised-substitution} applied identically to every component.

\subsection{Common shape and the product representation}

A general $N$-component Gamma mixture has $3N-1$ free parameters, so three moments do not identify it once $N\geq2$. Definition~\ref{def:gamma-mixture} therefore imposes a common shape. Writing $\vartheta_j=z_j/\alpha$, so that component $j$ has mean $z_j$, and using that $\operatorname{Gamma}(\alpha,z/\alpha)$ is the law of $zY$ with $Y\sim\operatorname{Gamma}(\alpha,1/\alpha)$, the common-shape mixture is exactly the law of a product,
\begin{equation}\label{eq:mixture-product-representation}
I_T\mid X_T=x\ \approx\ M\,Y,
\qquad M\perp\!\!\!\perp Y,
\qquad
\Q(M=z_j)=w_j,
\qquad
Y\sim\operatorname{Gamma}(\alpha,1/\alpha),
\end{equation}
with $\E[Y]=1$ and $\Var(Y)=1/\alpha$. The discrete factor $M$ is a regime level for realised variance (paths reaching the same terminal return $x$ may have travelled through high- or low-variance regions), and $Y$ is an independent normalised dispersion multiplier acting within a regime.

The representation makes the moment equations separable. Since
\begin{equation}\label{eq:Y-moments}
\E[Y^n]=e_n(\alpha),
\qquad
e_n(\alpha)=\frac{\Gamma(\alpha+n)}{\alpha^n\,\Gamma(\alpha)}
=\prod_{r=1}^{n-1}\Big(1+\frac{r}{\alpha}\Big),
\qquad e_1=1,
\end{equation}
independence gives $r_n=\nu_n\,e_n(\alpha)$ with $\nu_n=\E[M^n]$, so for a trial $\alpha$ the moments of the discrete factor are known explicitly:
\begin{equation}\label{eq:nu-from-r}
\nu_n(\alpha)=\frac{r_n}{e_n(\alpha)},
\qquad n=1,2,3.
\end{equation}
For $N=2$ the free parameters are $\alpha$, $z_1$, $z_2$, $w_1$, four in all. Matching $r_1,r_2,r_3$ pins $(z_1,z_2,w_1)$ in closed form at each trial $\alpha$ by two-point Gauss quadrature, leaving the shape $\alpha$ as the single remaining degree of freedom. The two-moment closure of Section~\ref{sec:conditional-gamma} is the case $N=1$.

\subsection{Two-component fit from three moments}

\begin{proposition}[Two-component common-shape fit]\label{prop:two-component-fit}
Fix $x$ at which $I_T\mid X_T=x$ is almost surely strictly positive with $v(x)>0$, and suppress $x$ from the notation. For $\alpha>0$ let $\nu_1,\nu_2,\nu_3$ be given by \eqref{eq:nu-from-r} and, whenever $D:=\nu_2-\nu_1^2\neq0$, set
\begin{equation}\label{eq:quadrature-bc}
b(\alpha)=\frac{\nu_3-\nu_1\nu_2}{D},
\qquad
c(\alpha)=\nu_1 b(\alpha)-\nu_2=\frac{\nu_1\nu_3-\nu_2^{2}}{D}.
\end{equation}
Let $z_1>z_2$ be the roots of $z^2-b z+c=0$ and put
\begin{equation}\label{eq:quadrature-weights}
w_1=\frac{\nu_1-z_2}{z_1-z_2},
\qquad
w_2=1-w_1,
\qquad
\vartheta_j=\frac{z_j}{\alpha}.
\end{equation}
Define
\begin{equation}\label{eq:mixture-A}
A=m\,r_3-r_2^{2},
\qquad
\alpha_c=\max\left\{\alpha_\Gamma,\ \frac{2r_2^{2}-m\,r_3}{A}\right\}.
\end{equation}
Then $A>0$ and $\alpha_c<\infty$, and the admissible set
\begin{equation}\label{eq:mixture-admissible-set}
\mathcal S=\big\{\alpha>0:\ z_1>z_2>0\ \text{and}\ w_1,w_2\in(0,1)\big\}
\end{equation}
is exactly the half-line $\mathcal S=(\alpha_c,\infty)$. For every $\alpha\in\mathcal S$ the mixture \eqref{eq:gamma-mixture-law} with parameters $(\alpha,\vartheta_j,w_j)$ is a genuine probability law on $(0,\infty)$ matching $r_1,r_2,r_3$ exactly. In particular a three-moment common-shape two-Gamma fit always exists.
\end{proposition}

\begin{proof}
\emph{Reduction.} Independence in \eqref{eq:mixture-product-representation} and \eqref{eq:Y-moments} makes matching $r_1,r_2,r_3$ equivalent to matching $\E[M^n]=\nu_n$ for $n=1,2,3$, that is to the system $w_1+w_2=1$ and $\sum_jw_jz_j^{\,n}=\nu_n$, $n=1,2,3$, in the four unknowns $z_1,z_2,w_1,w_2$. Since $\pi(z)=(z-z_1)(z-z_2)=z^2-bz+c$ vanishes at both atoms, $\E[\pi(M)]=\E[M\pi(M)]=0$, which reads
\[
\nu_2-b\nu_1+c=0,
\qquad
\nu_3-b\nu_2+c\,\nu_1=0 .
\]
This is linear in $(b,c)$ with determinant $\nu_2-\nu_1^{2}=D$, and Cramer's rule gives \eqref{eq:quadrature-bc}; \eqref{eq:quadrature-weights} is then the pair of remaining conditions $w_1+w_2=1$, $w_1z_1+w_2z_2=\nu_1$. Conversely, given \eqref{eq:quadrature-bc}--\eqref{eq:quadrature-weights}, the recursions $z^2=bz-c$ and $z^3=bz^2-cz$ valid at the roots give $\sum_jw_jz_j^{2}=b\nu_1-c=\nu_2$ and $\sum_jw_jz_j^{3}=b\nu_2-c\nu_1=\nu_3(\nu_2-\nu_1^{2})/D=\nu_3$, so the construction is exact wherever it is admissible.

\emph{Two of the three admissibility requirements are automatic.} Let $\Lambda=\nu_3-3\nu_1\nu_2+2\nu_1^{3}$ be the third central moment of $M$. Substituting $\nu_3=\Lambda+3\nu_1\nu_2-2\nu_1^{3}$ and $\nu_2=D+\nu_1^{2}$ into \eqref{eq:quadrature-bc} gives
\[
b=\frac{\Lambda}{D}+2\nu_1,
\qquad
c=\frac{\nu_1\Lambda}{D}+\nu_1^{2}-D,
\]
so the cross terms cancel in the discriminant,
\begin{equation}\label{eq:mixture-discriminant}
b^{2}-4c=\frac{\Lambda^{2}}{D^{2}}+4D,
\end{equation}
which is strictly positive whenever $D>0$; the roots are then real and distinct. Moreover $\pi(\nu_1)=\nu_1^{2}-b\nu_1+c=\nu_1^{2}-\nu_2=-D<0$, and a monic quadratic is negative only strictly between its roots, so $z_2<\nu_1<z_1$ and hence $w_1,w_2\in(0,1)$ by \eqref{eq:quadrature-weights}. Since $\nu_1=m>0$ this also gives $z_1>0$, and as $z_1z_2=c$ the support is positive if and only if $c>0$, in which case $b=z_1+z_2>0$. Admissibility therefore reduces to the two Stieltjes conditions $D>0$ and $\nu_1\nu_3-\nu_2^{2}>0$.

\emph{Translation into $\alpha$.} Write $\varepsilon=1/\alpha$, so that $e_1=1$, $e_2=1+\varepsilon$, $e_3=(1+\varepsilon)(1+2\varepsilon)$ and $\nu_1=m$. Then
\[
D>0
\iff\frac{r_2}{1+\varepsilon}>m^{2}
\iff v>m^{2}\varepsilon
\iff\alpha>\alpha_\Gamma .
\]
Multiplying $\nu_1\nu_3-\nu_2^{2}>0$ by $(1+\varepsilon)^{2}(1+2\varepsilon)>0$ gives the equivalent condition
\begin{equation}\label{eq:mixture-c-positive}
A+\varepsilon\,(A-r_2^{2})>0 .
\end{equation}
Because $I_T\geq0$, Cauchy--Schwarz applied to $I_T^{2}=I_T^{1/2}\cdot I_T^{3/2}$ gives $r_2^{2}\leq m\,r_3$, with equality only if $I_T$ is constant on $\{I_T>0\}$; strict positivity together with $v>0$ excludes that, so $A>0$. Hence \eqref{eq:mixture-c-positive} holds for every $\varepsilon>0$ when $A\geq r_2^{2}$, and for $\varepsilon<A/(r_2^{2}-A)$, that is $\alpha>(2r_2^{2}-m r_3)/A$, otherwise. Both conditions are lower bounds on $\alpha$, so $\mathcal S=(\alpha_c,\infty)$ with $\alpha_c$ as in \eqref{eq:mixture-A}, which is non-empty since $A>0$.
\end{proof}

\begin{remark}[Skewness range]\label{rem:mixture-skew-range}
In central moments $A=m\mu_3+v(m^{2}-v)$, so the existence condition $A>0$ of Proposition~\ref{prop:two-component-fit} is
\begin{equation}\label{eq:mixture-skew-bound}
\lambda_3(x)>\mathrm{cv}(x)-\frac{1}{\mathrm{cv}(x)} ,
\end{equation}
which is precisely the Cauchy--Schwarz constraint satisfied by every almost surely positive, non-degenerate law. The mixture therefore imposes no shape restriction beyond positivity of $I_T$ itself, in contrast to the two-moment Gamma, which pins $\lambda_3=2\,\mathrm{cv}$ exactly, and to the shifted fit of Section~\ref{sec:shifted-gamma}, which requires $\lambda_3\geq2\,\mathrm{cv}$. When $\mathrm{cv}(x)<1$, the usual regime for conditional realised variance, the bound \eqref{eq:mixture-skew-bound} is negative and steeply so, and negative conditional skewness is represented by a left-loaded discrete factor, most weight on the upper node $z_1$, which offsets the intrinsic right skew of $Y$. As $\lambda_3$ approaches the bound, $A\downarrow0$ and $\alpha_c\to\infty$: strongly left-skewed nodes require a large shape and the fit degenerates towards a pure two-point law.
\end{remark}

\begin{remark}[What the shape controls]\label{rem:mixture-alpha}
Because $\E[Y]=1$ and $\Var(Y)=1/\alpha$, the conditional variance splits as
\begin{equation}\label{eq:variance-decomposition}
v=\underbrace{\vphantom{\Big|}\Var(M)}_{\text{between regimes}}
+\underbrace{\vphantom{\Big|}\frac{\E[M^2]}{\alpha}}_{\text{within regime}},
\end{equation}
so $\alpha$ is exactly the between/within split of conditional realised-variance dispersion, and the three-moment fit leaves it free. The two endpoints of $\mathcal S$ are the two degenerate splits. As $\alpha\downarrow\alpha_c$ the fit collapses onto the two-moment Gamma of Section~\ref{sec:conditional-gamma} when $\alpha_\Gamma$ is the binding bound in \eqref{eq:mixture-A}, all dispersion becoming within-regime, and onto a node at the origin, $z_2\downarrow0$, when the Stieltjes bound binds instead. As $\alpha\to\infty$, $Y$ degenerates to $1$, all dispersion becomes between-regime, and the mixture tends to the purely discrete two-point quadrature of the target law.
\end{remark}

\subsection{Choosing the shape in practice}

Relative to the two-moment closure, the two-component mixture costs one further moment solve, for $M_3$ (the same solve the shifted closure of Section~\ref{sec:shifted-gamma} requires), and, per quadrature node in $x$, the closed-form evaluation \eqref{eq:quadrature-bc}--\eqref{eq:quadrature-weights} at the chosen shape. Pricing then reuses \eqref{eq:gamma-mixture-conditional-call} with two incomplete-gamma evaluations per node instead of one. The only quantity not fixed by the data is $\alpha(x)$.

The admissible range is known in closed form: $\alpha_c(x)$ is computed directly from $r_1,r_2,r_3$ by \eqref{eq:mixture-A}, and every $\alpha>\alpha_c(x)$ is valid. By Remark~\ref{rem:mixture-alpha}, neither end of that range is a good choice. Close to $\alpha_c(x)$ the fit is ill-conditioned: by \eqref{eq:mixture-discriminant} the node separation $z_1-z_2$ either blows up as $D\downarrow0$ or the lower node approaches the origin, and in both cases the fitted parameters are highly sensitive to small errors in $r_3$. At the other end, a very large $\alpha$ makes the within-regime component vanish and approximates a continuous conditional law by two atoms, which is a poor representation of $I_T$ regardless of how many moments it matches. The shape should therefore be chosen in the interior.

A simple default is a fixed multiple of the admissible lower bound,
\begin{equation}\label{eq:alpha-rule}
\alpha(x)=\varsigma\,\alpha_c(x),
\qquad \varsigma\in[1.5,3],
\end{equation}
applied with the same $\varsigma$ at every node and every strike. This keeps the fit away from the endpoint and requires no search. Since $\alpha_c(x)$ is smooth in $x$ wherever the moments are resolved, $\alpha(x)$ is smooth as well, so the integrand of \eqref{eq:gamma-mixture-price} inherits the regularity the outer quadrature needs. A node-by-node choice of $\alpha$ made independently at each $x$ should be avoided for the same reason: it can inject oscillation into an otherwise smooth conditional price. If a single shape for the whole smile is preferred, take $\alpha=\varsigma\sup_x\alpha_c(x)$ over the resolved region, which is admissible everywhere at the cost of over-smoothing the nodes where $\alpha_c$ is small.

The rule \eqref{eq:alpha-rule} can be monitored with two diagnostics. The between-regime share $\Var(M)/v$ from \eqref{eq:variance-decomposition} reports how the chosen shape splits the conditional dispersion; values close to $0$ or $1$ indicate that $\varsigma$ is too small or too large. The weight $w_2$ on the lower node reports conditioning; negligible weight or a lower node close to zero indicates a near-degenerate fit and suggests increasing $\varsigma$. Alternatively, one further solve for $M_4$ allows $\alpha$ to be chosen by matching the conditional fourth moment on $\mathcal S$, although that refinement has no analogous existence guarantee.

The fit is also a diagnostic for the baseline closure. The gap between the exact skewness \eqref{eq:cv-skew-def} and the two-moment Gamma value $2\,\mathrm{cv}(x)$ measures the shape restriction imposed by that closure. The family is not exact: two components with a common shape match three moments and nothing beyond, and matching more requires $N\geq3$ with the corresponding higher solves. The forward joint-law calculation of Section~\ref{sec:joint-law} remains the closure-free reference in the wings.

\begin{figure}[p]
\centering
\panel{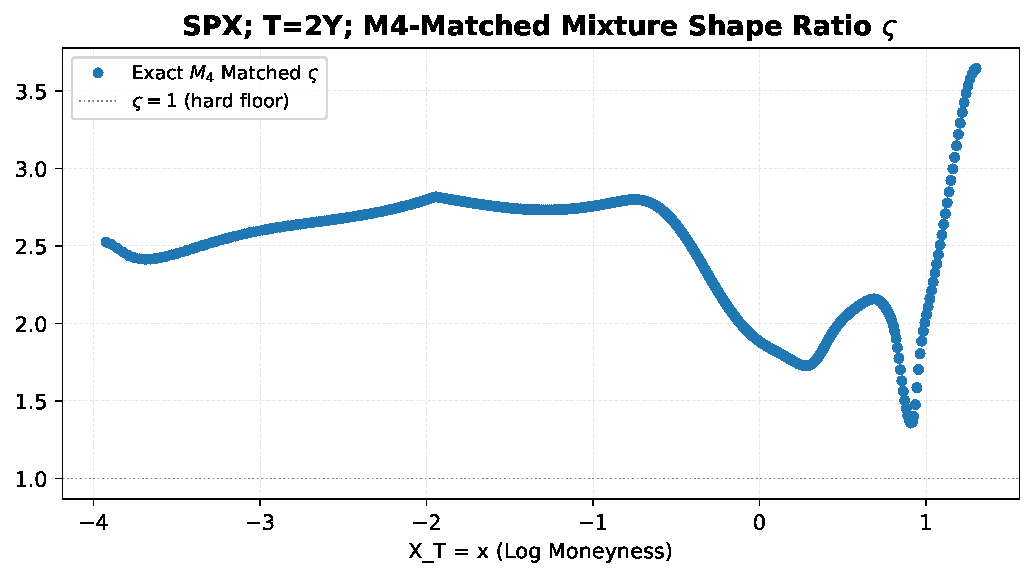}
\caption{Nodewise values of $\varsigma$ obtained by matching the fourth conditional moment for SPX.}
\label{fig:mix-gamma-varsigma}
\end{figure}

For SPX the numerical tests support the rule \eqref{eq:alpha-rule}. Choosing $\varsigma$ node by node to match $M_4$ exactly, while ignoring the requirement $\alpha(x)>\alpha_c(x)$, gives values mostly between 1.5 and 3.5 (Figure~\ref{fig:mix-gamma-varsigma}). The selected $\varsigma$ can, however, fall below 1, so a shape matching $M_4$ is not guaranteed to exist.

\begin{figure}[p]
\centering
\panel{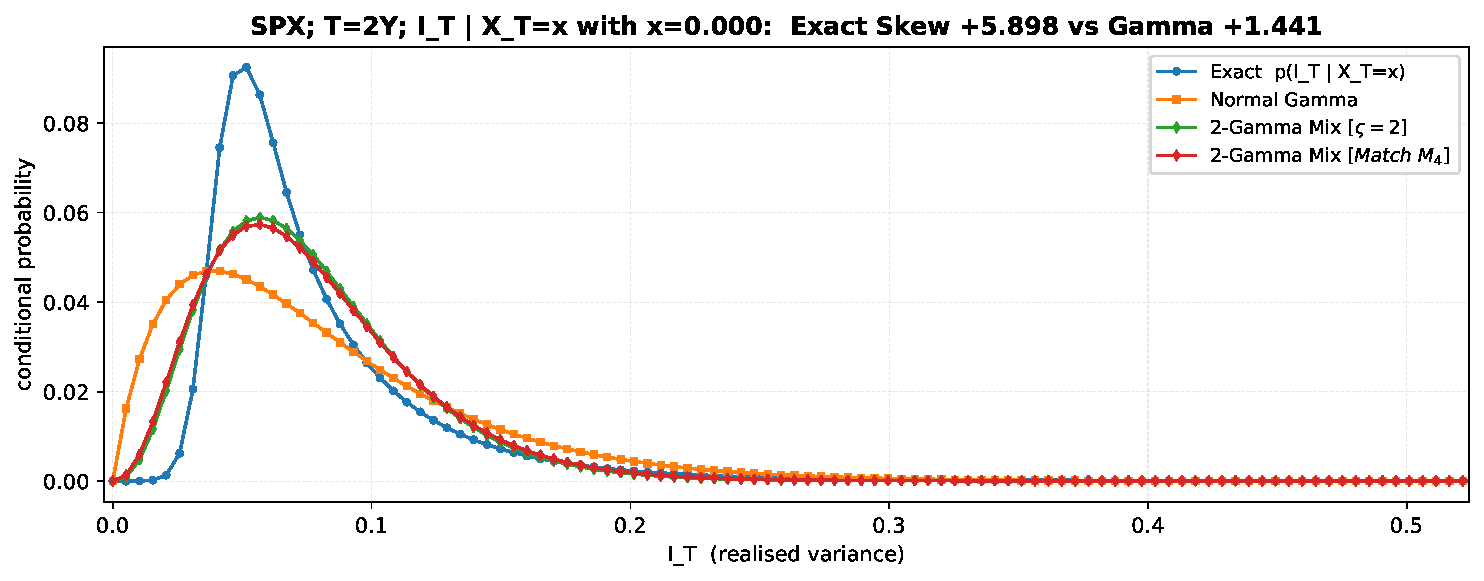}
\caption{Two-component common-shape Gamma-mixture fit at a terminal state with positive conditional skewness.}
\label{fig:mix-gamma-comp1}
\end{figure}

\begin{figure}[p]
\centering
\panel{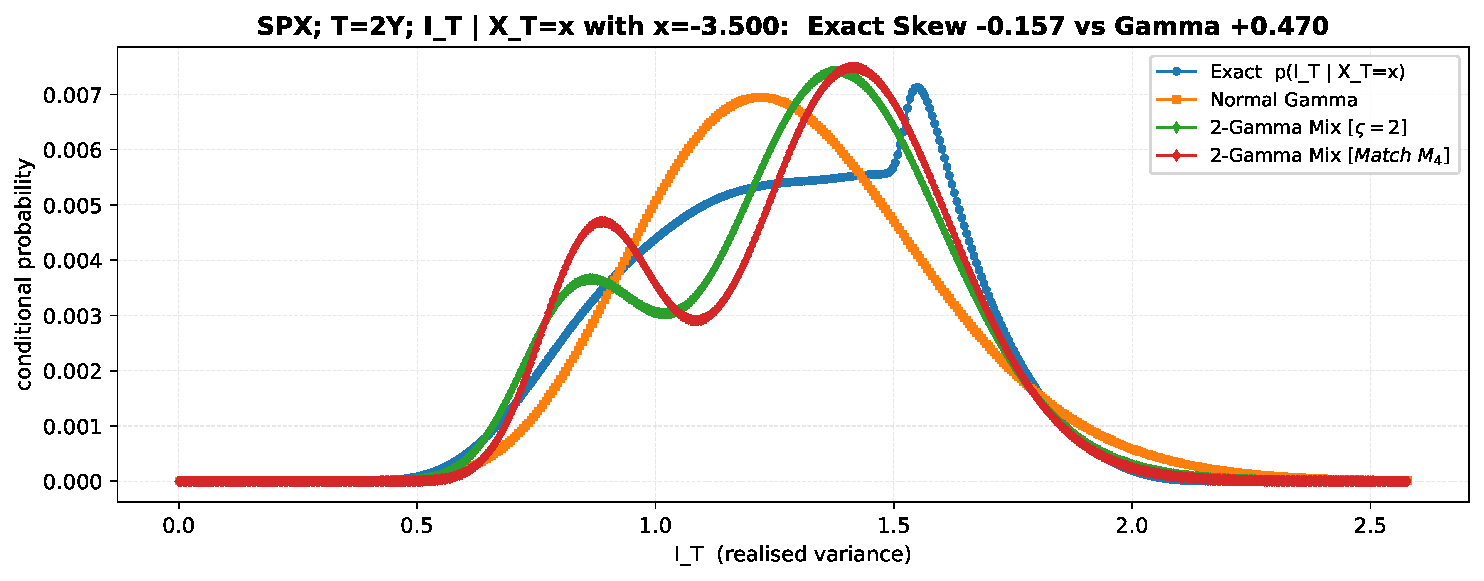}
\caption{Two-component common-shape Gamma-mixture fit at a terminal state with negative conditional skewness.}
\label{fig:mix-gamma-comp2}
\end{figure}

Figures~\ref{fig:mix-gamma-comp1} and~\ref{fig:mix-gamma-comp2} are examples for positive and negative skew, respectively.

\end{document}